\documentclass[12pt, letterpaper]{article}
\usepackage[utf8]{inputenc}

\title{No Screening is More Efficient with Multiple Objects%
\footnote{Authors are listed in alphabetical order. We are grateful to Itai Ashlagi, Yu Awaya, Nima Haghpanah, Shota Ichihashi, Yuichiro Kamada, Michihiro Kandori, Fuhito Kojima, Junpei Komiyama, Yukio Koriyama, Satoshi Kurihara, Fumio Ohtake, Wojciech Olszewski, Parag Pathak, Bruno Strulovici, Satoru Takahashi, Alex Teytelboym, Yuichi Yamamoto, Frank Yang, and all participants of the conference celebrating the 65th birthday of Michihiro Kandori (Tokyo), Summer Workshop on Economic Theory 2024 (Otaru), 7th International Workshop on Matching Under Preferences (Oxford), Japanese Economic Association 2024 Autumn Meeting (Fukuoka), Tokyo Conference on Market Design 2025 (Tokyo), Econometric Society World Congress 2025 (Seoul), Society of Social Choice and Welfare 2026 (Tokyo), and a seminar at Tokyo University of Science for helpful comments. This work has been supported by JST PRESTO Grant Number JPMJPR2368 and JST ERATO Grant Number JPMJER2301, Japan. All remaining errors are our own.}}
\author{Shunya Noda\thanks{Contact Author. Graduate School of Economics, the University of Tokyo, 7-3-1
Hongo, Bunkyo-ku, Tokyo, 113-0033, Japan. E-mail:
\href{mailto:shunya.noda@e.u-tokyo.ac.jp}{shunya.noda@e.u-tokyo.ac.jp}}
\and Genta Okada\thanks{Graduate School of Economics, the University of Tokyo, 7-3-1 Hongo,
Bunkyo-ku, Tokyo, 113-0033, Japan. E-mail: \href{mailto:ogenta0628@g.ecc.u-tokyo.ac.jp}{ogenta0628@g.ecc.u-tokyo.ac.jp}}}
\date{First Draft: August 20, 2024 \hspace{1em} This Version: \today}

\usepackage[utf8]{inputenc}
\usepackage[margin=1in]{geometry}
\usepackage{mathtools}
\usepackage{amsmath}
\usepackage{amsthm}
\usepackage{amssymb}
\usepackage{setspace}
\usepackage{booktabs}
\usepackage{tabularx}
\usepackage{amsmath} 
\usepackage{amssymb}
\usepackage{bm}
\usepackage{ascmac}
\usepackage{setspace}
\usepackage[at]{easylist}
\usepackage[normalem]{ulem}
\usepackage{comment}

\mathtoolsset{showonlyrefs}  

\makeatletter

\usepackage{amsthm}
\usepackage{amsfonts}
\usepackage{bbm}
\usepackage{graphics}
\usepackage{enumerate}
\usepackage{float}
\usepackage{caption}
\usepackage{subcaption}
\usepackage{natbib}
\usepackage{epigraph}
\theoremstyle{definition}
\newtheorem{theorem}{Theorem}
\newtheorem{lemma}{Lemma}
\newtheorem{definition}{Definition}
\newtheorem{proposition}{Proposition}
\newtheorem{corollary}{Corollary}

\theoremstyle{remark}

\newtheorem{remark}{Remark}

\usepackage[hyphens]{url}
\usepackage[colorlinks,urlcolor=blue, citecolor=blue, menucolor=blue]{hyperref}

\theoremstyle{plain}

\providecommand{\keywords}[1]
{
  \small	
  \textbf{Keywords:} #1
}

\usepackage{algorithm,algpseudocode}

\usepackage{amsmath}
\DeclareMathOperator*{\argmax}{arg\,max}

\DeclareMathOperator{\cov}{Cov}
\DeclareMathOperator{\var}{Var}
\DeclareMathOperator{\corr}{Corr}

\newcommand*{\mech}{\mathcal{M}}

\newcommand*{\bx}{\mathbf{x}}
\newcommand*{\by}{\mathbf{y}}
\newcommand*{\bu}{\mathbf{u}}
\newcommand*{\bw}{\mathbf{w}}
\newcommand*{\bp}{\mathbf{p}}
\newcommand*{\bv}{\mathbf{v}}
\newcommand*{\vmc}[1]{\textbf{VMC}(#1)}

\allowdisplaybreaks

\makeatother

\begin{document}
\maketitle

\begin{abstract}
We study the welfare-maximizing allocation of heterogeneous objects when screening uses costly effort rather than monetary transfers.
No-screening mechanisms perform well as object variety increases.
In a symmetric continuous market with i.i.d.\ values whose CDF is log-concave, the multidimensional problem reduces exactly to a single-dimensional problem in agents' best-option values. 
More options make low best-option values rarer, weakening the case for screening. We characterize when no screening is optimal and show it remains optimal as variety expands.
Large-variety limits and numerical results for finite, correlated markets support this pattern. 
We apply these results to propose an invitation-based vaccine appointment system.
\end{abstract} 



\pagebreak

\section{Introduction}
\label{sec: introduction}

Policymakers often face the challenge of allocating scarce resources under
conditions of information asymmetry. The goal is to allocate goods to those
with the highest valuations, but when tractable and socially costless
screening devices such as monetary transfers are unavailable, screening can
become costly. For example, when a scarce resource is released to all eligible
agents at a common time and allocated in the order of successful claims,
agents with high values have an incentive to monitor the release and act
immediately. These costly actions can improve sorting, but the effort is
wasted as an ordeal. Conversely, lottery-based mechanisms prevent wasted
effort but risk misallocating goods. Maximizing social welfare (defined as
\emph{residual surplus}, the agents' total payoffs from the realized allocation
minus wasted screening costs) requires balancing allocative efficiency against
wasted screening costs.

The single-object money-burning literature provides sharp characterizations of
when screening is useful \citep{Hartline2008}. Much less is known when the
planner allocates multiple heterogeneous objects to agents with
multidimensional values. Object heterogeneity can make screening more valuable:
screening may help the planner both select high-value agents and match them to
the objects they value most. But heterogeneity can also make screening less
valuable. When agents can choose among many distinct options, even an
unscreened agent may obtain an object that fits her preferences well. Whether
object variety strengthens or weakens the case for costly screening is
therefore theoretically ambiguous.

A prominent example of multi-object allocation from the COVID-19 pandemic is vaccine appointment scheduling: the downstream problem of deciding which participants within a policy-defined priority group obtain which appointment slots. Eligibility and priority rules are typically coarse, leaving many participants with the same formal priority for receiving a dose. Although vaccine doses are largely homogeneous, appointment slots are not: they differ in date, time, location, and, when relevant, vaccine type. Potential recipients may therefore have diverse preferences over available slots. In such settings, agents' effort influences not only whether they secure an appointment but also which appointment slot they obtain. 
Because the efficient allocation of a large number of heterogeneous goods had
not been sufficiently studied, economists had limited guidance to offer on
vaccine appointment scheduling. In practice, many jurisdictions opened a
common booking pool to all currently eligible recipients when slots were
released, with successful access determining the order in which recipients
chose among the remaining slots. This preserved recipient choice but made
costly response speed a de facto tie-breaker and often generated congestion
and confusion, as documented in
Section~\ref{sec: mass vaccination}.

We ask how object variety changes the residual-surplus tradeoff between screening and no screening. Agents have
multidimensional preferences over objects but demand at most one item. They
can signal these preferences through costly effort, which is burned rather
than transferred. The planner aims to generate high allocative value
while avoiding wasteful screening costs. Two prominent candidates are \emph{serial dictatorship with exogenous order} (SD), where agents sequentially choose their most preferred available goods without paying screening costs, and the \emph{Vickrey--Clarke--Groves mechanism} (VCG), which achieves allocative efficiency at the expense of high screening costs. Optimal mechanisms may lie between these extremes.

Our central finding is that object variety weakens the case for costly
screening. When an agent chooses her
preferred object from a set of heterogeneous options, the value relevant for
allocation is often the maximum of her object-specific values, and hence an
upper order statistic. An additional option raises this maximum only if its
value exceeds her current best, so the expected increase in the maximum is
larger when the current best is low. As the number of options increases, low
realizations of the maximum therefore become disproportionately less common,
which tends to reduce the additional surplus that costly screening can
generate by sorting agents according to their best-option values.
Consequently, when many heterogeneous objects are allocated, no-screening
mechanisms such as SD tend to outperform screening mechanisms such as VCG.

\begin{figure}[t!]
    \centering
    \begin{subfigure}[t]{0.48\textwidth}
        \centering
        \includegraphics[width=\textwidth]{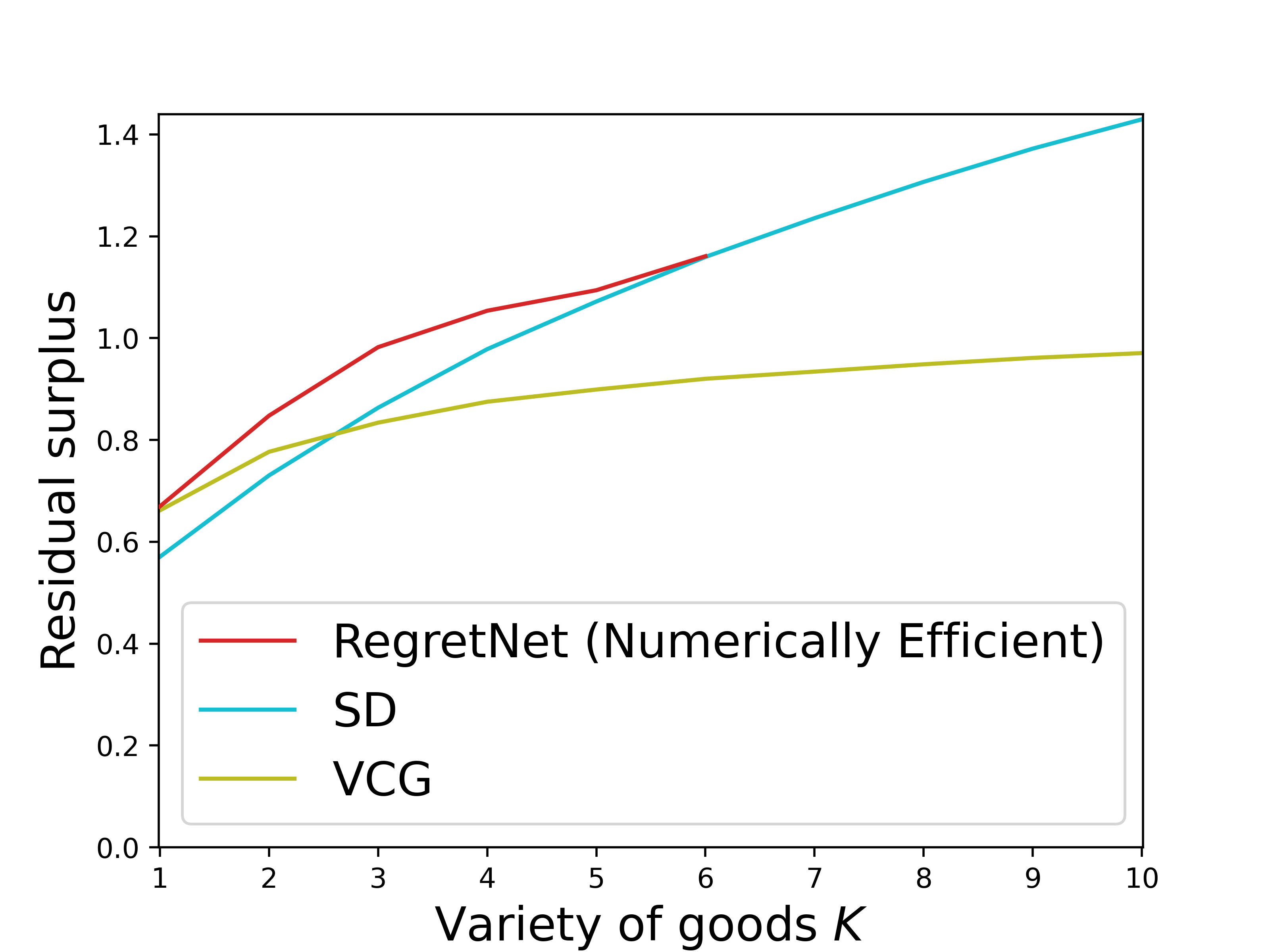}
        \subcaption{Finite Market (Weibull with Parameter $0.8$)}
        \label{fig: intro finite weibull08}
    \end{subfigure}
    \hspace{0.02\textwidth}
    \begin{subfigure}[t]{0.48\textwidth}
        \centering
        \includegraphics[width=\textwidth]{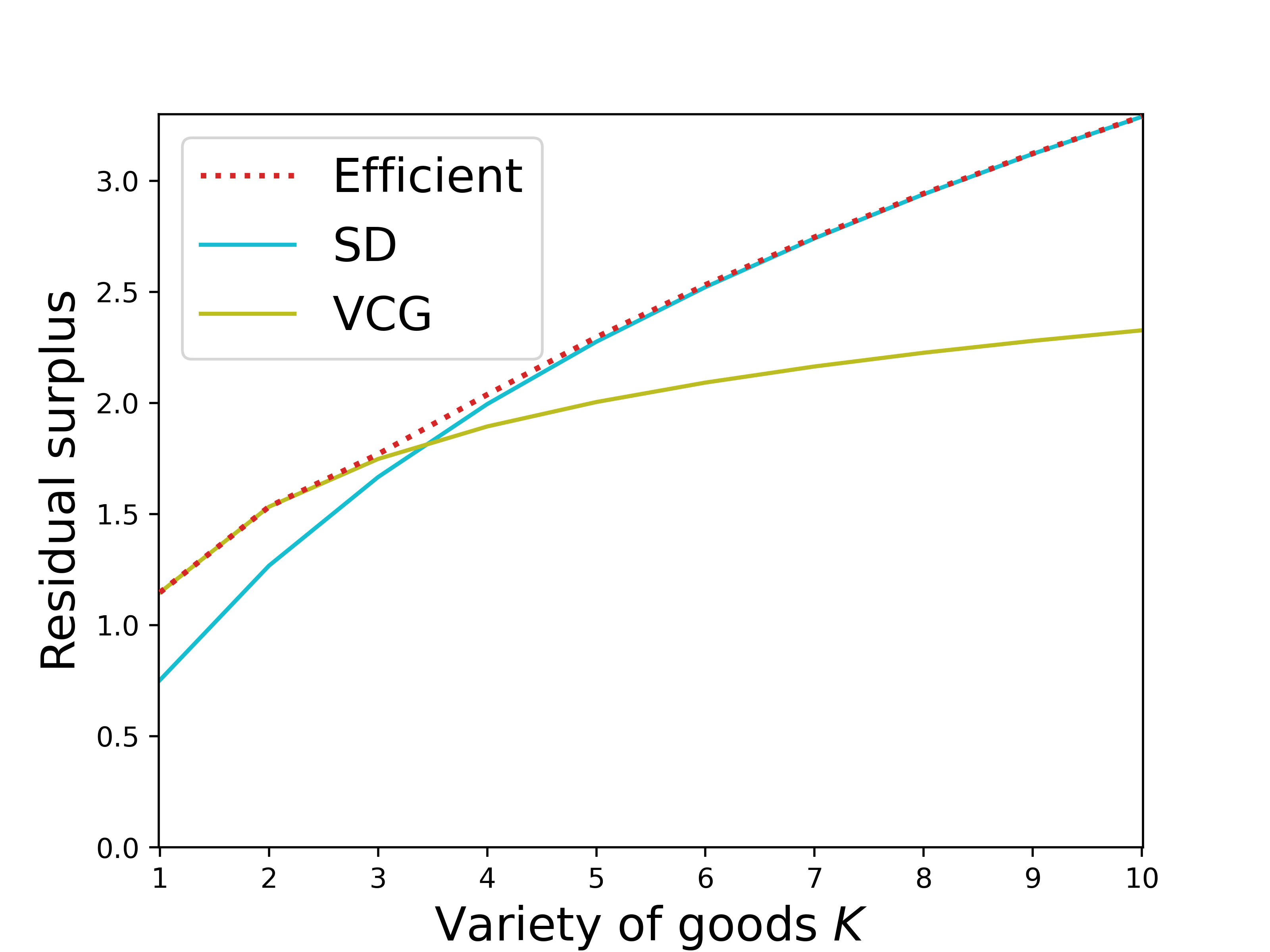}
        \subcaption{Continuous Market (Weibull with Parameter $0.6$)}
        \label{fig: intro continuous weibull06}
    \end{subfigure}
    \caption{The relationship between the object variety and the per-capita residual surplus achieved by SD, VCG, and the efficient mechanism (numerically computed in panel (a))}
    \label{fig: intro performance}
\end{figure}

Figure~\ref{fig: intro finite weibull08} illustrates this result. We consider a scenario where $2K$ agents each demand a single unit among $K$ heterogeneous objects, and each agent $i$'s value for object $k$, $v_k^i$, is drawn i.i.d.\ from a Weibull distribution with shape parameter 0.8. The horizontal axis of the figure represents the value of $K$, and the vertical axis shows the residual surplus per agent. Under this distribution, when $K = 1$ (i.e., the single-good case), the optimality of VCG has been proven by \citet{Hartline2008}. However, for $K \ge 3$, SD outperforms VCG, and when $K \ge 6$, the performance gap between SD and the mechanism learned by RegretNet \citep{dutting2019optimal}, an automated mechanism design method for discovering numerically efficient mechanisms, becomes negligible. As $K$ increases further, the performance of SD continues to improve, whereas the performance gains of VCG taper off.

Our first contribution is an exact reduction of the unrestricted multidimensional problem. We begin with a stylized large market with a continuum of agents and objects, called a \emph{continuous i.i.d.\ market}. Agents consume at most one object, objects are divided into $K < \infty$ types with equal capacities, and each agent's values for the $K$ types are independent draws from a common marginal distribution $G$. We prove that if $G$ is \emph{CDF log-concave}, i.e., $\log G$ is concave, then for every
$K$ and every aggregate capacity, the supremum of residual surplus over all
feasible multidimensional mechanisms is equal to the value of a reduced
single-dimensional problem. In the reduced problem, the resource being allocated is the right to receive one of the agent's favorite objects, and an agent's type is the value for her favorite object.
Equivalently, no unrestricted multidimensional mechanism can
outperform one that depends on the valuation vector only through its maximum
and, whenever it allocates, assigns a favorite object.
Without CDF log-concavity, an additional multidimensional screening
channel can arise. In our counterexample, the balance of an agent's valuation
vector is informative about her maximum value. A menu can exploit this signal
by combining a larger total allocation probability with the possibility of a
non-favorite assignment, thereby strictly increasing residual surplus.
Alternatively, the same reduction can be recovered without CDF
log-concavity under additional normative restrictions on the mechanism class.

Our second contribution is to use this reduction to identify how object variety changes optimal screening. The maximum of $K$ independent draws from $G$ has
distribution $G_K(v)=G(v)^K$.
Hence, once the exact reduction is established, increasing object variety
changes the mechanism-design problem by replacing $G_K$ with $G_{K+1}$. We
characterize when this transformation favors no screening. In the reduced
problem, a no-screening mechanism is
optimal if and only if $G_K$ satisfies the \emph{new better than used in
expectation} (NBUE) property. We then show that NBUE is preserved under taking
a higher-order maximum: if $G_K$ is NBUE, then $G_{K+1}$ is also NBUE.
Consequently, optimality of no screening at one capacity implies its
optimality at every capacity, both at the original variety and at
every larger variety.
Conversely, increasing object variety makes the decreasing-hazard-rate condition guaranteeing full-screening optimality progressively harder to satisfy.

Figure~\ref{fig: intro continuous weibull06} illustrates the observation for the continuous market. Parallel to the setting of Figure~\ref{fig: intro finite weibull08}, there are $K$ types of objects with equal mass and a mass of agents twice the total mass of objects. The type distribution is the product of $K$ independent Weibull marginal distributions with parameter $0.6$. 
We observe that the performance of SD improves as $K$ increases and that the
residual-surplus gap between SD and the exact optimum narrows rapidly over the
range shown. Section~\ref{subsec: continuous numerical analysis} derives
these curves analytically from the reduced single-dimensional problem.

We then use extreme value theory to characterize large-variety limits.
For any marginal value distribution, if the maximum of $K$ independent draws has a nontrivial limit distribution after suitable centering and scaling, the distribution must be either reverse-Weibull, Gumbel, or Fr\'echet. Broad classes of
familiar bounded-support, light-tailed, and power-law-tailed distributions
satisfy this convergence. Our bounded-support result covers the
reverse-Weibull case and more: residual surplus under no screening approaches
the first-best benchmark, whereas that under full screening converges to zero.
If $G$ satisfies the von Mises condition with parameter zero, so that the
normalized maximum converges smoothly to the Gumbel distribution, an efficient
allocation becomes constant on a set of types whose probability approaches
one, and hence pools rather than screens those types. For Fr\'echet limits with
a finite mean, some screening remains useful in the upper tail, but the
efficient mechanism pools the rest of the population. Thus, even when some screening is beneficial, departures from no screening are concentrated among
the highest-value types.

Our third contribution is to examine finite and correlated-value
environments, where the exact reduction is generally unavailable.
We leverage deep-learning techniques for automated mechanism design, which have recently been developed with a primary focus on revenue maximization \citep{dutting2019optimal}, to derive mechanisms that maximize residual surplus. 
The learned mechanisms confirm the comparative-static pattern from the
continuous benchmark: as the number of heterogeneous objects increases, SD
improves relative to both VCG and the numerically optimized mechanism, as depicted in Figure~\ref{fig: intro finite weibull08}.

The correlation analysis clarifies that the relevant notion is
effective variety, not merely the physical number of objects. Strong
within-agent correlation makes an agent's values for different objects move
together and therefore brings the environment closer to a homogeneous-good,
single-dimensional problem. It weakens the advantage of no screening.
Between-agent correlation has a different effect. When agents agree that some
objects are especially valuable, costly screening induces competition over a
common component of value and dissipates that component. In the pure
common-value limit, SD attains the first best while VCG dissipates the entire
surplus. Common popularity is therefore not a reason to allocate
access according to costly response speed; rather, it can make the associated race more wasteful.

Finally, we apply the analysis to the downstream appointment-scheduling layer
of mass vaccination. We use the theory to motivate a
\emph{register-invite-book system} (RIB). Recipients first register, the
authority determines an invitation order using the existing eligibility and
priority policy together with an exogenous tie-breaker, and each invitee then
chooses a preferred slot from the remaining menu. Unlike a common-time release
in which all currently eligible recipients compete to book, RIB uses policy
and an exogenous tie-breaker to determine when a recipient is invited, while
private preferences determine which remaining slot she chooses; costly
response speed determines neither. With sufficiently small invitation
batches, RIB approximates SD while retaining simple menu choice, low
communication costs, and immediate confirmation.

\section{Related Literature}
\label{sec: related literature}

Maximizing residual surplus is closely related to mechanism design with
burned transfers, often called \emph{money burning}. For homogeneous goods,
existing work studies allocation with burned transfers
\citep{Hartline2008,yoon2011optimal,condorelli2012money,
chakravarty2013optimal,adachi2026ascending}. Recent work compares screening
devices and studies equitable screening and the allocation of money through
costly screening
\citep{Akbarpour2023comparison,tokarski2025equitable,
dworczak2025allocate}. Related work without monetary transfers studies random
allocation with cardinal preferences, heterogeneous services, targeting, and
common-value settings
\citep{miralles2012cardinal,Ashlagi2016,tokarski2026targeting,
sato2026common}.

Multidimensional revenue maximization is analytically difficult. The literature has therefore developed structural
characterizations and tractable cases
\citep{mcafee1988multidimensional,pavlov2011optimal,
Bikchandani2022,ghili2023characterization}, algorithmic and duality methods
\citep{cai2012algorithmic,daskalakis2017strong,
giannakopoulos2018duality,cai2019duality}, and robust or worst-case mechanisms
\citep{carroll2017robustness,deb2024multi,che2024robustly}. A more closely
related line identifies conditions under which simple or sharply structured
solutions suffice, including separate sales, favorite-value projections, pure
or nested bundling, rank-preserving reductions, and low-complexity pricing
\citep{manelli2006bundling,Hart2017approximate,alaei2013simple,haghpanah2014virtual,
Haghpanah2020bundling,bikhchandani2024rank,yang2025nested,
cai2015extreme}. \citet{Yang2024costly} provides a related
dimension-reduction result, identifying conditions under which additional
costly screening dimensions can be discarded. Most of these results concern
revenue maximization, where payments benefit the seller; in our problem,
burned payments reduce residual surplus one-for-one.

Parallel difficulties arise under residual- or consumer-surplus objectives.
\citet{tokarski2025damages} shows that degrading goods can provide an
additional screening instrument in a deterministic assignment problem. Existing work obtains approximation or
prior-free guarantees in multi-object and more general environments
\citep{Fotakis2016EfficientMB,ezra2026priorfree}. The closest papers by
objective are \citet{goldner2024simple,ezra2024multi}; both derive
approximation guarantees relative to welfare. Our result instead gives an
exact reduction of the unrestricted problem: 
in a continuum market with equal capacities and i.i.d.\ object values
from a CDF-log-concave distribution, the optimum over all feasible
multidimensional mechanisms equals the optimum of a single-dimensional
problem indexed by the favorite-object value.
Lower
valuation coordinates and non-favorite assignments provide no additional
screening value. To our knowledge, this is the first such exact reduction for
heterogeneous objects under a residual-surplus objective.

Our result adds a residual-surplus counterpart to the
literature on simple mechanisms. The contrast is especially relevant for
implementation. Under a revenue objective, optimal multidimensional mechanisms
can involve complex menus combining bundles and stochastic allocations that depend sensitively on underlying environments 
\citep{thanassoulis2004haggling,manelli2007multidimensional,
pavlov2011optimal}. In some environments, even a continuum of lotteries is
necessary \citep{daskalakis2017strong}. These results concern revenue
maximization, but they illustrate the implementation complexity that can arise
in unrestricted multidimensional design. By contrast, our reduction theorem
and the NBUE characterization of the reduced problem provide conditions under
which SD, a simple and transparent mechanism, is itself optimal among all feasible multidimensional mechanisms.

Indeed, SD is well studied for its desirable strategic and allocative properties.
\citet{Pycia2024random} characterize random SD in sequential form through
Pareto efficiency, symmetry, and obvious strategy-proofness
\citep{li2017obviously}. \citet{bade2015serial} shows robustness when agents can acquire preference information, and \citet{hakimov2023costly} provide empirical support from school choice.
SD also guarantees
at least $1-1/e$ of the first-best matching size
\citep{Krysta2014size,Noda2020size} and prevents scalping
\citep{Hakimov2021appointment}.
Our results add a residual-surplus rationale for using SD when many heterogeneous goods are allocated without monetary transfers.

Finally, research on medical resource allocation during crises studies
production incentives
\citep{Ahuja2021preparing,castillo2021market,Athey2022expanding}, capacity
design \citep{Noda2018large}, priorities
\citep{Gans2022allocation,Akbarpour2023prioritization}, and reserves
\citep{Pathak2022reserve,Pathak2023Vaccine}. We complement this literature by
studying downstream appointment-slot allocation among participants whom
upstream policy treats alike.

\section{Continuous Market}
\label{sec: continuous market}
We first consider a large market where a continuum of agents demands a continuum of goods, which can be classified into a finite variety of types. By imposing a few simplifying assumptions, this market can be reduced to a well-characterized single-dimensional environment, allowing us to analytically derive the efficient mechanism. We demonstrate that as the object variety increases, the relative performance of the no-screening mechanism improves. Then, we characterize the properties of the efficient mechanism in the limit, using the insights from extreme value theory.

\subsection{Model}
\label{sec: model}

There is a unit mass of agents, represented by $I \coloneqq [0, 1]$, each with unit demand. Each object has a type $k \in K$, where $K$ is a finite set of object types. The mass of type-$k$ objects is denoted by $m_{k} \in \mathbb{R}_{+}$. We denote the total mass of objects by $\bar{m}\coloneqq \sum_{k \in K}m_{k}$.

Each agent is specified by their valuation vector, $\bv = (v_{k})_{k \in K}\in V \subset \mathbb{R}_{+}^{K}$, where $v_{k}$ denotes this agent's valuation for a type-$k$ object. For notational simplicity, we omit the index specifying the agent's identity; thus, $\bv$ represents the valuation vector of a single agent, not their profile. Agents can choose a payment (effort cost) level, and the central planner can observe it. Since efforts (ordeals) are used just like payments, we refer to them simply as ``payments,'' whereas we emphasize that all payments are burned without enriching the central planner as revenue. Accordingly, when this agent obtains object $k$ with probability $x _{k}\in [0, 1]$ for each $k$ while making a payment of $p$, her payoff is $\sum_{k \in K}v_{k}x_{k} - p$.

Let $F\in\Delta(V)$ denote the distribution of valuation vectors. Equivalently, $F$ is the limit of empirical type distributions in finite economies in which agents' valuation vectors are drawn independently from a common distribution. We condition on this aggregate distribution and suppress any aggregate state.\footnote{A formal construction of a continuum of i.i.d.\ random variables on the usual Lebesgue agent space requires additional measure-theoretic care. See, e.g., \citet{Sun2006ExactLLN} for a treatment and the corresponding law of large numbers.}

Throughout this section, we consider direct mechanisms that determine an allocation and payment based only on the agent's own report and the distribution of reported types.\footnote{This property is obtained in the continuous limit of a sequence of mechanisms satisfying \citeauthor{liu2016ordinal}'s~(\citeyear{liu2016ordinal}) regularity condition.}
This restriction has two effects. First, we can drop an index representing the agent's identity because an agent's identity does not alter the agent's outcome. Second, by the continuous-limit interpretation, idiosyncratic type uncertainty is averaged out at the aggregate level. Thus, in equilibrium, the reported-type distribution is deterministic. Accordingly, we can omit other agents' type reports, or their realized aggregate distribution, from the arguments of the mechanism. A mechanism $\mech$ consists of an allocation rule $\bx : V \to [0,1]^{K}$ and a payment rule $p: V \to \mathbb{R}_{+}$, where $\bx(\bv)$ and $p(\bv)$ are the respective allocation and payment when the agent has a valuation $\bv$. Our main measure of social welfare, \emph{residual surplus} from a mechanism $\mech = (\bx, p)$, is defined as
\begin{equation}
    \label{eq: defn social welfare}RS(\mech) \coloneqq \int_{V}\left(\sum_{k
    \in K}v_{k}x_{k}(\bv) - p(\bv)\right) dF(\bv).
\end{equation}
In our setting, because payments are burned, the objective function \eqref{eq: defn social welfare} differs from the \emph{gross surplus}, the social welfare considered in a setting with monetary transfers: 
\begin{equation}
    \label{eq: defn allocative welfare}GS(\bx) \coloneqq \int_{V}\sum_{k \in K}
    v_{k}x_{k}(\bv) dF(\bv).
\end{equation}
An allocation rule $\bx$ is \emph{allocatively efficient} if it maximizes the gross surplus.

Below, we list several constraints a mechanism must satisfy. A mechanism $\mech$ satisfies the \emph{resource constraint} if, for each $k$, there is at most a mass $m_{k}$ of agents who receive object $k$:
\begin{equation}
    \label{eq: defn resource}\int_{V}x_{k}(\bv) dF(\bv) \le m_{k}\text{ for all
    }k \in K.
\end{equation}
A mechanism $\mech$ is \emph{strategy-proof} if there is no gain from misreporting:\footnote{Note that \emph{Bayesian incentive compatibility}, which requires a truthful strategy profile to constitute a Bayesian Nash equilibrium, is identical to strategy-proofness in a continuous i.i.d.\ market because there is no aggregate uncertainty. The distinction becomes substantive only in the finite market, where we retain strategy-proofness as a more robust and practical implementation requirement.}
\begin{equation}
    \label{eq: defn strategy proofness}\sum_{k \in K}v_{k}x_{k}(\bv)-p(\bv)\ge
    \sum_{k \in K}v_{k}x_{k}(\bv')-p(\bv') \text{ for all }\bv, \bv' \in V.
\end{equation}
A mechanism $\mech$ is \emph{individually rational} if no agent obtains a negative payoff:
\begin{equation}
    \label{eq: defn individual rationality}\sum_{k \in K}v_{k}x_{k}(\bv)-p(\bv
    )\ge 0 \text{ for all }\bv \in V.
\end{equation}
A mechanism $\mech$ satisfies the \emph{unit demand} condition if an agent obtains at most one object:
\begin{equation}
    \label{eq: defn unit demand}\sum_{k \in K}x_{k}(\bv) \le 1 \text{ for all
    }\bv \in V.
\end{equation}
We call a mechanism \emph{feasible} if it satisfies the resource
constraint, strategy-proofness, individual rationality, and unit demand.
A feasible mechanism is \emph{efficient} if it maximizes residual surplus.



\begin{remark}
For applications with observable eligibility or priority classes, the model
is naturally applied conditional on a class. Upstream policy determines which
agents are considered together and, where relevant, the order in which
classes are served; $\bv$ captures their private values over the objects
available within the resulting allocation problem. Costly screening in the
model is therefore a potential additional ranking device among agents whom
the upstream policy treats alike, rather than a substitute for eligibility or
substantive priority.
\end{remark}

\subsection{Reduction to a Single-Dimensional Environment}\label{subsec: reduction to a single-dimensional environment}

For theoretical analyses, we make several additional simplifying assumptions to enable a closed-form characterization of an efficient mechanism. Specifically, we assume that the set of all possible valuations can be written as $V = [0, \bar{v})^{K}$, where $\bar{v}\in \mathbb{R}_{++}\cup \{+ \infty\}$, and there exists a \emph{marginal value distribution} $G: [0, \bar{v}) \to [0, 1]$ such that the type distribution is the product measure: $F = \bigotimes_{k \in K}G$. Equivalently, writing $F$ and $G$ for the associated cumulative distribution functions, for all $\bv = (v_k)_{k \in  K}$, $F(\bv) = \prod_{k \in K} G(v_k)$. Throughout this paper, we assume that $G$ has a finite mean and is twice continuously differentiable, with $g(v) > 0$ for all $v \in (0, \bar{v})$. The first- and second-order derivatives of $G$ are written as $g$ and $g'$, respectively. We further assume that all object types have the same capacity; i.e., we set $m_{k}= \bar{m}/K$ for all objects $k \in K$. We focus on the case of $\bar{m} \in (0, 1)$ to make the problem non-trivial. We refer to this environment as a \emph{continuous i.i.d.\ market}, which is parametrized by its marginal value distribution $G$, the cardinality of object types, $|K|$, and the capacity parameter $\bar{m}$. Slightly abusing notation, we use $K$ to denote its cardinality $|K|$ unless it causes confusion.\footnote{Appendix~\ref{subsubsec: asymmetric capacity} studies the case of asymmetric capacity.}

These assumptions imply that both the demand for each object, as determined by the value distribution, and the supply of each object, as determined by capacities, are perfectly symmetric. Moreover, because we consider a continuous market without aggregate uncertainty, the aggregate allocation can be made deterministic. To increase gross surplus, whenever an object is assigned to an agent, it is desirable to assign the object that the agent values most highly, and this condition can be satisfied with probability one. Because demand and supply are symmetric, if some agents receive non-favorite objects with positive probability, one can reallocate assignment probabilities across agents to increase every agent's probability of receiving her favorite object while continuing to satisfy the capacity constraints. Accordingly, the class of \emph{favorite-only mechanisms}, which never assign an agent an object other than her favorite, is a natural candidate for residual surplus maximization.

In the following, we provide a sufficient condition for guaranteeing that a favorite-only mechanism is residual-surplus maximizing. This allows us to reduce the environment to a single-dimensional one. The \emph{reduced market} is a single-dimensional environment in which an agent's type is $v= \max_{k \in K}v_k \sim G_K$, where $G_K(v) = (G(v))^K$. We denote its density function as $g_K(v) \coloneqq G_K'(v) = K (G(v))^{K-1}g(v)$. The single resource is the right to receive the agent's favorite object and has total capacity $\bar m$. A reduced direct mechanism is a pair $(x,p)$ with $x:[0,\bar v)\to[0,1]$ and $p:[0,\bar v)\to\mathbb R_+$. Its residual-surplus problem is
\begin{align}
                  & \sup_{(x,p)}
                    \int_0^{\bar v}\bigl(vx(v)-p(v)\bigr)dG_K(v)
                    \label{eq: reduced objective}\\
    \text{s.t. }  & \int_0^{\bar v}x(v)dG_K(v)\le\bar m,
                    \label{eq: reduced resource}\\
                  & vx(v)-p(v)\ge vx(v')-p(v')
                    \quad\text{for all }v,v'\in[0,\bar v),
                    \label{eq: reduced sp}\\
                  & vx(v)-p(v)\ge0
                    \quad\text{for all }v\in[0,\bar v).
                    \label{eq: reduced ir}
\end{align}
The range restriction $x\in[0,1]$ is the reduced unit-demand condition. The value of the reduced problem is unchanged if equality is imposed in \eqref{eq: reduced resource}: whenever $\int x\,dG_K<\bar m$, raising the allocation floor from $x$ to $\max\{x,c\}$ for an appropriate $c$ weakly raises residual surplus and exhausts the resource.

\begin{definition}[CDF Log-Concavity]
    The marginal value distribution $G$ is \emph{CDF log-concave} if $\log G$ is a concave function, with the convention $\log 0 = - \infty$.
\end{definition}

Since we assume $G$ is twice continuously differentiable, CDF log-concavity of $G$ is equivalent to the condition that the reverse hazard rate $g/G$ is nonincreasing. Note that this condition is weaker than the usual \emph{PDF log-concavity}, which requires $\log g$ to be concave. Log-concavity of $g$ implies log-concavity of $G$, but the converse does not hold. This distinction is important for this paper's research question because PDF log-concavity is a sufficient condition for a no-screening mechanism to be efficient in a single-dimensional environment (see Section~\ref{subsubsec: characterizations for single-dimensional environment}). 

CDF log-concavity is satisfied by a broad class
of familiar distributions, including the uniform, power, exponential, Gamma, Weibull, lognormal, Pareto, Lomax, and Fr\'echet families, subject to the finite-mean restrictions imposed elsewhere.  This class is stable under positive location-scale transformations, truncation, convolution, increasing convex transformations of the random variable, and finite products and positive powers of distribution functions.

\begin{theorem}[Reduction to a Single-Dimensional Problem]
\label{thm: exact reduction}
In a continuous i.i.d.\ market, if the marginal value distribution $G$ is CDF log-concave, then for every $K\in\mathbb Z_{++}$ and every $\bar m\in(0,1)$, the supremum of residual surplus in the unrestricted multidimensional problem \eqref{eq: defn social welfare}--\eqref{eq: defn unit demand} is equal to the value of the reduced problem \eqref{eq: reduced objective}--\eqref{eq: reduced ir}.

Equivalently, there is no loss in restricting attention to mechanisms that depend on $\bv$ only through $R(\bv) = \max_{k\in K} v_k$ and, whenever they allocate, allocate an agent's favorite object.
\end{theorem}

A feasible reduced mechanism $(x,p)$ lifts to the original market by assigning the favorite object with probability $x(R(\bv))$ and charging $p(R(\bv))$. Because favorite labels are uniformly distributed, each object type uses exactly a fraction $1/K$ of the aggregate resource. A misreport can direct the allocation only to an object whose true value is at most $R(\bv)$, so the reduced incentive constraints imply the original multidimensional constraints.


The nontrivial direction is proved by decomposing the multidimensional type
space into single-dimensional paths.
Conditional on the favorite value being $v$, each remaining value is
distributed as a draw from $G$ truncated at $v$, and can therefore be written
as $G^{-1}(ZG(v))$ for $Z\sim\mathrm{Unif}[0,1]$. Holding the realizations of
these auxiliary variables fixed amounts to comparing types whose non-favorite
values occupy the same conditional quantile ranks as the favorite value
varies. CDF log-concavity then ensures that every non-favorite value rises
with $v$, but by no more than the increase in $v$. That is, for every $z\in[0,1]$ and
$s<t$,
\[
    0\le G^{-1}(zG(t))-G^{-1}(zG(s))\le t-s.
\]
Consequently, along each path, increasing the favorite value from $s$ to $t$
raises the value of any allocation by at most $t-s$ times its total
allocation probability. Incentive compatibility in the multidimensional environment therefore
implies the upward incentive constraints of a single-dimensional mechanism whose
type is $v$ and whose allocation is the total probability of receiving an
object. The appendix shows that imposing only these upward constraints does
not increase the optimal residual surplus. Bounding each path by the
corresponding single-dimensional problem and then averaging across paths, using
concavity of the reduced value in capacity, gives the desired upper bound.

\subsection{Scope and Alternative Foundations of the Reduction}

CDF log-concavity is not a necessary condition for the exact reduction in
Theorem~\ref{thm: exact reduction}. Some restriction is nevertheless required: there is an example
in which assigning non-favorite objects provides an additional screening
instrument and strictly increases residual surplus.

\begin{theorem}[Non-Favorite Assignments Can Improve Residual Surplus]
\label{thm:nonfavorite-screening}
There exists a continuous i.i.d.\ market with a full-support, twice
continuously differentiable marginal value distribution for which the
supremum of residual surplus over unrestricted multidimensional mechanisms
strictly exceeds the supremum over favorite-only mechanisms.
\end{theorem}

Appendix~\ref{subsec:ex_post_efficiency_is_not_without_loss}
proves Theorem~\ref{thm:nonfavorite-screening} by constructing a counterexample. Its construction relies on a marginal
distribution concentrated near three value levels, with substantially less
mass near the middle one.
The mechanism uses the balance
of an agent's valuation vector as a screening instrument: types whose values
are high for both objects select an option with a larger total allocation
probability that sometimes assigns a non-favorite object. Thus, lower
coordinates of the valuation vector can contain screening information that
is lost in the reduced market. The resulting gain is quantitatively modest
and relies on a subtle screening channel, so its practical significance is
unclear.

There is also an independent normative case
for restricting attention to favorite-object assignments.
A mechanism is \emph{neutral} if relabeling the objects only relabels the allocation
and leaves payments unchanged. An allocation rule is \emph{ex post efficient} if no
feasible reallocation weakly raises the value delivered to every type and
strictly raises it on a set of positive measure.
These are natural normative requirements: in particular, ex post efficiency
rules out deliberately assigning less-preferred objects merely to create an
additional screening instrument, a practice that would also be difficult to
justify in implementation. Together, neutrality and ex post efficiency force
favorite-object assignment almost surely.


\begin{theorem}[Reduction under Neutrality and Ex Post Efficiency]
\label{thm:neutrality-ex-post-reduction}
Consider a continuous i.i.d.\ market.
Among feasible mechanisms that are neutral and ex post efficient, the supremum of residual surplus is equal to the value of the
reduced problem
\eqref{eq: reduced objective}--\eqref{eq: reduced ir}.
\end{theorem}

Appendix~\ref{subsec:neutrality-ex-post-reduction} formalizes this normative route:
within the class of mechanisms satisfying these two requirements, the original
continuous i.i.d.\ problem is exactly equivalent in residual-surplus terms to
the reduced single-dimensional problem, without any CDF-log-concavity
assumption.
Thus, the single-dimensional analysis below remains applicable beyond the
CDF-log-concave class whenever favorite-object assignment is imposed directly
or follows from additional admissibility requirements.



\subsection{Mechanisms}

This section describes how our benchmark mechanisms are represented in the original multidimensional environment and the reduced single-dimensional environment.

\paragraph{No Screening/Serial Dictatorship}

For a reduced market, a \emph{no-screening mechanism} is defined as a mechanism that uniformly assigns the right to obtain the most preferred object with probability $\bar{m} \in (0, 1)$, requiring no payment. That is, $x(v) = \bar{m}$ and $p(v) = 0$ for all $v \in [0, \bar{v})$. The residual surplus achieved under the no-screening mechanism, $\mech_{SD}$, is
\begin{equation}
    RS(\mech_{SD}) = \bar{m}\mathbb{E}_{v \sim G_K}[v] = \bar{m}\int_{0}^{\bar{v}}v dG_{K}(v).
\end{equation}
The revenue equivalence theorem of \citet{myerson1981optimal} implies that the gross surplus $\bar{m}\mathbb{E}[v]$ cannot be improved further without burning payments.

This result is achieved in a continuous i.i.d.\ market if we apply \emph{serial dictatorship (SD)}, in which agents take their most preferred objects sequentially following an exogenous priority order. In SD, the first mass $\bar{m}$ of agents in the priority order, facing all goods available, obtain their favorite goods. Once a mass $\bar m$ of agents has finished choosing, all goods are simultaneously exhausted, and the remaining $1 - \bar{m}$ agents receive nothing.\footnote{Formally, the outcome of SD in a continuous environment is specified by an eating mechanism. While we skip it because this paper does not intensively analyze the general continuous multidimensional environment, interested readers should refer to \citet{Noda2018large}. The asymptotic equivalence of (random) SD and an eating mechanism is established by \citet{che2010asymptotic}.}

Note that, besides SD, various no-screening mechanisms (e.g., the random favorite mechanism discussed in Appendix~\ref{subsubsec: asymmetric capacity}) yield a reduced mechanism equivalent to that induced by SD in a continuous i.i.d.\ market.

\paragraph{Full Screening/VCG}
For a reduced market, a \emph{full-screening mechanism} is defined as a mechanism that identifies agents' values to achieve allocative efficiency, even at the cost of screening. In the reduced market, the allocatively efficient allocation rule is to assign the right to obtain the most preferred object with probability $1$ to the mass $\bar{m}$ of agents with the highest values, while not allocating the good to any other agents; i.e., $x(v) = 1$ if $v \ge q$ and $x(v) = 0$ otherwise, where $q \coloneqq G_K^{-1}(1 - \bar{m})$ is the $(1 - \bar{m})$-quantile of the reduced value distribution $G_{K}$. To implement this allocation rule, the mechanism should charge allocated agents a price of $q$. The residual surplus achieved under VCG, $\mech_{VCG}$, is
\begin{equation}
    RS(\mech_{VCG}) = \bar{m} \mathbb{E}_{v \sim G_K}[v - q| v > q] = \int_{q}^{\bar{v}}\left(v - q\right)dG_{K}(v).
\end{equation}
The revenue equivalence theorem of \citet{myerson1981optimal} implies that to achieve the gross surplus $\bar{m}\mathbb{E}[v|v \ge q]$, each allocated agent needs to burn $q$; thus, the total expected burned payment is $\bar{m} q$.

For general environments, including finite environments and environments with continuous agents and objects but non-i.i.d.\ values, the \emph{Vickrey--Clarke--Groves mechanism} (VCG), which requires each agent to pay their externality, achieves allocative efficiency. Furthermore, \citet{green1977characterization,green1979incentives,holmstrom1979groves} show that, under certain conditions, VCG is the unique direct mechanism that achieves allocative efficiency even in environments with multidimensional types.

Direct mechanisms such as VCG are rarely used literally in money-burning
applications. A practically relevant indirect counterpart is a
first-come-first-served (FCFS) race. Suppose that all eligible agents may act
when allocation opens, higher effort yields an earlier position in the access
order, each agent chooses her most preferred object among those remaining,
and effort is sunk regardless of whether she receives an object. In the
reduced single-dimensional market, this is an all-pay contest for the right
to receive a favorite object. In the standard monotone equilibrium, the
agents with the highest reduced values obtain the available capacity, which
is the full-screening allocation rule. Revenue equivalence then implies the
same expected utilities, and hence the same residual surplus, as VCG.

Thus, within the reduced benchmark, an FCFS race is an
indirect implementation of full screening, whereas SD implements no
screening. This equivalence is specific to the reduced environment: in finite
or unrestricted multidimensional markets, FCFS need not reproduce the VCG
allocation. We therefore use VCG as the canonical full-screening benchmark
rather than claim that every real FCFS system implements VCG.

\subsection{Characterizations for the Reduced Market}
\label{subsubsec: characterizations for single-dimensional environment}

\citet{myerson1981optimal} proves that (i) an allocation rule can constitute a strategy-proof mechanism if and only if $x$ is \emph{monotonic} (i.e., $x(v') \ge x(v)$ for all $v' \ge v$), and (ii) if an allocation is nondecreasing, strategy-proofness uniquely identifies the payment rule (up to a constant). Using this approach, \citet{Hartline2008} show that the expectation of the residual surplus is equal to the expectation of the \emph{virtual valuation for utility} $\vartheta$ under strategy-proofness (see their Lemma~2.6):
\begin{equation}
    \mathbb{E}[v x(v) - p(v)] = \int_{0}^{\bar{v}}\left(v x(v) - p(v)\right)
    dG_{K}(v) = \int_{0}^{\bar{v}}\vartheta(v; G_{K})x(v) dG_{K}(v) = \mathbb{E}
    [\vartheta(v; G_{K})x(v)],
\end{equation}
where
\begin{equation}
    \vartheta(v; G_{K}) \coloneqq \frac{1 - G_{K}(v)}{g_{K}(v)}.
\end{equation}
In the context of revenue maximization, we exploit the relation of $\mathbb{E}[p(v)] = \mathbb{E}[(v - \vartheta(v; G_{K}))x(v)]$ to characterize an optimal mechanism. The value, $v - \vartheta(v; G_{K})$, is defined as the virtual valuation \emph{for payment}. Since this paper's focus is on residual-surplus maximization, we instead refer to $\vartheta(v; G_{K})$ as the virtual valuation. Note that $r(v; G_K) = 1/\vartheta(v; G_K) = g_K(v)/(1- G_K(v))$ is the \emph{hazard rate} function of the distribution $G_K$.

The problem reduces to the maximization of $\mathbb{E}[\vartheta(v)x(v)]$ subject to the resource constraint, individual rationality, the unit demand condition, and monotonicity of $x$. To do so, we apply a procedure called ``ironing'' \citep{myerson1981optimal} to obtain a nondecreasing \emph{ironed virtual value} function $\bar{\vartheta}(v; G_K)$. Then, to derive an efficient mechanism, starting from higher values of $v$, we greedily increase the allocation probabilities $x(v)$ while assigning equal allocation probabilities within intervals where $\bar{\vartheta}(v; G_K)$ is constant, until the resource constraint $\mathbb{E}[x(v)] \le \bar{m}$ becomes binding.

Below, we establish a primitive condition on the reduced value distribution $G_K$ that is necessary and sufficient for a no-screening mechanism to be efficient. Specifically, we prove that the condition is that $G_K$ satisfies the \emph{new better than used in expectation} (NBUE) property. To our knowledge, this paper is the first to characterize the equivalence between NBUE and constancy of the ironed virtual value, thereby establishing conditions for the optimality of no screening.

\begin{definition}
    A distribution function $G$ satisfies the \emph{new better than used in expectation (NBUE)} property if for all $t \in (0, \bar{v})$, we have the following:
    \begin{equation}\label{eq: defn NBUE}
        \mathbb{E}_G[v] \ge \mathbb{E}_G[v - t| v > t]. 
    \end{equation}
\end{definition}

Applied to $G_K$, NBUE compares residual surplus per unit of resource:
$\mathbb E_{G_K}[v]$ under random allocation and
$\mathbb E_{G_K}[v-t\mid v>t]$ under a threshold offer requiring a
burned payment of $t$. It requires that no threshold offer improve
this per-unit return. The terminology comes from survival analysis,
where the inequality compares the expected remaining lifetimes of a
new and a used product.

\begin{theorem}\label{thm: NBUE iff no screening}
    The following are equivalent:
    \begin{enumerate}[(i)]
        \item The reduced value distribution $G_K$ satisfies NBUE.
        \item For every capacity $\bar{m} \in (0, 1)$, a no-screening mechanism with $x(v) \equiv \bar{m}$ is an efficient mechanism.
    \end{enumerate}
    Moreover, if $G_K$ is not NBUE, then a no-screening mechanism is inefficient for every $\bar{m}$.
\end{theorem}


Every nondecreasing allocation rule is a nonnegative mixture of
constant and threshold allocations. NBUE therefore bounds the residual
surplus of every feasible mechanism by that of no screening. Conversely,
if a threshold has a higher per-unit return, a small shift from a constant
allocation toward that threshold raises residual surplus while preserving
resource use. Because $\bar m\in(0,1)$ leaves room for such a shift,
this improvement is available at every capacity.

\begin{remark}
NBUE is implied by the \emph{increasing hazard rate} (IHR) condition,
which requires the hazard rate $r(\cdot; G)$ to be nondecreasing.
Indeed, IHR implies that the mean residual life is nonincreasing
\citep[Corollary~4]{bagnoli2005logconcave}; this yields the NBUE inequality in our setting.
In their Corollary 2.11, \citet{Hartline2008} provide IHR as a tractable
sufficient condition for no screening to be efficient in a finite market.
However, since IHR is stronger than NBUE, IHR is not necessary.
Alternatively, while IHR is equivalent to a nonincreasing virtual value,
this is only a sufficient condition for a constant ironed virtual value,
a property equivalent to NBUE.

PDF log-concavity implies both CDF log-concavity and IHR
\citep[Theorem~1 and Corollary~2]{bagnoli2005logconcave}.
Since the density of $G_K$ is $g_K(v)=K G(v)^{K-1}g(v)$,
the concavity of both $\log G$ and $\log g$ implies that $\log g_K$ is
concave for every $K$. Accordingly, PDF log-concavity of the marginal
distribution $G$ guarantees that a no-screening mechanism is efficient
for all $K \ge 1$.

By contrast, CDF log-concavity does not imply either IHR or NBUE.
The Weibull family provides a counterexample. Its CDF is log-concave for
every $\alpha>0$, whereas, for $\alpha\in(0,1)$, its density is
log-convex, its hazard rate is decreasing, and its mean residual life is
increasing
\citep[Theorem~1 and Section~6.3]
{bagnoli2005logconcave}.
The last property implies that the distribution is not NBUE.
\end{remark}

Conversely, the following theorem demonstrates that the \emph{decreasing hazard rate} (DHR) property of the reduced value distribution $G_K$, that is, its hazard rate $r(\cdot; G_K)$ is nonincreasing, is necessary and sufficient for the efficiency of full-screening mechanisms.

\begin{theorem}\label{thm: DHR iff full screening}
    The following two are equivalent:
    \begin{enumerate}[(i)]
        \item The reduced value distribution $G_K$ has a DHR.
        \item For all $\bar{m} \in (0, 1)$, a full-screening mechanism is efficient.
    \end{enumerate}
\end{theorem}

As previously described, an efficient mechanism is obtained by increasing $x(v)$ from the highest values of $v$. If, at the moment the resource constraint $\mathbb{E}[x(v)] \le \bar{m}$ becomes binding, the current $v$ does not belong to an interval where the ironed virtual value $\bar{\vartheta}(v; G_K)$ is flat, then full screening is efficient. Thus, the absence of such intervals---equivalently, a nondecreasing virtual value---is a sufficient condition for full screening efficiency.\footnote{For a finite market, \citet{Hartline2008} propose this sufficient condition as Corollary 2.12.} This condition corresponds exactly to the distribution $G_K$ satisfying DHR. Conversely, if $G_K$ does not satisfy DHR, there must exist intervals where the ironed virtual value is flat, implying the existence of some $\bar{m}$ for which full screening is inefficient. Note that full screening may still be efficient even when the ironed virtual value is flat in the low-$v$ region. Thus, DHR is not a necessary condition for full screening to be efficient for some fixed $\bar{m}$.

\subsection{Effects from Increasing Variety}

This subsection isolates the distributional channel behind our comparative
static. In the reduced market, object variety enters through the
transformation $G\mapsto G_K=G^K$. The relevant question is therefore how variety changes
the mean residual life, hazard rate, and ironed virtual value of the
largest-order-statistic distribution, the distributional objects that determine the
residual-surplus return to screening.

First, we can show that NBUE becomes easier to satisfy as the variety of objects $K$ increases.

\begin{theorem}\label{thm: NBUE expands}
    Suppose that $G_K$ satisfies NBUE. Then, $G_{K+1}$ also satisfies NBUE.
\end{theorem}

The intuition behind Theorem~\ref{thm: NBUE expands} is that the
conditioning event $v>t$ becomes less selective as $K$ increases. Since
$v=\max\{v_1,\dots,v_K\}$, this event reveals only that at least one of the
$K$ underlying draws exceeds $t$. Equivalently, it rules out the event that
all $K$ draws lie below $t$, whose probability is $G(t)^K$. As $K$ grows,
this excluded event becomes increasingly rare, so conditioning on $v>t$
carries progressively less favorable information about the residual value
$v-t$. In the survival interpretation, adding products makes it more likely
that at least one remains functioning at $t$, without implying that the other products have survived.
Thus, although $\mathbb{E}[v-t\mid v>t]$ may also increase with $K$, the
selection effect generated by the conditioning becomes weaker, whereas the
unconditional mean $\mathbb{E}[v]$ directly benefits from maximizing over
more independent draws. This shifts the NBUE comparison in favor of the
unconditional mean.


Combining Theorems~\ref{thm: exact reduction},
\ref{thm: NBUE iff no screening}, and \ref{thm: NBUE expands} gives a
comparison across both variety and capacity.

\begin{corollary}
\label{cor: no screening variety capacity}
Suppose that $G$ is CDF log-concave. If no screening is efficient in
$(G,K_0,\bar m_0)$ for some $K_0\ge1$ and $\bar m_0\in(0,1)$, then
it is efficient in $(G,K,\bar m)$ for every $K\ge K_0$ and every
$\bar m\in(0,1)$.
\end{corollary}

Efficiency at $(G,K_0,\bar m_0)$ implies that $G_{K_0}$ is NBUE.
This property persists for every $K\ge K_0$, giving efficiency at every
capacity by the NBUE characterization and the exact reduction.
Thus, no screening can be efficient even under severe scarcity:
its optimality is determined by the distribution of favorite-object
values, not by a minimum level of supply per agent.



Second, we can also show that as the object variety $K$ increases, the region in which the reduced value distribution has an IHR expands.

\begin{theorem}\label{thm: hazard rate increase} 
    Let $r'(\cdot; G_{K})$ be the derivative of $r(\cdot; G_{K})$ with respect to $v$. Then, $r'(v ; G_{K + 1}) \ge 0$ if $r'(v; G_{K}) \ge 0$. Furthermore, for all $v \in (0, \bar{v})$, there exists $K_{0}$ such that for all $K > K_{0}$, $r'(v; G_{K}) > 0$.
\end{theorem}

We describe the intuition using the terminology of survival analysis.
The hazard rate
$r(v;G_K)$ is the instantaneous rate of complete system failure,
conditional on at least one of the $K$ products surviving until $v$.
Complete failure is a first-order event only when exactly
one product remains. At small $v$, especially when $K$ is large,
several products are likely still functioning, so conditioning on system
survival provides little evidence that only one product remains. At large
$v$, by contrast, individual survival is rare, so conditional on system
survival, the surviving set is much more likely to consist of a single
product. Adding another product therefore reduces the hazard rate by a larger
proportion at small $v$ than at large $v$. This uneven reduction tilts the
hazard function upward, expanding the region over which it is increasing.


Theorem~\ref{thm: hazard rate increase} implies that, as the object variety increases, DHR becomes harder to satisfy, making full screening less likely to be an efficient mechanism.

\begin{corollary}
    For every marginal distribution function $G$, there exist $K$ and $\bar{m}$ such that full screening is inefficient for a continuous i.i.d.\ market $(G, K, \bar{m})$.
\end{corollary}

The formula $G_K=G^K$ is exact in the continuous i.i.d.\ market, but the
order-statistic channel has a broader reach. Whenever an agent chooses from
the options remaining at her turn, her realized value is the maximum over
that menu; when the remaining menu is rich, this value is typically an upper
order statistic relative to her original valuation vector. Depletion,
capacity asymmetries, and correlation determine the precise distribution, but
the economically relevant object remains the distribution of upper order statistics. The theory therefore yields a comparative-static prediction for other
environments. A wider variety of options (object types)
tends to reduce the incremental residual-surplus gain from screening agents using
costly effort.

\begin{figure}[t!]
    \centering
    \begin{subfigure}[t]{0.48\textwidth}
        \centering
        \includegraphics[width=\textwidth]{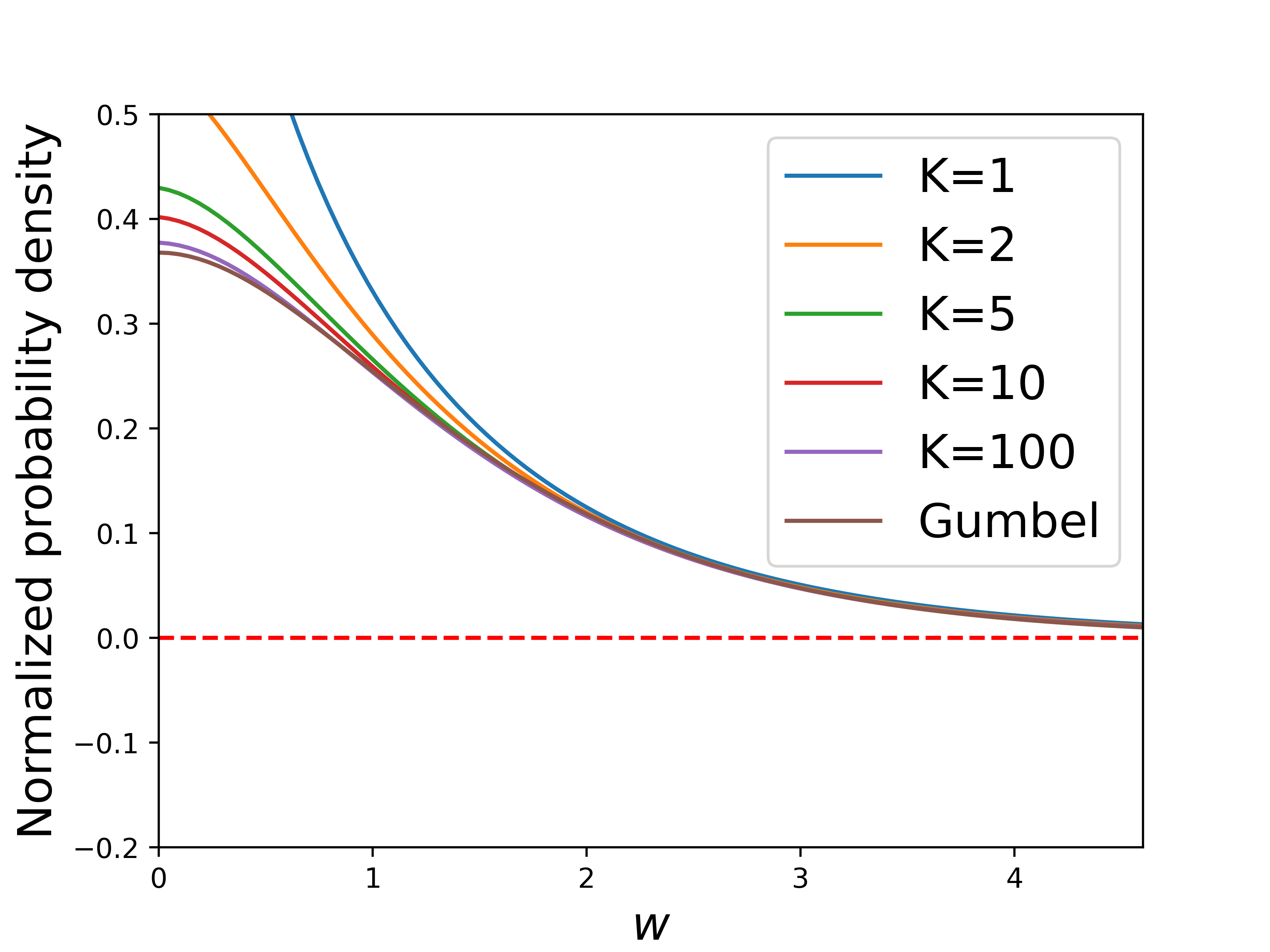}
        \subcaption{Normalized probability density}
        \label{fig: weibull example probability density}
    \end{subfigure}
    \hspace{0.02\textwidth}
    \begin{subfigure}[t]{0.48\textwidth}
        \centering
        \includegraphics[width=\textwidth]{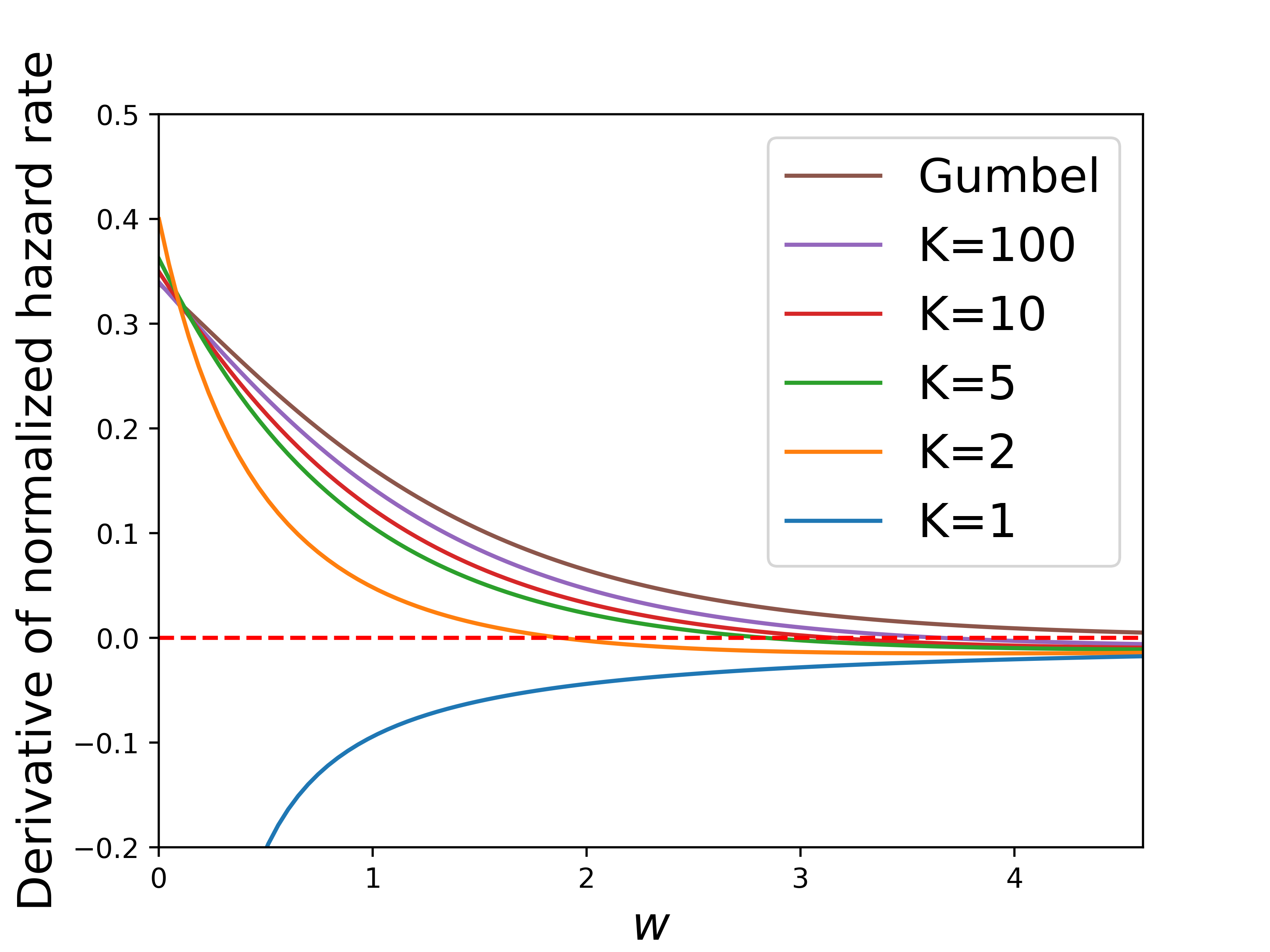}
        \subcaption{Derivative of the normalized hazard rate}
        \label{fig: weibull example derivative of HR}
    \end{subfigure}
    \caption{The behavior of the distribution of the normalized largest-order statistic ($w = (v - b_K)/a_K$) when the value distribution is the Weibull distribution with shape parameter $0.9$. The sequence $(a_{K}, b_{K})$ is shown in Appendix~\ref{sec: Weibull example}.}
    \label{fig: Weibull example}
\end{figure}

The effect of increasing the object variety $K$ on whether $G_K$ satisfies NBUE or IHR is substantial. Figure~\ref{fig: Weibull example} illustrates the case of a Weibull distribution with parameter $0.9$, which is globally DHR. As shown in Panel~\ref{fig: weibull example derivative of HR}, the derivative of the hazard rate is negative everywhere when $K=1$, but becomes positive over a large region even with $K = 2$. At this point, a no-screening mechanism already outperforms a full-screening mechanism. As $K$ increases further, this region continues to expand, and in the limit $K \to \infty$, the distribution of the largest-order statistic $G_K$ converges to the Gumbel distribution, which is globally IHR.

\subsection{Large-Variety Limits}

We present the large-market results based on the asymptotic properties of the reduced value distribution $G_{K}$ in the limit of $K \to \infty$. The limiting behavior of the largest-order statistic has been extensively analyzed in the literature of extreme value theory. We utilize these results to characterize efficient mechanisms in the large-market limit.

\begin{definition}
    Let $G$, $H$ be distribution functions. We say that $H$ is an \emph{extreme value distribution} and that $G$ lies in the \emph{domain of attraction} of $H$ if $G$ and $H$ are nondegenerate and there exists a sequence of constants $(a_{K}, b_{K})_{K = 1}^{\infty}$ with $a_{K}> 0$ for all $K$, such that $\hat{G}_{K}(w) \coloneqq G_{K}(a_{K}w + b_{K}) \to H(w)$ as $K \to \infty$ for all $w$ at which $H$ is continuous.
\end{definition}

For notational simplicity, we define the normalized value as $w = (v - b_K)/a_K$. We denote its value distribution $\hat{G}_{K}$ and allocation rule $\hat{x}$ as $\hat{G}_{K}(w) \coloneqq G_{K} (a_{K}w + b_{K})$ and $\hat{x}(w) \coloneqq x(a_{K}w + b_{K})$. 
This change of coordinates preserves monotonicity and resource use. Furthermore,
\[
    r'(w;\hat{G}_{K})
    =
    a_{K}^{2}r'(a_{K}w+b_{K};G_{K}),
\]
implying that the sign of the hazard-rate derivative can be studied in normalized coordinates and then translated back to the original value scale.

When the number of object types $K$ becomes sufficiently large, the shape of the reduced value distribution $G_{K}$ becomes less dependent on the shape of the marginal value distribution, $G$. Consequently, the limit distribution can only be one of the Gumbel, Fr\'echet, or reverse-Weibull distributions.

\begin{proposition}[\citet{fisher1928limiting,Gnedenko1943}]
    \label{prop: fisher-tippett-gnedenko theorem} Only the Gumbel, Fr\'echet, and reverse-Weibull distributions,
    \begin{align}
        \Lambda(w)       & = \exp(- \exp(- w)), \hspace{1em}-\infty < w < \infty \tag*{(Gumbel)}                                                \\
        \Phi_{\alpha}(w) & = \exp( - w^{-\alpha}), \hspace{1em}w \ge 0, \alpha > 0, \tag*{(Fr\'echet)}                                         \\
        \Psi_{\alpha}(w) & = \begin{cases}\exp(-(-w)^{\alpha})&w < 0\\ 1&w \ge 0,\end{cases} \hspace{1em}\alpha > 0 \tag*{(reverse-Weibull)}
    \end{align}
    can be an extreme value distribution.
\end{proposition}

For the proof of Proposition~\ref{prop: fisher-tippett-gnedenko theorem}, see Theorems~1.1.3 and 1.1.6 of \citet{haan2006extreme}.\footnote{\citet{haan2006extreme} employ the representation of the generalized extreme value distribution, introduced on pp.~9-10, which is equivalent to Proposition~\ref{prop: fisher-tippett-gnedenko theorem}.} Theorem 1.1.6 of \citet{haan2006extreme} additionally implies that, whenever a distribution function $G$ belongs to the domain of attraction of Gumbel, $\hat{G}_{K}(w) \coloneqq G_{K}(a_{K}w + b_{K}) \to \Lambda(w)$ with
\begin{equation}
    a_{K}= s(G^{-1}(1 - 1/K)),\: b_{K}= G^{-1}(1 - 1/K), \text{ where }s(v) = \frac{\int_{v}^{\bar{v}}(1 - G(v')) dv'}{1 - G(v)}.  \label{eq: a_K b_K gumbel}
\end{equation}
Furthermore, whenever a distribution function $G$ belongs to the domain of attraction of Fr\'echet with shape parameter $\alpha$, $\hat{G}_{K}(w) \coloneqq G_{K}(a_{K} w + b_{K}) \to \Phi_{\alpha}(w)$ with 
\begin{equation}
    a_{K}= G^{-1}(1 - 1/K),\: b_{K}= 0. \label{eq: a_K b_K frechet}
\end{equation}
Throughout the paper, when we refer to $(a_{K}, b_{K})_{K = 1}^{\infty}$, these constants are defined by \eqref{eq: a_K b_K gumbel} or \eqref{eq: a_K b_K frechet}.

We omit the condition for reverse-Weibull because the convergence occurs only if $\bar{v}< + \infty$, and in that case, we do not need extreme value theory for characterizing an efficient mechanism in the limit (see Theorem~\ref{thm: bounded support}).

\begin{table}[t]
    \centering
    \caption{Extreme Value Distributions of Common Probability Distributions}
    \label{tab: extreme value distribution}
    \begin{tabular}{cc}
        \toprule Limit  & Marginal Value Distribution                                                                              \\
        \midrule Gumbel & \begin{tabular}{c}Exponential, Normal, Gamma, Log-Normal, Logistic, Weibull, Gumbel, etc.\end{tabular} \\
        Fr\'echet       & Pareto, Cauchy, $t$, $F$, Burr, Log-Gamma, Zipfian, Fr\'echet, etc.                                     \\
        Reverse-Weibull & Uniform, Beta, Reverse-Weibull, etc.                                     \\
        \bottomrule
    \end{tabular}
\end{table}

Proposition~\ref{prop: fisher-tippett-gnedenko theorem} does not state that the distribution of the normalized maximum always converges.\footnote{For example, the binomial, Poisson, and geometric distributions do not have an extreme value distribution.} Nevertheless, it is also known that many continuous distribution functions have an extreme value distribution.\footnote{\citet{pickands1975statistical} mentions that \textit{``Most `textbook' continuous families of distribution functions are discussed in \citet{gumbel1958statistics}. They all lie in the domain of attraction of some extremal distribution function.''} (page~119)} Table~\ref{tab: extreme value distribution} lists extreme value distributions of common probability distributions. The limit is identical as long as distributions have the same right-tail property, even when their overall shape differs significantly.

The extreme value theory literature has also characterized the conditions under which a marginal value distribution $G$ belongs to the domain of attraction of each extreme value distribution. However, this paper focuses not on the convergence of the distribution function $\hat{G}_K$, but on the derivative of the hazard rate, i.e.,
\begin{equation}
    r'(w; \hat{G}_{K}) = \frac{\hat{g}'_{K}(w)(1 - \hat{G}_{K}(w)) + (\hat{g}_{K}(w))^{2}}{\left(1
    - \hat{G}_{K}(w)\right)^{2}}.
\end{equation}
For convergence of $r'(\cdot; \hat{G}_{K})$, we also need convergence of $\hat{g}_{K}$ and $\hat{g}'_{K}$, which are the first and second derivatives of $\hat{G}_{K}$. The following is the necessary and sufficient condition for it.

\begin{definition}
    Let $G$ and $H$ be distribution functions. We say that $G$ lies in the
    \emph{twice differentiable domain of attraction} of $H$ if $G$ and $H$ are
    nondegenerate and twice differentiable for all sufficiently large $w$ and
    $\hat{G}_{K}^{(l)}(w) \to H^{(l)}(w)$ as $K \to \infty$ uniformly on every finite interval for all $l = 0, 1, 2$.
\end{definition}

\begin{definition}[von Mises condition]
    \label{definition: von Mises condition} A distribution function $G$ satisfies
    the \emph{von Mises condition with parameter $\gamma$} if $G$ is twice
    differentiable for all sufficiently large $v$ and
    \begin{equation}
        \label{eq: von Mises condition}\vartheta'(v ; G) = \left(\frac{1 - G(v)}{g(v)}
        \right)' \to \gamma \text{ as }v \to \infty.
    \end{equation}
    We denote the set of all distribution functions satisfying \eqref{eq: von Mises condition}
    by \vmc{$\gamma$}.
\end{definition}

The von Mises condition \eqref{eq: von Mises condition} was introduced by
\citet{VonMises1936} as a tractable sufficient condition for normalized
maxima to converge to an extreme-value distribution.
\citet[Theorem~5.2]{Pickands1986continuous} later showed that it exactly
characterizes the twice differentiable domain of attraction. The condition
is satisfied by standard smooth parametric families and mainly excludes
distributions with irregular tail derivatives, such as persistently
oscillatory tails.

\begin{proposition}[Theorem 5.2 of \citet{Pickands1986continuous}]
    \label{prop: Pickands twice diff convergence}\hspace{1em}
    \begin{enumerate}[(i)]
        \item $G$ is in the twice differentiable domain of attraction of the
            Gumbel distribution $\Lambda$ if and only if $G \in \vmc{0}$.

        \item $G$ is in the twice differentiable domain of attraction of the
            Fr\'echet distribution with shape parameter $\alpha$,
            $\Phi_{\alpha}$, if and only if $G \in \vmc{1/\alpha}$.
    \end{enumerate}
\end{proposition}


Based on these results, we characterize efficient mechanisms in a large market limit of the continuous i.i.d.\ market.

\paragraph{Bounded Support}
We start with the case of bounded support, i.e., $\bar{v}< + \infty$. In this case, no screening becomes asymptotically first-best, whereas the residual surplus from full screening vanishes. When $K$ is large, almost every agent has at least one good that they value close to $\bar{v}$. Therefore, even with no screening, the gross surplus approaches $\bar{v}$, and there is little need to screen the intensity of preferences. On the other hand, if screening is conducted, almost all the surplus is burned due to competition.

\begin{theorem}
    \label{thm: bounded support} Suppose that $\bar{v}< + \infty$. Then, for
    any $\epsilon > 0$ and $\delta > 0$, there exists $K_{0}> 0$ such that
    for all $K > K_{0}$, we have both $RS(\mech_{SD}) > \bar{m}(1 - \delta)(\bar
    {v}- \epsilon)$ and $RS(\mech_{VCG}) < \bar{m}\epsilon$.
\end{theorem}

\paragraph{Gumbel Case}

Next, we consider the case where the marginal value distribution $G$ satisfies \vmc{0}, and thus belongs to the twice differentiable domain of attraction of the Gumbel distribution $\Lambda$. The hazard rate function of the Gumbel is given by
\begin{equation}
    r(w; \Lambda) = \frac{\lambda(w)}{1 - \Lambda(w)}= \frac{\exp(- w)}{\exp(\exp(-
    w)) - 1},
\end{equation}
and thus the Gumbel distribution $\Lambda$ has an IHR. This limiting shape therefore implies eventual pooling on every normalized interval.

Proposition~\ref{prop: Pickands twice diff convergence} ensures that whenever $G \in \vmc{0}$, $\hat{G}_{K}(w)$ converges to $\Lambda(w)$ up to the second derivative, uniformly on any finite interval. Accordingly, for any finite interval $[\underline{c}, \bar{c}]$, the hazard rate becomes increasing on $[\underline{c}, \bar{c}]$ with sufficiently large $K$ (see Lemma~\ref{lem: gumbel case hazard rate convergence} in the Appendix). Furthermore, by standard arguments in optimal mechanism design, there exists an efficient allocation rule that is constant over the interval where the hazard rate is nondecreasing. (For clarity, we formally present and prove this result as Lemma~\ref{lem: flat allocation} in the Appendix.) 

Accordingly, for any interval $[\underline{c}, \bar{c}]$, there exists $K_{0}$ such that for all $K > K_{0}$, an efficient allocation rule becomes constant in the interval: $\hat{x}(w) = \bar{x}$ for $w \in [\underline{c}, \bar{c}]$. Furthermore, as $\hat{G}_K(\bar{c} ) - \hat{G}_K(\underline{c}) \to \Lambda(\bar{c}) - \Lambda(\underline{c})$, by taking small $\underline{c}$, large $\bar{c}$, and large $K$, we can make $\hat{G}_{K}(\bar{c}) - \hat{G}_{K}(\underline{c})$ arbitrarily close to one. Since the allocation is flat on the interval $[\underline{c},\bar{c}]$ and this interval can be taken arbitrarily large to carry probability mass arbitrarily close to one, the resource constraint forces this constant allocation level to be arbitrarily close to $\bar{m}$, with any residual allocation absorbed by the negligible outside mass. In this sense, an efficient mechanism conducts no screening asymptotically.

\begin{theorem}
    \label{cor: Gumbel conclusion 2} If $G \in \vmc{0}$, then for all
    $\epsilon > 0$, there exists $K_{0}$ such that for all $K > K_{0}$, there exists a constant $\bar{x}_K \in [0, 1]$ such that there is an efficient allocation rule $\hat{x}$ for $(G, K, \bar{m})$ which satisfies
    $\hat{G}_K(\{w: \hat{x}(w) = \bar{x}_K\}) > 1 - \epsilon$. Furthermore,  $\bar{x}_K\in ((\bar{m}-\epsilon) /(1 - \epsilon), \bar{m}/(1 - \epsilon))$.
\end{theorem}

\paragraph{Fr\'echet Case}

Finally, we consider the case where the marginal value distribution $G$ satisfies \vmc{$1/\alpha$} for some $\alpha > 0$. When $\alpha \in (0, 1]$, the right tail of Fr\'echet, $\Phi_\alpha$, is extremely heavy, and the distribution does not have a finite mean; thus, we focus on the case of $\alpha > 1$, where agents' expected payoffs are well-defined. The hazard rate function of Fr\'echet with parameter $\alpha$ is given by
\begin{equation}
    r(w; \Phi_{\alpha}) = \frac{\phi_{\alpha}(w)}{1 - \Phi_{\alpha}(w)}= \frac{\alpha
    w^{-\alpha - 1}\exp(- w^{- \alpha})}{1 - \exp(- w^{-\alpha})}.
\end{equation}
We say that a distribution has an \emph{increasing-decreasing hazard rate (IDHR)} if there exists $w^{*}$ such that the hazard rate is increasing for $w < w^{*}$ and decreasing for $w > w^{*}$. We can verify that Fr\'echet has an IDHR.\footnote{For Fr\'echet with parameter $\alpha$, $w^{*}$ is the unique solution of
\begin{equation}
    \frac{w^{-\alpha}}{1 - \exp(- w^{-\alpha})}= \frac{\alpha + 1}{\alpha}
\end{equation}
with $w > 0$.}

Because the Fr\'echet distribution has neither an IHR nor a DHR, we apply the
ironing procedure of \citet{myerson1981optimal} to characterize the efficient
mechanism. Under IDHR, this procedure yields a cutoff
$w^{**}\in(w^{*},+\infty]$: the efficient mechanism pools all types below
$w^{**}$ and uses full screening only above it. If the distribution is NBUE,
$w^{**}=+\infty$, so the pooling region covers the entire support
and no screening is efficient.\footnote{IDHR does not preclude NBUE: under
the standard parameterization, an inverse Gaussian distribution
$\operatorname{IG}(\mu,\lambda)$ has an IDHR and is NBUE whenever
$\lambda\ge 2\mu$ 
\citep[see, e.g.,][]{chhikara1989inverse,tang1999mean}.}
The Fr\'echet distribution considered here, however, has a mean residual life
that diverges as $w\to\infty$ and hence is not NBUE, implying
$w^{**}<+\infty$.%
\footnote{If $w^{**}<+\infty$, the threshold $w^{**}$ is characterized by \eqref{eq: vdstar} in Appendix~\ref{subsec: IDHR efficient mechanism}.
The mean residual life function $\mathbb{E}[v - t| v > t]$ of Fr\'echet diverges as $t$ increases, and thus it is not NBUE. See Section 6.2 of \citet{embrechts2013modelling}, for example.} If the supply $\bar{m}$ is insufficient to allocate all agents with $w > w^{**}$, then only those with the highest values will win an item. Otherwise, an efficient mechanism first allocates goods with probability one to agents with values $w > w^{**}$. The remaining goods are then distributed equally among the other agents without payment.

\begin{proposition}\label{prop: IDHR efficient mechanism}
    When the reduced value distribution function $\hat{G}_{K}$ has an IDHR,
    there exists $w^{**}\in (w^{*}, +\infty]$ such that an efficient allocation rule
    can be represented as follows:
    \begin{enumerate}[(i)]
        \item If $1 - \hat{G}_{K}(w^{**}) \ge \bar{m}$, then
            \begin{equation}
                \hat{x}(w) =
                \begin{cases}
                    1 & \text{ for }w \in [\hat{G}_{K}^{-1}(1 - \bar{m}), +\infty), \\
                    0 & \text{ for }w \in [0, \hat{G}_{K}^{-1}(1 - \bar{m})).
                \end{cases}
            \end{equation}

        \item If $1 - \hat{G}_{K}(w^{**}) < \bar{m}$, then
            \begin{equation}
                \hat{x}(w) =
                \begin{cases}
                    1 & \text{ for }w \in [w^{**}, + \infty), \\
                    \dfrac{\bar{m} - (1 - \hat{G}_K(w^{**}))}{\hat{G}_K(w^{**})} & \text{ for }w \in [0, w^{**}).
                \end{cases}
            \end{equation}
    \end{enumerate}
    In particular, if $w^{**} = +\infty$, then \textup{(ii)} implies $\hat{x}(w) \equiv \bar{m}$ (i.e., no screening).
\end{proposition}

\begin{figure}[t!]
    \centering
    \begin{subfigure}[t]{0.48\textwidth}
        \includegraphics[width=\textwidth]{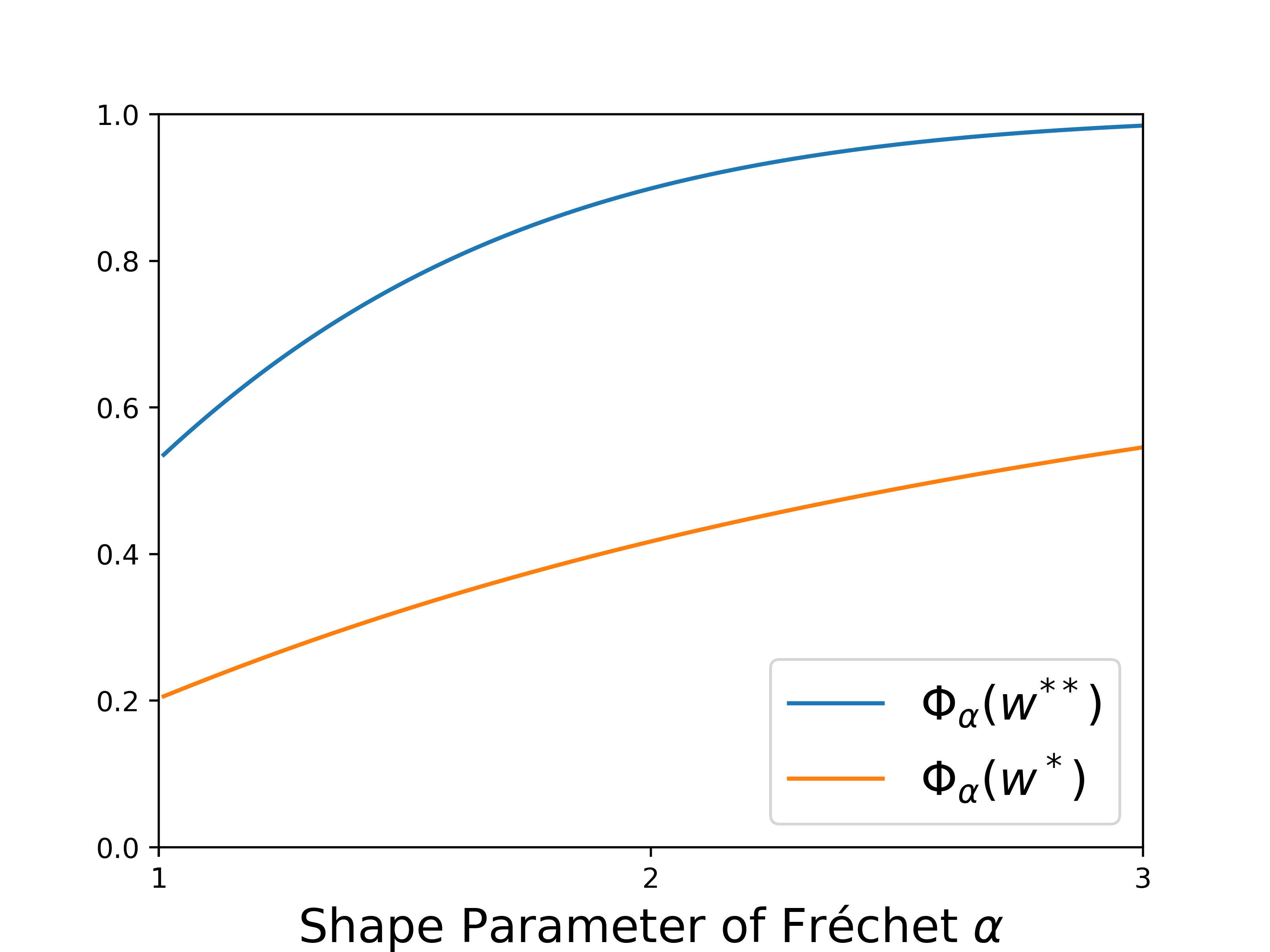}
        \subcaption{Cumulative probabilities of regions in which the extreme value distribution has an IHR ($\Phi_\alpha(w^*)$) and the efficient mechanism uses a constant allocation ($\Phi_\alpha(w^{**})$)}\label{fig: frechet threshold}
    \end{subfigure}
    \hspace{0.02\textwidth}
    \begin{subfigure}[t]{0.48\textwidth}
        \includegraphics[width=\textwidth]{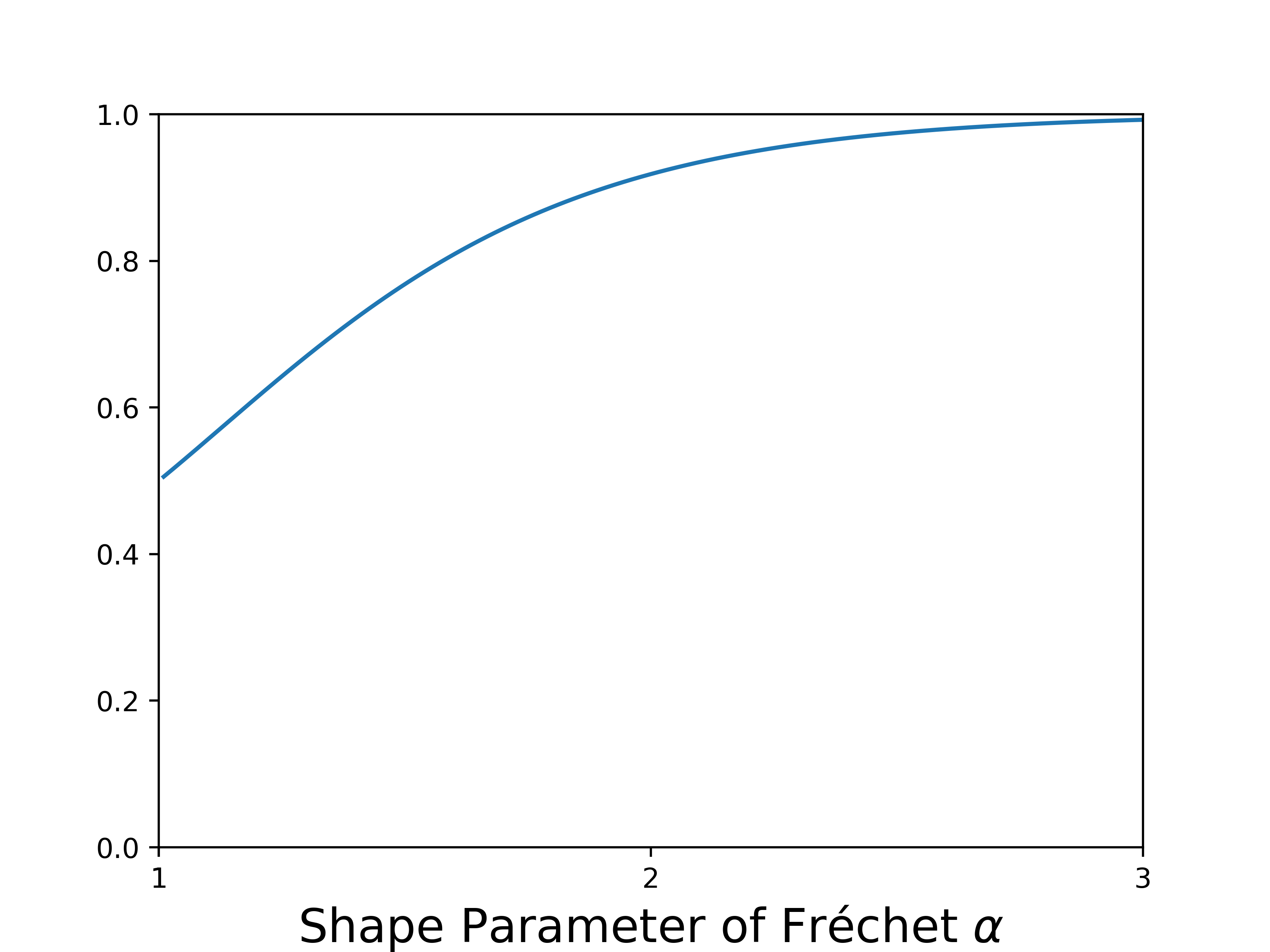}
        \subcaption{Performance of the no-screening mechanism relative to the efficient one}\label{fig: frechet SD performance}
    \end{subfigure}
    \caption{How close the no-screening mechanism is to the efficient mechanism in the large market limit in the Fr\'echet case}
\end{figure}

Figure~\ref{fig: frechet threshold} shows the values of $\Phi_{\alpha}(w^{*})$ and $\Phi_{\alpha}(w^{**})$ for various shape parameters $\alpha$.\footnote{For Fr\'echet with parameter $\alpha$, $w^{**}$ is a solution of
\begin{equation}
    \label{eq: frechet vdstar equation}(1 - \exp(- w^{-\alpha})) w + \Gamma\left
    (\frac{\alpha - 1}{\alpha}, w^{-\alpha}\right) - \frac{1}{\alpha}w^{\alpha
    + 1}(1 - \exp(- w^{-\alpha})) = 0,
\end{equation}
where $\Gamma(s, x) = \int_{x}^{\infty}t^{s-1}e^{-t}dt$ is the upper incomplete gamma function. See Appendix~\ref{subsec: derivation of frechet vdstar equation} for the derivation of \eqref{eq: frechet vdstar equation}.} 
As $\alpha$ increases, the Fr\'echet distribution becomes increasingly concentrated around one, and under the local normalization $x = \alpha (w - 1)$, its limiting shape is Gumbel.
Accordingly, as $\alpha$ increases, both $\Phi_{\alpha}(w^{*})$ and $\Phi_{\alpha}(w^{**})$ increase and converge to one. The convergence of $\Phi_{\alpha}(w^{*})$ is relatively slow. By contrast, the fraction of agents with $w > w^{**}$ diminishes rapidly even with relatively small $\alpha$, implying that under Fr\'echet with relatively small $\alpha$, an efficient mechanism conducts no screening for most agents. For example, when $\alpha = 3$ and $\bar{m}= 0.5$, an efficient mechanism requires a substantial amount of money burning to obtain a good with probability $1$, which only agents with the highest $1.5\%$ of values will choose. The remaining units, which can cover $48.5\%$ of the population, are distributed to the remaining $98.5\%$ of agents with no screening, just as SD would. In addition, Figure~\ref{fig: frechet SD performance} illustrates that the performance of SD rapidly approaches that of the efficient mechanism as the shape parameter $\alpha$ increases.

Finally, we characterize the efficient mechanism when $G \in \vmc{1/\alpha}$ and $K$ is large but finite. The condition $G \in \vmc{1/\alpha}$ implies that for any finite interval $[\underline{c}, \bar{c}]$, if $K$ is sufficiently large, $\hat{G}_{K}$ is close to Fr\'echet $\Phi_{\alpha}$ within that interval. By choosing $\underline{c}$ close to zero and $\bar{c}$ larger than $w^{**}$, and accordingly taking a large $K$, the efficient allocation rule $\hat{x}$ for $\hat{G}_{K}$ becomes constant over most of the interval $[0, w^{**}]$.

\begin{theorem}
    \label{thm: frechet conclusion} If $G \in \vmc{1/\alpha}$, then for all $\epsilon \in (0, 1)$, there exists $K_{0} \in \mathbb{Z}_{++}$ such that for all $K > K_{0}$, there exists a constant $\bar{x}_K \in [0, 1]$ such that an efficient allocation rule $\hat{x}$ for $(G, K, \bar{m})$ satisfies $\hat{x}(w) = \bar{x}_K$ for all $w\in (\epsilon, w^{**}- \epsilon)$, where $w^{**}$ is a solution of the equation \eqref{eq: frechet vdstar equation}.
\end{theorem}

\subsection{Finite Variety (with Continuum Copies)}
\label{subsec: continuous numerical analysis} 
Figure~\ref{fig: intro continuous weibull06} in the introduction illustrates how residual surplus changes with the number of object types $K$ under SD (or no screening), VCG (or full screening), and the efficient mechanism. This visualization is based on a continuous i.i.d.\ market, where the total mass of objects $\bar{m}$ is $1/2$, and the marginal value distribution $G$ is Weibull with parameter $0.6$. The performance of each mechanism is derived analytically using the reduction to a single-dimensional environment.

The marginal value distribution has a DHR; thus, when $K = 1$, i.e., the single-good case, full screening is efficient. However, since the distribution also belongs to $\vmc{0}$, the efficient allocation becomes asymptotically pooled on a set whose probability approaches one. For intermediate, finite $K$, the no-screening mechanism outperforms full screening for $K \ge 4$ in this setting. Furthermore, beyond this point, the performance of the no-screening mechanism rapidly converges to that of the efficient mechanism. This observation suggests that the key policy implications of our study, the superiority of SD in multi-object allocation without money, likely hold for practical settings.

\section{Finite Market}\label{sec: finite market}

This section considers markets in which a finite number of objects are allocated to a finite number of agents (i.e., no continuum copies of agents and objects). The analysis of such finite markets is widely believed to be considerably more challenging than the analysis of continuous markets. In this study, instead of characterizing efficient mechanisms analytically, we adopt the deep-learning-based approach developed by \citet{dutting2019optimal}, originally for numerically constructing revenue-maximizing auction mechanisms.

\subsection{Model}
We consider a finite market with a finite set of agents $I$ and a finite set of object types $K$. Slightly abusing notation, $I$ and $K$ also represent the cardinality of those sets. The endowment of each object type $k$ is $m_{k} \in \mathbb{Z}_{++}$. We assume that the goods are scarce: $I > \sum_{k \in K} m_k$. Agents' utilities are identical to those of the continuous market: When agent $i$ obtains object $k$ with probability $x_{k}^{i}$ for each $k$ while making a payment of $p^{i}$, her payoff is $\sum_{k \in K}v_{k}^{i}x_{k}^{i}- p^{i}$.

Let $V^{i}\subset \mathbb{R}_{+}^{K}$ be the set of all possible valuation vectors for agent $i$, and $\bv^{i}$ be its generic element. We define the set of all valuation profiles as $V \coloneqq \prod_{i \in I}V^{i}$, and $\bv = (\bv^{i})_{i \in I}$ as its generic element. Here, we abandon equal treatment of equals and a direct mechanism $\mech$ consists of an allocation rule $\bx : V \to [0,1]^{I \times K}$ and a payment rule $\bp: V \to \mathbb{R}_{+}^{I}$, where $\bx^{i}(\bv)$ and $p^{i}(\bv)$ are the respective allocation and payment for agent $i$ under valuation profile $\bv$. Let $F$ be the distribution of the valuation profile $\bv$.

The (per-capita) \emph{residual surplus} from a mechanism $\mech = (\bx, \bp)$ is
\begin{equation}
    \label{eq: defn social welfare_finite}
    RS(\mech) \coloneqq \int_{V}\frac{1}{I}\sum_{i\in I}\left(\sum_{k \in K}v^{i}_{k}x^{i}_{k}(\bv) - p^{i}(\bv)\right) dF(\bv).
\end{equation}
A mechanism $\mech$ satisfies the \emph{resource constraint} if
\begin{equation}
    \label{eq: defn resource_finite}
    \sum_{i \in I}x^{i}_{k}(\bv) \leq m_{k}\text{ for all }k \in K, \bv \in V.
\end{equation}
A mechanism $\mech$ is \emph{strategy-proof} if
\begin{equation}
    \label{eq: defn strategy proofness_finite}\sum_{k \in K}v^{i}_{k}x^{i}_{k}(\bv^{i},\bv^{-i})-p^{i}(\bv^{i},\bv^{-i})\ge \sum_{k \in K}v^{i}_{k}x^{i}_{k}(\hat{\bv}^{i},\bv^{-i})-p^i(\hat{\bv}^{i},\bv^{-i}) \text{ for all }i\in I, \bv^{i}, \hat{\bv}^{i}\in V^{i}, \bv^{-i}\in V^{-i}.
\end{equation}
A mechanism $\mech$ is \emph{individually rational} if
\begin{equation}
    \label{eq: defn individual rationality_finite}
    \sum_{k \in K}v^{i}_{k}x^{i}_{k}(\bv)-p^{i}(\bv)\ge 0 \text{ for all }i \in I, \bv \in V.
\end{equation}
A mechanism $\mech$ satisfies the \emph{unit demand} condition if
\begin{equation}
    \label{eq: defn unit demand_finite}
    \sum_{k \in K}x^{i}_{k}(\bv) \le 1 \text{ for all }i \in I, \bv \in V.
\end{equation}
We call a mechanism \emph{feasible} if it satisfies the resource constraint,
strategy-proofness, individual rationality, and unit demand. A feasible
mechanism is \emph{efficient} if it maximizes residual surplus.

\subsection{RegretNet}
To search numerically for residual-surplus-maximizing mechanisms in finite markets, we adopt \emph{RegretNet}, a deep learning-based method for designing revenue-maximizing auction mechanisms proposed by \citet{dutting2019optimal}. In RegretNet, a mechanism, which is a function that takes a valuation profile as input and returns an allocation and a payment profile, is represented by a neural network, which can approximate a wide range of functions using a relatively small number of parameters. In \citet{dutting2019optimal}, the loss function used for learning consists of the negated empirical expected revenue of the seller and a penalty term for violating strategy-proofness. RegretNet uses the augmented Lagrangian method to find a mechanism (i.e., the parameters) minimizing such a loss.
We adapt RegretNet by replacing the revenue objective with residual
surplus while retaining the allocation and payment constraints and the
regret penalty. The learned mechanisms let us assess whether numerical
optimization finds substantial improvements over SD.

In principle, when the type space is finite, the efficient mechanism can also be computed by solving a linear program (LP), because both the residual-surplus objective and constraints are linear in the allocation and payment variables. However, this approach scales poorly in the finite multi-object environments studied here. To see this, suppose that each of the $K$ value dimensions is discretized into $n$ grid points, so that each agent has $N=n^{K}$ possible types. A direct-mechanism LP must specify allocations and payments for every valuation profile, requiring $O(IKN^{I})$ variables, and strategy-proofness imposes constraints for every agent, every pair of own types, and every profile of opponents' types, yielding $O(IN^{I+1})=O(In^{(I+1)K})$ incentive constraints. Thus, the LP size grows exponentially in both the number of agents and the number of object types. While such LPs are useful for very small diagnostic examples, they become computationally infeasible even for moderately sized finite markets, and hence are not suitable for our purpose of studying how mechanisms behave as market size and object variety increase. RegretNet provides a tractable alternative by representing the mechanism with a neural network and optimizing it over sampled valuation profiles, allowing us to numerically explore approximately efficient mechanisms in environments beyond the reach of exact LP methods.

\subsection{I.I.D.\ Case}
\label{subsubsec: finite iid case}

We examine the extent to which the continuous-market model approximates a finite market well. We first consider a market with $4c$ agents and two object types ($K = 2$), each with capacity $c \in \mathbb{Z}_{++}$. Each agent $i$'s value for object $k$, $v_{k}^{i}$, is drawn i.i.d.\ from a Weibull distribution.

\begin{figure}[t!]
    \centering
    \begin{subfigure}[t]{0.48\textwidth}
        \includegraphics[width=\textwidth]{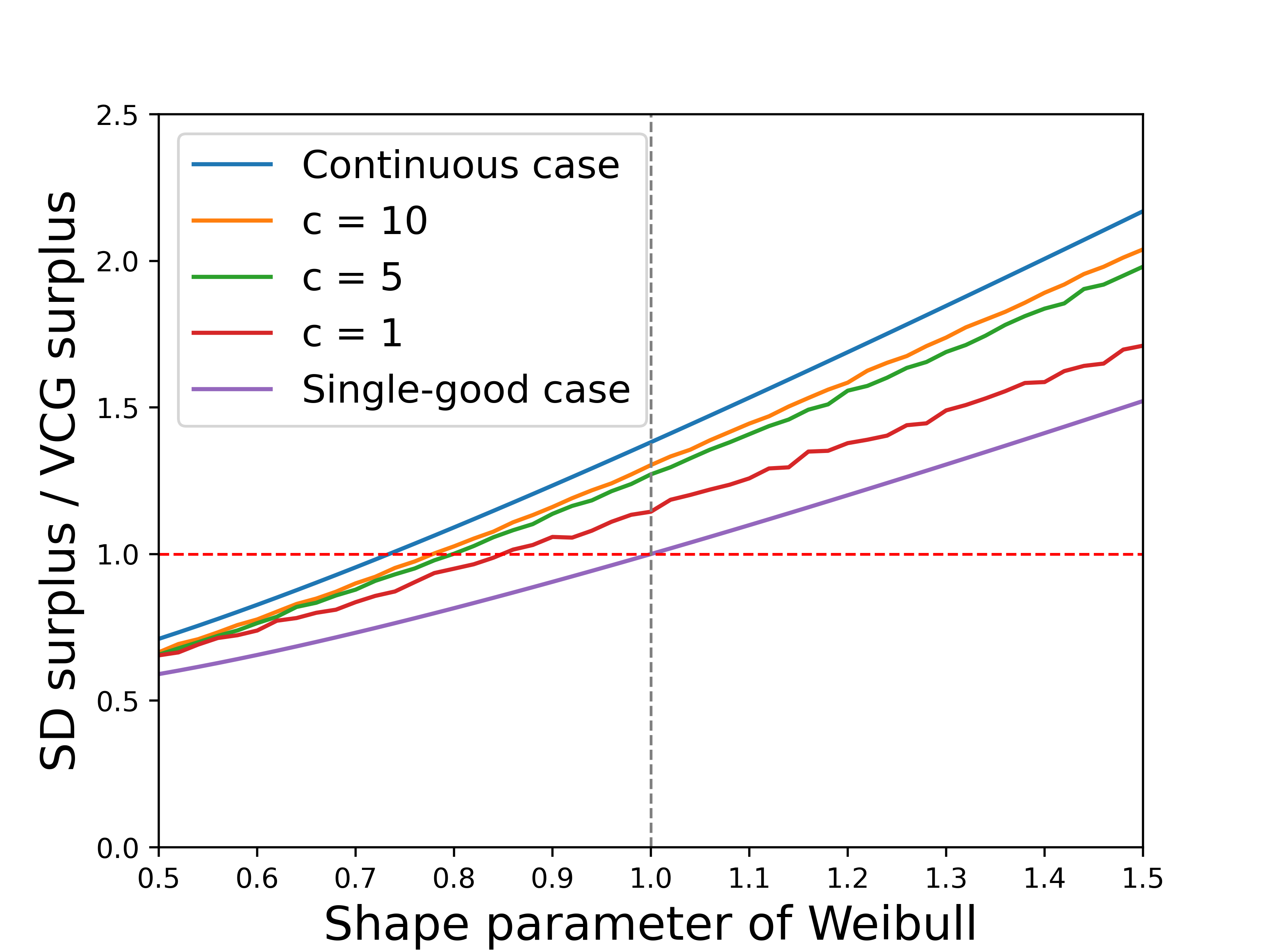}
        \subcaption{The ratio of the residual surplus achieved by SD and VCG, $(RS(\mech_{SD}) / RS(\mech_{VCG}))$}
        \label{fig: finite weibull welfare ratios}
    \end{subfigure}
    \hspace{0.02\textwidth}
    \begin{subfigure}[t]{0.48\textwidth}
        \includegraphics[width=\textwidth]{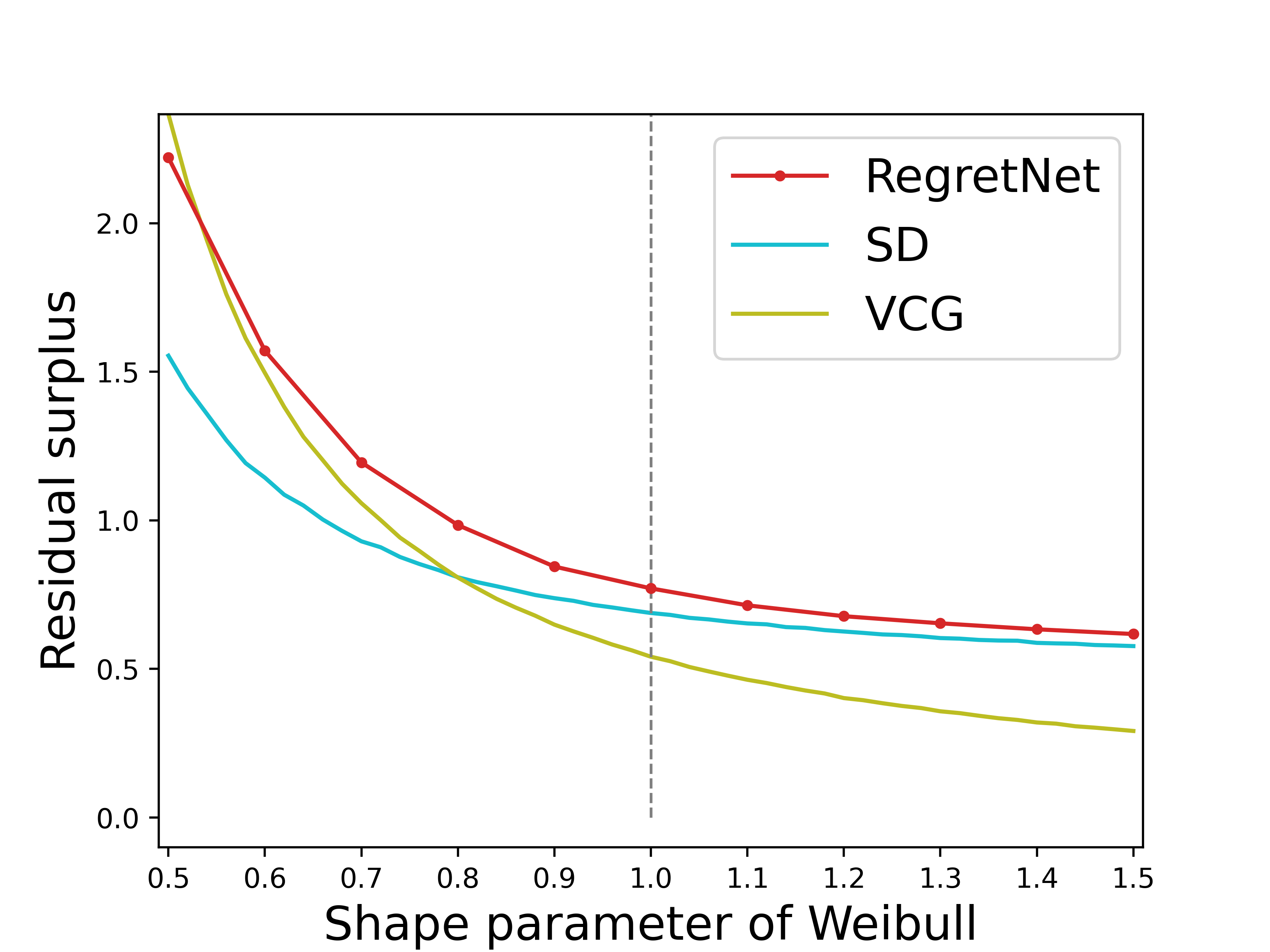}
        \subcaption{The performance of SD, VCG, and the mechanism developed by RegretNet for $c = 5$}
        \label{fig: finite weibull c5}
    \end{subfigure}
    \caption{Comparison of the performance of SD, VCG, and the numerically efficient mechanism in a finite market}
\end{figure}

In Figure~\ref{fig: finite weibull welfare ratios}, we generate $10{,}000$ samples and compare the ratio of the residual surplus achieved by SD and VCG, i.e., $RS(\mech_{SD})/RS(\mech_{VCG})$ for $c = 1, 5, 10$. We also plot (i) the ratio in a continuous market with a unit mass of agents and two object types with $m_{k}= 0.25$ units of endowments (i.e., $\bar{m}= 0.5$), which corresponds to a finite market in the limit of $c \to \infty$, and (ii) the ratio in a finite market with two agents and one object, which corresponds to the single-good case.

As characterized by \citet{Hartline2008}, SD outperforms VCG if and only if $\alpha > 1$ in the single-good case. In the continuous case, SD achieves larger residual surplus than VCG as long as the shape parameter is larger than roughly $0.7$, given $K = 2$. We observe a similar pattern for finite markets: Even with $c = 1$, the threshold at which SD starts to outperform VCG is strictly smaller than $1.0$, and as $c$ increases, the market becomes closer to the continuous case.

Figure~\ref{fig: finite weibull c5} shows the residual surplus of SD, VCG, and the numerically efficient mechanism returned by RegretNet for the case of $c = 5$. As we consider the case of $K = 2$, where the object variety is minimal, RegretNet strictly outperforms both SD and VCG under some parameters, particularly around $\alpha = 0.8$, where SD surpasses the performance of VCG. However, based on the analysis of the continuous market, this gap is expected to diminish as $K$ and $c$ increase.\footnote{In Appendix~\ref{subsec: performance of SD VCG and Efficient appendix}, we present figures corresponding to Figure~\ref{fig: finite weibull c5} for $c = 1$, single-good, and continuous cases.}

\subsection{Correlation}
\label{subsubsec: correlation}

Next, we relax the i.i.d.\ assumption and investigate cases where values are correlated. It is useful to distinguish two economically different departures from the i.i.d.\ benchmark. Within-agent correlation means that an agent who values one object highly also tends to value other objects highly. This captures variation in agents' overall need for receiving any object. In the limiting case of perfect within-agent correlation, the objects become effectively homogeneous from each agent's perspective, and the problem approaches a single-dimensional allocation problem. Between-agent correlation, by contrast, means that different agents tend to agree on the value of a given object. This captures object popularity or a common-value component. For example, in appointment scheduling, earlier or more convenient slots may be commonly attractive to many recipients.  In the limiting case of perfect between-agent correlation, the objects have pure common values.

Throughout this subsection, we consider an environment with $8$ agents and $4$ objects. We employ a \emph{Gaussian copula}, constructed by applying a probability integral transform to a multivariate normal distribution. This allows us to generate multiple joint distributions $F$ of $\bv = (v_{k}^{i})_{k \in K, i \in I}$ in which the marginal distribution of each $v_{k}^{i}$ is the same while the correlation between $v_{k}^{i}$ and $v_{l}^{j}$ for $(i, k) \neq (j, l)$ varies. Using these joint distributions, we analyze how the performance of SD, VCG, and RegretNet changes as the strength of within-agent correlation and between-agent correlation varies. The detailed method is described in Appendix~\ref{subsec: supp finite market}.

\paragraph{Within-Agent Correlation}

We first evaluate the effect of the \emph{within-agent correlation}, the correlation between $v_{k}^{i}$ and $v_{l}^{i}$ for two distinct objects $k \neq l$ (i.e., some agents need any object more strongly than other agents). Using the Gaussian copula, we generate a joint distribution $F$ whose marginal distribution is Weibull with a shape parameter $\alpha$ while the correlation coefficients between each pair of $v_{k}^{i}$ and $v_{l}^{i}$, $\corr(v_{k}^{i}, v_{l}^{i})$, vary. For $i \neq j$, $\corr(v_{k}^{i}, v_{l}^{j}) = 0$ for all $k, l \in K$. By construction, when $\corr(v_{k}^{i}, v_{l}^{i}) = 0$, values are drawn i.i.d., and when $\corr(v_{k}^{i}, v_{l}^{i}) = 1$, $v_{k}^{i}= v_{l}^{i}$ holds for all $k, l \in K$ with probability one, i.e., goods are perfectly homogeneous.

\begin{figure}[t]
    \begin{subfigure}{.49\linewidth}
        \centering
        \includegraphics[width=\linewidth]{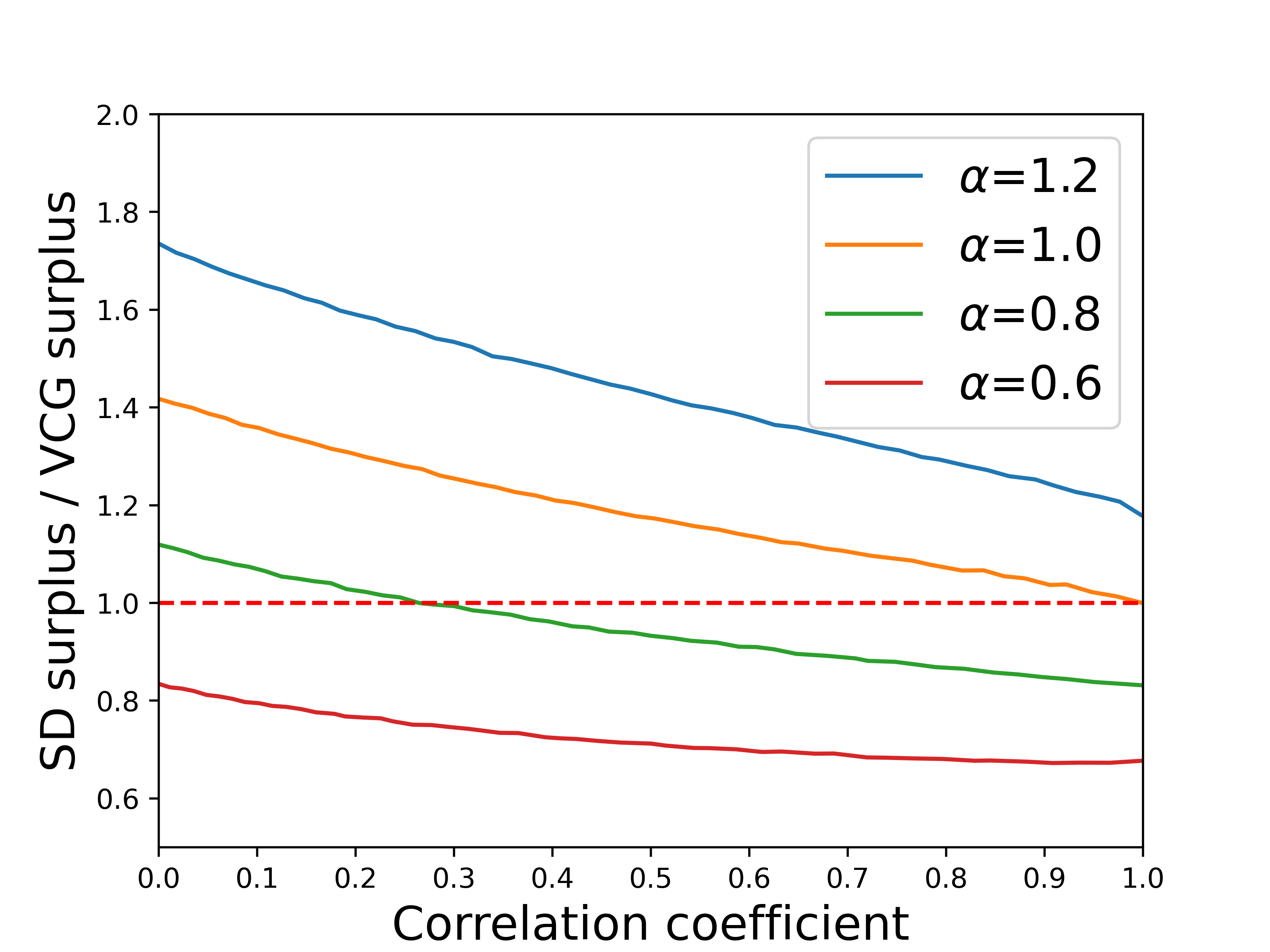}
        \subcaption{The performance ratio between SD and VCG: $(RS(\mech_{SD}) / RS(\mech_{VCG}))$}
        \label{fig: correlation_within_ratios}
    \end{subfigure}
    \begin{subfigure}{.49\linewidth}
        \centering
        \includegraphics[width=\linewidth]{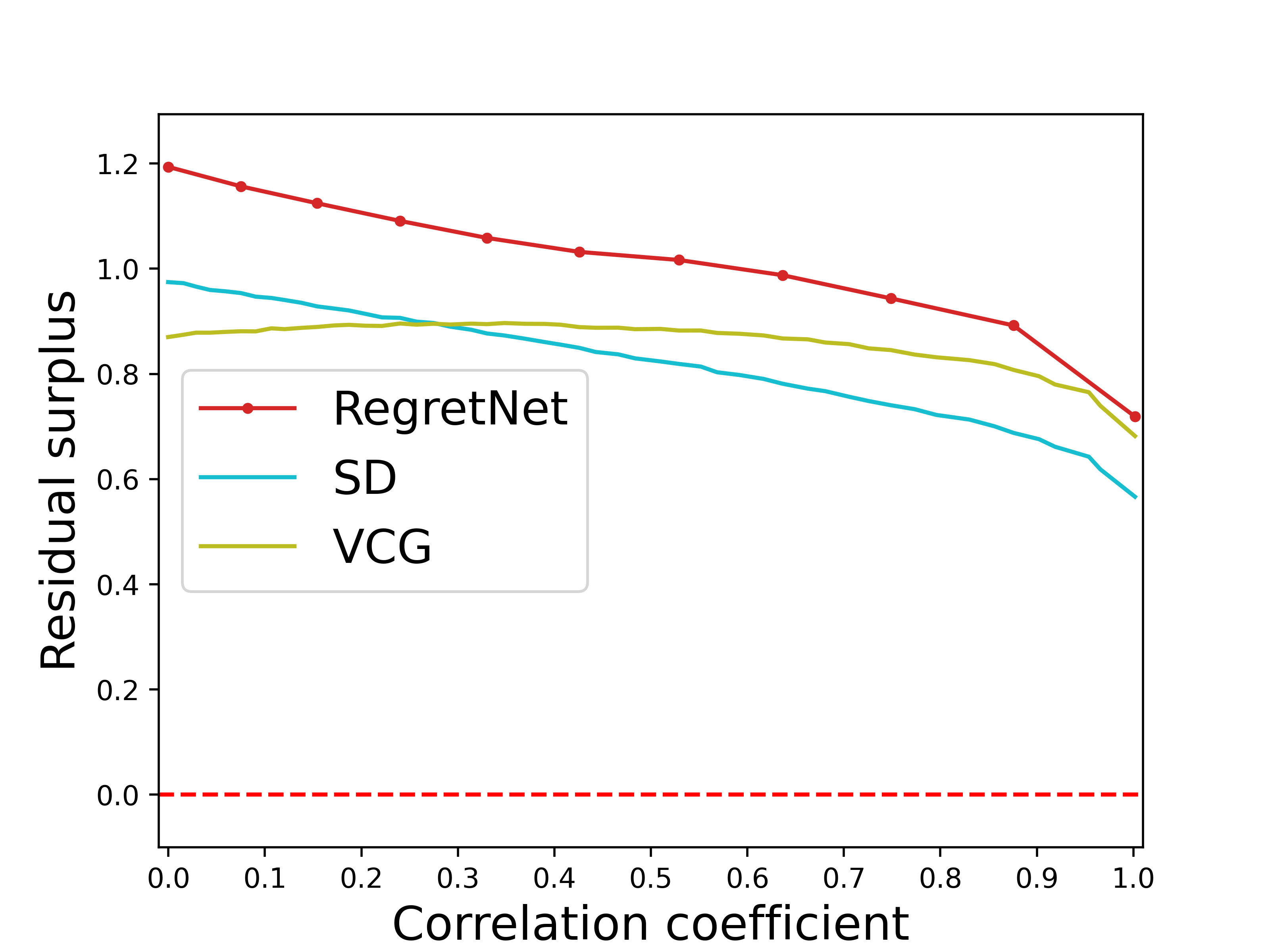}
        \subcaption{The performance of SD, VCG, and RegretNet given $\alpha = 0.8$}
        \label{fig: correlation_within_weibull08}
    \end{subfigure}
    \caption{The performance of mechanisms under within-agent correlation. The horizontal axis represents $\corr(v_{k}^{i}, v_{l}^{i})$.}
    \label{fig: within-agent correlation}
\end{figure}

Figure~\ref{fig: correlation_within_ratios} shows the performance ratio of SD and VCG, whereas Figure~\ref{fig: correlation_within_weibull08} shows the residual surplus achieved by SD, VCG, and RegretNet for the case of $\alpha = 0.8$. The results confirm the effective-variety interpretation. 
As within-agent correlation becomes stronger, the relative performance of VCG improves because the multi-object environment moves closer to a homogeneous-good problem: agents differ more in their overall need and less in their relative preferences across objects. 
At the endpoint of perfect correlation, the environment collapses to the single-dimensional benchmark. 
For intermediate correlations, however, the outcome is not the same as the single-object case. 
The residual idiosyncratic variation across objects continues to create gains from letting agents choose among heterogeneous objects without costly screening, and the multiplicity of objects still improves the performance of SD relative to VCG. 
Thus, within-agent correlation attenuates, but does not by itself eliminate, the force identified in the i.i.d.\ analysis.


\paragraph{Between-Agent Correlation}

\begin{figure}[t]
    \begin{subfigure}{.49\linewidth}
        \centering
        \includegraphics[width=\linewidth]{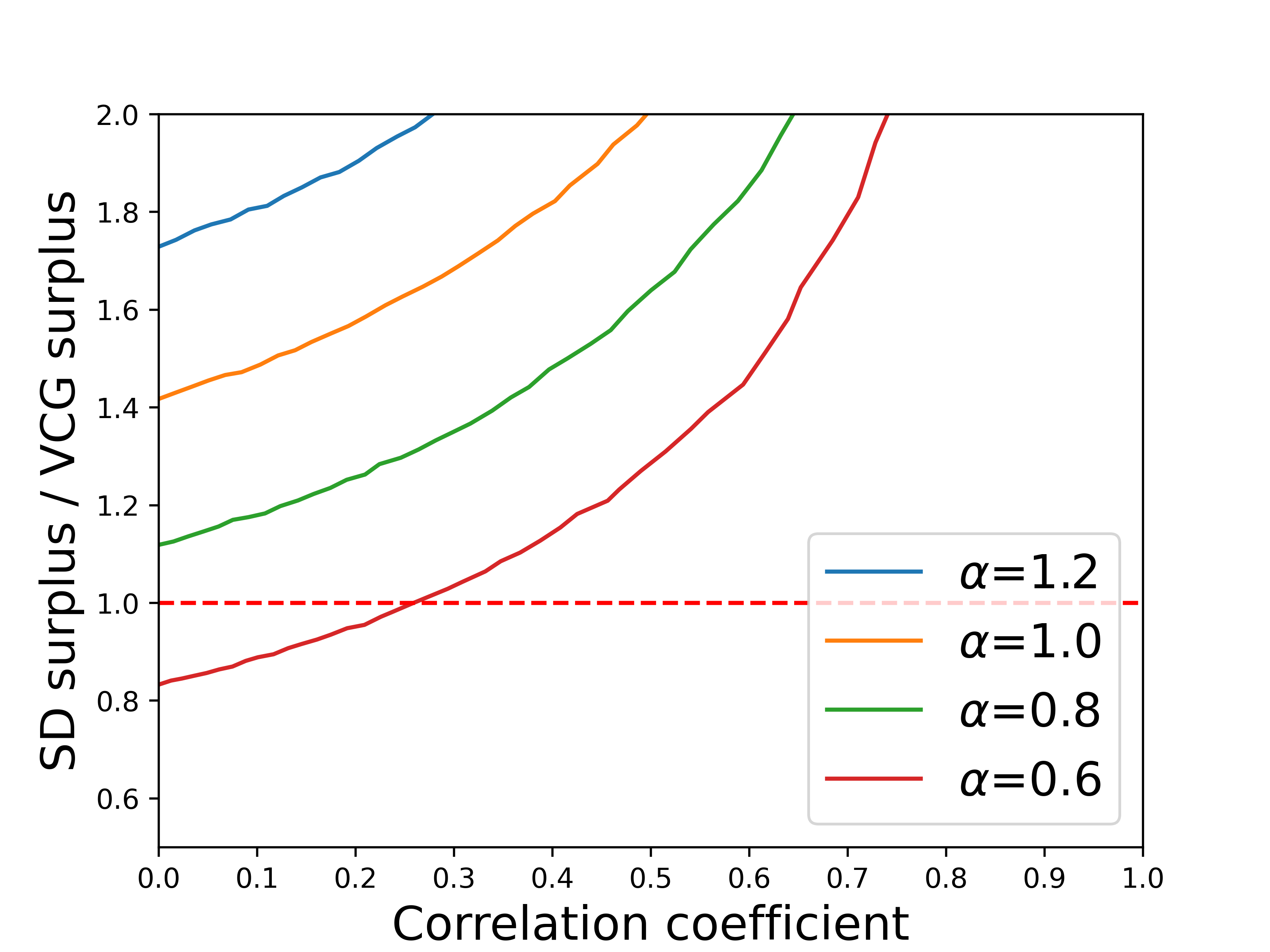}
        \subcaption{The performance ratio between SD and VCG: $(RS(\mech_{SD}) / RS(\mech_{VCG}))$}
        \label{fig: correlation_between_ratios}
    \end{subfigure}
    \begin{subfigure}{.49\linewidth}
        \centering
        \includegraphics[width=\linewidth]{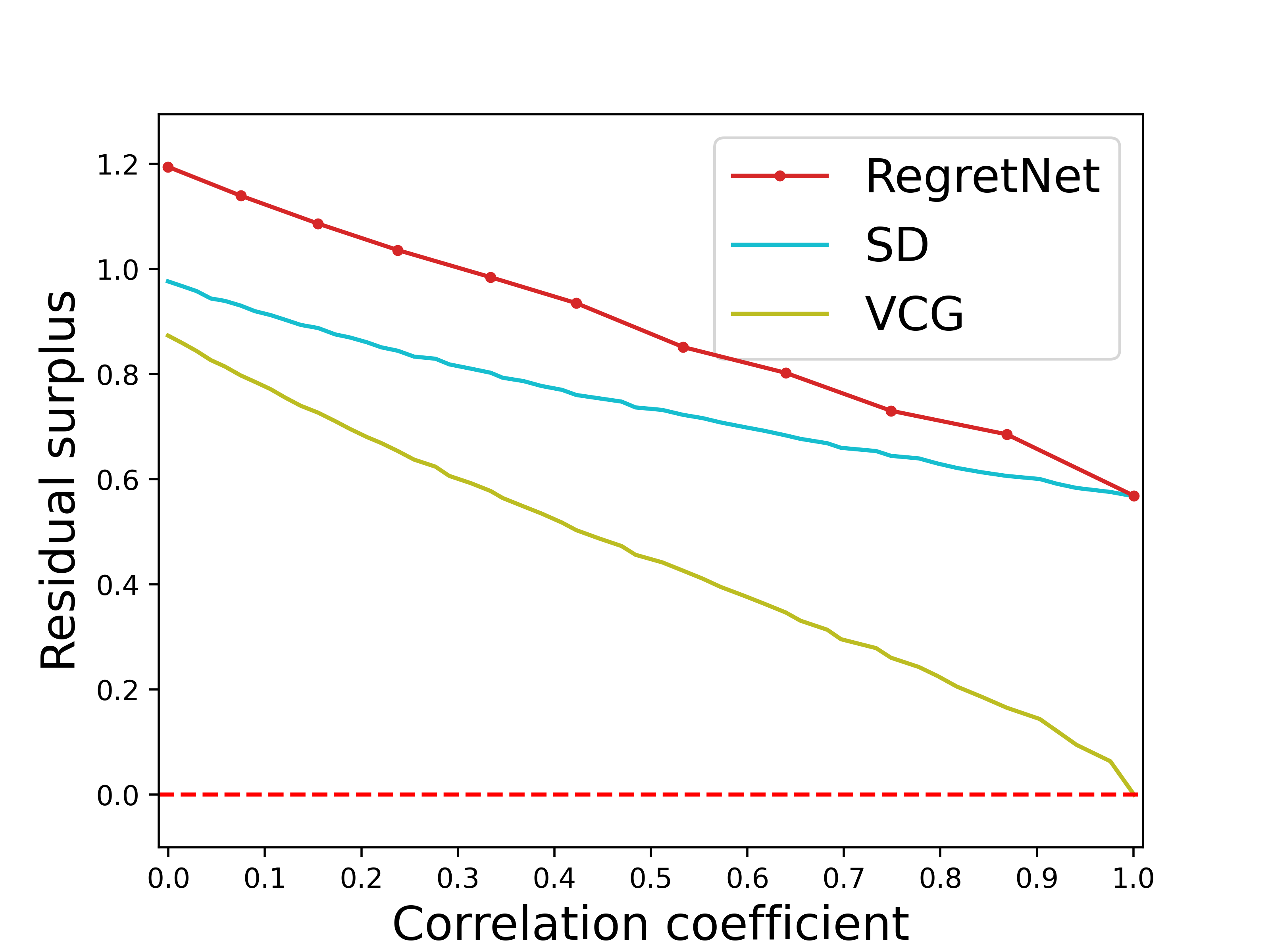}
        \subcaption{The performance of SD, VCG, and RegretNet given $\alpha = 0.8$}
        \label{fig: correlation_between_weibull08}
    \end{subfigure}
    \caption{The performance of mechanisms under between-agent correlation. The horizontal axis represents $\corr(v_{k}^{i}, v_{k}^{j})$.}
    \label{fig: between-agent correlation}
\end{figure}

Next, we evaluate the effect of the \emph{between-agent correlation}, the correlation between $v_{k}^{i}$ and $v_{k}^{j}$ for two distinct agents $i \neq j$ (i.e., some objects are more popular than other objects). When such a correlation becomes stronger, the environment approaches a common-value scenario, where goods with high common value generate high value regardless of who receives them. In such cases, the performance of SD improves, and SD becomes the first best in the case of pure common values. In contrast, adopting a mechanism that involves screening, such as VCG, would lead to intense competition for popular goods. Consequently, the common value would be entirely lost due to money burning. The following proposition formalizes this observation.

\begin{proposition}\label{thm: between-agent correlation first best}
    Consider a profile of marginal distributions $(F_k)_{k \in K}$, where $\mathbb{E}_{v \sim F_k}[v] = \mu_k > 0$ and $\mathrm{Var}_{v \sim F_k}(v) = \sigma_k^2 \in (0, + \infty)$. Consider a sequence of joint distribution functions $F^\rho$ such that for all $\rho$, (i) for every agent $i \in I$ and object $k \in K$, the marginal distribution of $v_k^i$ is $F_k$, and (ii) for all distinct agents $i, j \in I$ and object $k \in K$, $\mathrm{Corr}(v_k^i,v_k^j) \ge \rho$. Then, as $\rho \to 1$, we have (a) $RS(\mech_{SD}; F^\rho)/RS(\mech_{FB}^\rho; F^\rho) \to 1$, and (b) $RS(\mech_{VCG}; F^\rho) \to 0$, where $\mech_{FB}^\rho$ is a mechanism that maximizes residual surplus \eqref{eq: defn social welfare_finite} subject to the resource constraint \eqref{eq: defn resource_finite} and the unit demand condition \eqref{eq: defn unit demand_finite}.
\end{proposition}

Figure~\ref{fig: correlation_between_ratios} shows the performance ratio of SD and VCG, whereas Figure~\ref{fig: correlation_between_weibull08} shows the residual surplus achieved by SD, VCG, and RegretNet for the case of $\alpha = 0.8$. The simulation setting parallels that for the within-agent correlation cases, except that we vary $\corr(v_{k}^{i}, v_{k}^{j})$ instead of $\corr(v_{k}^{i}, v_{l}^{i})$. We set $\corr(v_{k}^{i}, v_{l}^{j}) = 0$ for all $k \neq l$. When $\corr(v_{k}^{i}, v_{k}^{j}) = 0$, values are drawn i.i.d., and when $\corr(v_{k}^{i}, v_{k}^{j}) = 1$, we have pure common values. For every parameter $\alpha$, as the between-agent correlation becomes stronger, the ratio of SD to VCG performance increases, indicating that the relative performance of SD improves. Furthermore, in the case of perfect correlation, the residual surplus under VCG becomes zero.

\section{Application: Downstream Appointment Scheduling in Mass Vaccination}
\label{sec: mass vaccination}

\subsection{Downstream Appointment Scheduling}

In this section, we propose downstream appointment scheduling as an
application. Mass vaccination combines several allocation decisions.
Authorities procure doses, allocate supply across locations, define
eligibility, and rank recipients according to medical risk, occupation, and
other public objectives. We take these upstream decisions as given. We focus
on the downstream appointment scheduling problem that begins once a set of
recipients is eligible to book: which recipient obtains which available
appointment slot.

This downstream appointment scheduling problem is of significant economic
importance. Priority rules in mass campaigns are typically coarse, and thus many
recipients are eligible at the same time and share the same policy priority.
Appointment slots, meanwhile, differ in date, time, location, and, where
relevant, vaccine type, and potential recipients may therefore have diverse
preferences over them. While appointment assignment is only one component of
vaccination policy, it is downstream in sequence rather than small in scale.
Mass vaccination is an extremely large-scale problem, and every vaccination
requires coordination of a recipient, a dose, a time, and a site. Even modest
losses at each booking can therefore become economically important in
aggregate.

The mechanism-design question is how appointment slots are assigned within
the policy hierarchy. Agents are potential recipients who are currently
eligible under the given upstream policy. The value $v_k^i$ can include both
recipient $i$'s value of receiving the vaccine and the fit of slot $k$ with
recipient $i$'s schedule, location, and other constraints, whereas $v_k^i$
does \emph{not} reflect the planner's social priority for vaccinating
recipient $i$. Throughout this section, no screening means no screening
through costly booking effort within the policy hierarchy, not the absence of
medical eligibility or priority rules.

The value of an appointment slot can vary for three distinct reasons. First,
recipient--slot-specific factors, such as work schedules, travel costs, and accessibility constraints, generate
idiosyncratic relative preferences over a large menu of dates, times, and sites. 
This creates effective object variety, which enhances the relative performance of no-screening mechanisms. Second, substantial between-agent correlation is natural in this
application: earlier dates or convenient times and sites are commonly
attractive. Our analysis reveals that a race for commonly preferred slots dissipates part of the common value attached to those slots, and therefore, between-agent correlation improves the performance of no-screening mechanisms even further.

Third, recipients differ in their overall value of vaccination, including the
value of receiving it promptly, which generates within-agent correlation
across slots. This component generally strengthens the case for
screening. In this application, however, much of its observable component is
already incorporated into the upstream hierarchy. Medical vulnerability,
occupational exposure, age, and similar observable characteristics related to the value of vaccination are natural inputs
into eligibility and priority, and the authority can use them directly rather
than require recipients to burn effort to signal them. The hierarchy does not
eliminate all variation in overall value, but it narrows the remaining margin
on which effort screening could be useful. Conditional on this upstream use
of observable need, the downstream problem therefore combines substantial
slot variety and common slot popularity with a narrower residual role for
effort screening. Accordingly, our analysis predicts that a no-screening mechanism should perform well in downstream appointment scheduling in mass vaccination.

\subsection{Open FCFS}
\label{subsec: open fcfs vaccination}

During the COVID-19 rollout, many jurisdictions used an \emph{open first-come-first-served system} (open FCFS). Under open FCFS, the authority releases a pool of
appointments and allows all recipients who are currently eligible under the
policy hierarchy to compete for access at the same time. The order in which
recipients successfully reach the booking interface determines the order in
which they choose among the remaining slots. Variants of this architecture
appeared in physical queues, online portals, and telephone booking systems in
Florida, Massachusetts, Washington, D.C., Cyprus, and many Japanese
municipalities
\citep{fox_leecounty_2021,routefifty_eventbrite_2021,
nbcboston_massportal_2021,dcist_dc_2021,cyprusmail_portal_2021,
asahi2021B,asahi2021A,yomiuri2021}.

Open FCFS turns response speed into an endogenous priority. Recipients can
move forward by monitoring announcements, arriving or logging in before a
release, repeatedly refreshing a website, remaining in physical or virtual
queues, or making repeated telephone calls. These activities are ``burned
payments'' in the sense that they are costly; they affect both whether a recipient receives an appointment and which slot
she obtains, but they neither expand vaccination capacity nor produce revenue
for the authority. 

When effort advances the access order and recipients choose effort in
response to their values, the resulting race resembles an all-pay
contest for early access. In the reduced benchmark, its monotone
equilibrium implements the full-screening allocation and, by revenue
equivalence, yields the same expected utilities as VCG. This
correspondence need not hold in a general multidimensional market.

The case against open FCFS does not depend on effort being a poor
signal of value. Even in the reduced benchmark, where the race sorts
recipients efficiently, the additional gross surplus can be outweighed
by burned effort. Greater object variety weakens the case for this
costly sorting, while common slot popularity adds value that competition
can dissipate. The design implication is to preserve recipients' choice
among heterogeneous slots while replacing effort-based access priority
with an exogenous order within the existing policy hierarchy.



The historical record indicates that the associated effort was substantial,
not merely a theoretical possibility. In Lee County, Florida, residents
camped overnight and waited for hours before officials moved away from the
walk-up system \citep{fox_leecounty_2021}. In other Florida counties, online
appointments disappeared within minutes, and call centers were overwhelmed
\citep{routefifty_eventbrite_2021}. When Massachusetts expanded eligibility,
hundreds of thousands of residents attempted to book and the state portal
failed; Washington, D.C., and Cyprus experienced similar crashes and jammed
phone lines
\citep{nbcboston_massportal_2021,dcist_dc_2021,
cyprusmail_portal_2021}. Japanese municipalities likewise faced overloaded
websites and telephone lines, including a suspension of reservations in
Yokohama \citep{asahi2021B,asahi2021A,yomiuri2021}. The disruption was large
enough that Minnesota replaced its previous release with a 24-hour
preregistration lottery, and San Luis Obispo County announced a similar move
\citep{startribune_minnesota_2021,mustangnews_slo_2021}; local reporting on
the Minnesota lottery described it as fairer and easier than the preceding
FCFS process \citep{postbulletin_minnesota_2021}.

Recipient time and attention spent waiting, monitoring, and repeatedly
attempting to book correspond to the effort costs in our model. Open FCFS
also generated additional costs outside the model. Long physical queues
created congestion at vaccination sites, while simultaneous online access
and repeated calls overloaded portals, servers, and call centers. These are
system-level costs imposed by the common opening itself, rather than merely
private effort borne by the recipients who compete. They provide an
additional practical reason to avoid an open race for access.

\subsection{Register-Invite-Book}
\label{subsec: register invite book}

There are several ways to implement a no-screening allocation, and they differ
substantially in communication and timing. A literal direct-revelation
implementation of SD would ask each recipient to rank a potentially large set
of appointment slots before the allocation is computed. Such rankings are
costly to communicate and become stale as new slots are added. Until the
assignment is run and communicated, a recipient may also need to keep several
possible dates open. Open FCFS avoids these problems: a recipient names only
one currently available slot and receives immediate confirmation. These
interface advantages plausibly help explain its widespread use, even when the
open race used to determine access is undesirable.\footnote{Kakogawa City
initially prepared a direct implementation of random SD, but switched to open
FCFS before vaccination expanded. The accompanying account emphasizes the
operational burden of eliciting and processing detailed preferences
\citep{Tada2021Kakogawa}.} Practical mechanism design should therefore
preserve the simple choice and immediate confirmation offered by FCFS while
removing costly competition for the right to choose.

We propose the \emph{register-invite-book system} (RIB) as a mechanism that approximately implements no screening with a convenient interface. At \emph{registration},
recipients provide contact information and the observable information
required by the existing eligibility and priority policy. They need not
report any preference upon registration. The authority then orders
registered recipients using the policy hierarchy and an exogenous
tie-breaker, such as a lottery, among recipients whom the policy treats
alike. The system sequentially and gradually sends \emph{invitations}
in that order, adjusting the invitation rate to current inventory. Upon receiving an invitation, a recipient
accesses the menu, selects one preferred remaining slot, and completes the
\emph{booking} with immediate confirmation. Booking may be FCFS among the
recipients in the same invitation batch, but the batch is deliberately kept
small.

An invitation corresponds to the arrival of a recipient's turn in SD. Before
receiving it, a recipient cannot accelerate her turn by monitoring the site,
refreshing a page, or waiting in a queue. With one active invitee at a time
and a fixed inventory, RIB implements SD in sequential form. A practical
system will generally invite a small batch, in which case it approximates SD:
response speed can move a recipient ahead only of other members of the same
batch, rather than the entire eligible population. Batch size, invitation pace, and the response
window are therefore mechanism-design parameters, not merely technical
details.

RIB retains the interface advantages of open FCFS. After receiving an
invitation, a recipient reports only one choice and learns the assignment
immediately. Her decision problem is also transparent: choose the favorite
slot among those currently remaining. In the exact one-at-a-time benchmark,
this local simplicity has a formal counterpart in obvious
strategy-proofness \citep{li2017obviously}. Indeed, \citet{Pycia2024random} show that random SD in
sequential form is the unique mechanism satisfying Pareto efficiency,
symmetry, and obvious strategy-proofness. RIB preserves this sequential-choice logic while
using policy and an exogenous tie-breaker, rather than costly response speed,
to determine when the choice menu becomes available.

British Columbia and Singapore used closely related systems during the
COVID-19 rollout. British Columbia asked residents to register once and sent
invitations to book as age eligibility and supply advanced; Singapore
collected expressions of interest and sent SMS booking links as appointments
became available
\citep{bc_getvaccinated_2021,govinsider_singapore_2021}. These cases
demonstrate that the register-and-invite architecture is operationally
feasible at mass scale without a single opening at which the entire eligible
population competes. The Singapore account specifically attributes the
avoidance of booking-site surges to separating registration from invitation
and booking \citep{govinsider_singapore_2021}.

RIB offers several further advantages in a setting with policy priorities and
gradual supply. The authority can place recipients into policy-determined
priority groups, break ties exogenously within each group, and place a
late-registering high-priority recipient ahead of lower-priority recipients
who have not yet been invited. Persistent registration also means that new
supply does not require unsuccessful recipients to return whenever
appointments are released, or require the authority to elicit a new ranking
over each new menu; a recipient can typically interact with the system only
at registration and booking.\footnote{In New York City, new appointments
were released around midnight, leading recipients to check repeatedly at
that time \citep{verge_nyc_2021}.} Finally, the system can disclose progress
through the invitation queue, allowing recipients to form a rough expectation
of when their turn will arrive. Supply uncertainty and new registrations
prevent exact prediction, but RIB gives recipients more guidance about when
to keep time available than repeated open releases. These implementation
benefits are not part of our formal residual-surplus comparison, but they
reinforce its design implication: preserve recipient choice among
heterogeneous slots while removing costly competition for access to the menu.

\section{Concluding Remarks}

We study residual-surplus-maximizing mechanisms for allocating
multiple heterogeneous objects when agents can signal their values only by
taking costly actions. Our findings reveal a clear trend: as
object variety increases, the case for costly screening weakens, and
no-screening mechanisms such as SD become more attractive.

We first establish this result in a stylized continuous i.i.d.\ market. 
When the marginal value distribution is CDF log-concave,
the multidimensional problem reduces to a single-dimensional one in which an
agent's value is her highest value across object types. The relevant
distribution is therefore the largest-order-statistic distribution
$G_K$. No screening is optimal if and only if $G_K$ is NBUE, and this
property, once satisfied, is preserved as the number of object types
increases. The region over which $G_K$ has an IHR also
expands with variety. With bounded support, no screening approaches the
first best; in the Gumbel domain, an efficient allocation becomes constant
for almost all types; and under Fr\'echet limits, screening is concentrated
among agents in the upper tail. These results identify the distributional
force behind the comparative static: increasing variety changes not only the
value of an agent's best attainable match, but also the shape of the value
distribution that governs the return to screening.

Second, we use automated mechanism design to study finite and correlated
environments in which the exact reduction is unavailable. The numerical
results show that the same comparative-static pattern remains visible in
finite markets. The correlation exercises clarify that what matters is
effective rather than nominal variety. Within-agent correlation weakens the
force by making the objects closer to homogeneous from each agent's
perspective, whereas between-agent correlation can strengthen the case
against screening by making competition over commonly popular objects more
dissipative. 
These exercises do not imply that SD is universally optimal,
but they indicate that the force identified in the continuous benchmark is
not merely an artifact of the continuum or i.i.d.\ assumptions. Combined
with SD's simplicity, strategy-proofness, and limited informational
requirements, these results make SD a natural benchmark for multi-object allocation
without transfers.

Finally, the analysis suggests a broader design principle for practical
allocation systems. In the benchmark model, the value relevant for screening
is exactly a maximum; more generally, when recipients choose among many
sufficiently differentiated options, their realized value is often governed
by an upper order statistic. An increase in effective variety can therefore
reduce the incremental benefit of screening agents by costly effort even when
the exact assumptions of the model do not hold. The same logic may be
relevant to reservations for municipal sports facilities, community halls,
public campsites, and cultural venues; appointments at public clinics,
health-screening centers, passport or licensing offices, and other
administrative services; and time- and location-specific places in publicly
subsidized educational, care, or recreational programs. Eligibility rules
and substantive policy priorities may remain essential in all of these
settings. The design question is whether waiting, repeated applications,
paperwork, or response speed should create an additional ranking among agents
who are otherwise similarly prioritized. Before using such screening
devices, a planner should account for how much of the desired sorting is
already achieved by allowing agents to choose among heterogeneous options.

\bibliographystyle{ecta}
\bibliography{references,news}

\pagebreak

\appendix

\renewcommand{\thefigure}{\thesection.\arabic{figure}}
\setcounter{figure}{0}

\part*{Appendix}

\section{Exact Reduction and Its Scope}

This section contains the proofs of the three main-text results concerning
the exact reduction. Section~\ref{subsec: proof exact reduction} proves
Theorem~\ref{thm: exact reduction}.
Section~\ref{subsec:ex_post_efficiency_is_not_without_loss} proves
Theorem~\ref{thm:nonfavorite-screening} by constructing and smoothing a
counterexample.
Section~\ref{subsec:neutrality-ex-post-reduction} proves
Theorem~\ref{thm:neutrality-ex-post-reduction} through a stronger
mechanism-level correspondence.

\subsection{Exact Reduction}
\label{subsec: proof exact reduction}

This subsection proves that, under CDF log-concavity, allowing a mechanism to use
non-favorite objects as screening devices cannot increase the supremum of residual
surplus.  The proof proceeds in four steps.  We first symmetrize the mechanism and
rewrite it as a mechanism over ranked values.  We then couple the lower order
statistics conditional on the highest value.  CDF log-concavity makes every
coordinate of this coupling $1$-Lipschitz in the highest value.  Finally, the
multidimensional incentive constraints reduce pathwise to a single-dimensional system
that retains only upward incentive constraints, and we show that this relaxation has
the same value as the usual single-dimensional problem.

Throughout the proof, all allocation and payment rules are taken to be Borel
measurable.  For a permutation $\sigma$ of the object labels, define
$(\sigma\bv)_k\coloneqq v_{\sigma^{-1}(k)}$, and use the same action on allocation
vectors.

\subsubsection{Symmetrization and Rank Representation}

\begin{definition}[Neutrality]
\label{defn:neutrality}
A mechanism $\mech=(\bx,p)$ is \emph{neutral} if, for every permutation
$\sigma$ of the object labels and every $\bv\in V$,
\begin{equation}
    \bx(\sigma\bv)=\sigma\bx(\bv)
    \quad\text{and}\quad
    p(\sigma\bv)=p(\bv).
\end{equation}
\end{definition}

\begin{lemma}[Symmetrization]
\label{lem: exact reduction symmetrization}
Every feasible mechanism in the unrestricted multidimensional problem has a
neutral feasible symmetrization with the same residual surplus.
\end{lemma}

\begin{proof}
Let $\mathfrak S_K$ denote the set of permutations of the object labels.  For each
$\sigma\in\mathfrak S_K$, define
\begin{equation}
    \bx^{\sigma}(\bv)\coloneqq \sigma^{-1}\bx(\sigma\bv),
    \qquad
    p^{\sigma}(\bv)\coloneqq p(\sigma\bv).
\end{equation}
For a true type $\bv$ and a report $\bv'$, we have
\begin{equation}
    \sum_{k\in K}v_k x_k^{\sigma}(\bv')-p^{\sigma}(\bv')
    =
    \sum_{k\in K}(\sigma\bv)_k x_k(\sigma\bv')-p(\sigma\bv').
\end{equation}
Hence $(\bx^{\sigma},p^{\sigma})$ inherits strategy-proofness, individual
rationality, and unit demand from $(\bx,p)$.  Exchangeability of $F$ and equality of
the capacities imply
\begin{equation}
    \int_V x_k^{\sigma}(\bv)dF(\bv)
    =
    \int_V x_{\sigma(k)}(\bv)dF(\bv)
    \le \frac{\bar m}{K}
    \qquad\text{for every }k,
\end{equation}
so the resource constraints are also preserved.  The same change of variables
shows that $(\bx^{\sigma},p^{\sigma})$ has the same residual surplus as
$(\bx,p)$.

Now average these mechanisms over $\mathfrak S_K$:
\begin{equation}
    \bar\bx(\bv)
    \coloneqq
    \frac{1}{K!}\sum_{\sigma\in\mathfrak S_K}\bx^{\sigma}(\bv),
    \qquad
    \bar p(\bv)
    \coloneqq
    \frac{1}{K!}\sum_{\sigma\in\mathfrak S_K}p^{\sigma}(\bv).
\end{equation}
All feasibility constraints are linear, so $(\bar\bx,\bar p)$ is feasible and
has the same residual surplus.  Reindexing the average gives
$\bar\bx(\tau\bv)=\tau\bar\bx(\bv)$ and
$\bar p(\tau\bv)=\bar p(\bv)$ for every $\tau\in\mathfrak S_K$.
\end{proof}

Let
\begin{align}
    E_K
    &\coloneqq
    \{\eta\in[0,\bar v)^K:\eta_1\ge\eta_2\ge\cdots\ge\eta_K\},\\
    Y_K
    &\coloneqq
    \left\{y\in[0,1]^K:
    y_1\ge y_2\ge\cdots\ge y_K\ge0,
    \ \sum_{j=1}^K y_j\le1\right\}.
\end{align}
For $\bv\in V$, let $\eta(\bv)$ denote the decreasing rearrangement of $\bv$,
and let $\mu_K$ denote the distribution of $\eta(\bv)$ under $F$.

\begin{lemma}[Sorted allocations at sorted types]
\label{lem: exact reduction sorted allocations}
If a neutral mechanism satisfies strategy-proofness and unit demand, then
$\bx(\eta)\in Y_K$ for every $\eta\in E_K$.
\end{lemma}

\begin{proof}
Fix $i<j$ and let $\tau$ exchange coordinates $i$ and $j$.  If
$\eta_i>\eta_j$, strategy-proofness against the report $\tau\eta$, together
with neutrality, gives
\begin{equation}
    \bigl(x_i(\eta)-x_j(\eta)\bigr)(\eta_i-\eta_j)\ge0.
\end{equation}
If $\eta_i=\eta_j$, then $\tau\eta=\eta$, and neutrality gives
$x_i(\eta)=x_j(\eta)$.  Thus the allocation probabilities are nonincreasing
in rank.  Unit demand then implies $\bx(\eta)\in Y_K$.
\end{proof}

A \emph{rank mechanism} is a pair of Borel maps
$y:E_K\to Y_K$ and $p^r:E_K\to\mathbb R_+$ such that, for all
$\eta,\eta'\in E_K$,
\begin{align}
    \sum_{j=1}^K \eta_j y_j(\eta)-p^r(\eta)
    &\ge
    \sum_{j=1}^K \eta_j y_j(\eta')-p^r(\eta'),
    \label{eq: exact reduction rank IC}\\
    \sum_{j=1}^K \eta_j y_j(\eta)-p^r(\eta)
    &\ge0,
    \label{eq: exact reduction rank IR}
\end{align}
and
\begin{equation}
    \int_{E_K}\sum_{j=1}^K y_j(\eta)d\mu_K(\eta)\le\bar m.
    \label{eq: exact reduction rank resource}
\end{equation}
Here, $y_j(\eta)$ is the probability that a type with ordered values $\eta$
receives her $j$-th ranked object.  For a rank mechanism, write
\begin{equation}
    RS(y,p^r)
    \coloneqq
    \int_{E_K}
    \left(\sum_{j=1}^K\eta_j y_j(\eta)-p^r(\eta)\right)d\mu_K(\eta).
\end{equation}
Let $V^{\mathrm{rank}}$ denote the supremum of $RS(y,p^r)$ over rank
mechanisms, and let $V^{\mathrm{full}}$ denote the supremum in the unrestricted
multidimensional problem.

\begin{lemma}[Rank representation]
\label{lem: exact reduction rank representation}
We have $V^{\mathrm{full}}=V^{\mathrm{rank}}$.
\end{lemma}

\begin{proof}
By Lemma~\ref{lem: exact reduction symmetrization}, it is without loss to start
with a neutral feasible mechanism.  Restrict it to $E_K$ and define
$y(\eta)\coloneqq\bx(\eta)$ and $p^r(\eta)\coloneqq p(\eta)$.
Lemma~\ref{lem: exact reduction sorted allocations} gives $y(\eta)\in Y_K$.
The original strategy-proofness and individual-rationality constraints imply
\eqref{eq: exact reduction rank IC} and
\eqref{eq: exact reduction rank IR}.  Moreover,
\begin{equation}
    \int_{E_K}\sum_{j=1}^K y_j(\eta)d\mu_K(\eta)
    =
    \int_V\sum_{k\in K}x_k(\bv)dF(\bv)
    \le\bar m.
\end{equation}
Neutrality also implies that the truthful utility depends only on the ordered
vector $\eta(\bv)$, so residual surplus is preserved.  Hence
$V^{\mathrm{full}}\le V^{\mathrm{rank}}$.

Conversely, take a rank mechanism $(y,p^r)$.  Fix a Borel rule
$\pi(\bv)=(\pi_1(\bv),\ldots,\pi_K(\bv))$ that lists the object labels in
weakly decreasing order of value, using any deterministic rule to break ties.
Define a mechanism in the original market by
\begin{equation}
    x_{\pi_j(\bv)}(\bv)\coloneqq y_j(\eta(\bv))
    \quad (j=1,\ldots,K),
    \qquad
    p(\bv)\coloneqq p^r(\eta(\bv)).
    \label{eq: exact reduction rank lift}
\end{equation}
For a true type $\bv$ and a report $\bv'$, the rearrangement inequality and
\eqref{eq: exact reduction rank IC} imply
\begin{align}
    \sum_{k\in K}v_kx_k(\bv')-p(\bv')
    &\le
    \sum_{j=1}^K \eta_j(\bv)y_j(\eta(\bv'))-p^r(\eta(\bv'))\\
    &\le
    \sum_{j=1}^K \eta_j(\bv)y_j(\eta(\bv))-p^r(\eta(\bv))\\
    &=
    \sum_{k\in K}v_kx_k(\bv)-p(\bv).
\end{align}
Thus the induced mechanism is strategy-proof; individual rationality and unit
demand are immediate.  Since $G$ is continuous, ties have probability zero.
Away from the tie set, exchangeability implies that every object has the same
expected load.  Therefore,
\begin{equation}
    \int_Vx_k(\bv)dF(\bv)
    =
    \frac{1}{K}
    \int_{E_K}\sum_{j=1}^K y_j(\eta)d\mu_K(\eta)
    \le\frac{\bar m}{K}
\end{equation}
for every $k$.  Truthful utility, and hence residual surplus, is the same in
the two representations.  This proves $V^{\mathrm{rank}}\le
V^{\mathrm{full}}$.
\end{proof}

\subsubsection{Single-Dimensional Value}

Let $H$ be a continuous distribution function on $[0,\bar v)$ with finite
mean, where $\bar v$ may equal $+\infty$.  For $b\in[0,1]$, define
\begin{equation}
    \mathcal R(H,b)
    \coloneqq
    \sup_x
    \left\{
    \int_0^{\bar v}(1-H(t))x(t)dt:
    \begin{array}{l}
        x:[0,\bar v)\to[0,1]\text{ is Borel and nondecreasing},\\[-0.1em]
        \displaystyle\int_0^{\bar v}x(t)dH(t)\le b
    \end{array}
    \right\}.
    \label{eq: exact reduction one dimensional value}
\end{equation}

\begin{lemma}[Single-dimensional envelope representation]
\label{lem: exact reduction one dimensional representation}
The value of the single-dimensional residual-surplus problem with value
distribution $H$ and capacity $b$ is $\mathcal R(H,b)$.  Moreover,
$b\mapsto\mathcal R(H,b)$ is finite, nondecreasing, and concave on $[0,1]$.
In particular, the value of the reduced problem
\eqref{eq: reduced objective}--\eqref{eq: reduced ir} is
$\mathcal R(G_K,\bar m)$.
\end{lemma}

\begin{proof}
By the standard single-dimensional incentive-compatibility and envelope
characterization, feasible mechanisms are equivalent to nondecreasing
allocation rules $x:[0,\bar v)\to[0,1]$ satisfying the resource constraint,
with truthful utility $U(t)=\int_0^t x(s)ds$.
Tonelli's theorem then gives
\[
    \int_0^{\bar v}U(t)dH(t)
    =
    \int_0^{\bar v}(1-H(t))x(t)dt,
\]
so the value is $\mathcal R(H,b)$.

Monotonicity in $b$ is immediate, while concavity follows by taking convex
combinations of feasible allocation rules.  Finally,
\[
    0\le\mathcal R(H,b)
    \le\int_0^{\bar v}(1-H(t))dt
    =\mathbb E_{T\sim H}[T]<+\infty.
\]
Setting $H=G_K$ and $b=\bar m$ proves the final claim.
\end{proof}

\subsubsection{Coupling and Pathwise Reduction}

Use the quantile convention
\begin{equation}
    G^{-1}(q)
    \coloneqq
    \inf\{t\in[0,\bar v):G(t)\ge q\},
    \qquad q\in[0,1).
\end{equation}
Because $G(t)<1$ for every $t<\bar v$, every argument
$Z_{(j-1)}G(t)$ used below lies in this quantile's domain.
Let $Z_1,\ldots,Z_{K-1}$ be i.i.d.\ uniform random variables on $[0,1]$,
independent of a random variable $T\sim H\coloneqq G_K$.  Let
$Z_{(1)}\ge\cdots\ge Z_{(K-1)}$ denote their decreasing order statistics, and
write $\omega=(Z_{(1)},\ldots,Z_{(K-1)})$.  Define
\begin{equation}
    \eta_1(t,\omega)\coloneqq t,
    \qquad
    \eta_j(t,\omega)
    \coloneqq
    G^{-1}\bigl(Z_{(j-1)}G(t)\bigr),
    \quad j=2,\ldots,K.
    \label{eq: exact reduction coupling}
\end{equation}

\begin{lemma}[Conditional-order-statistic coupling]
\label{lem: exact reduction coupling}
The random vector $\eta(T,\omega)$ has distribution $\mu_K$, and $T$ is
independent of $\omega$.
\end{lemma}

\begin{proof}
Conditional on the maximum of $K$ i.i.d.\ draws being $t$, the remaining
$K-1$ values are i.i.d.\ with the distribution $G$ truncated to $[0,t]$.
Their conditional CDF is $G(s)/G(t)$ for $s\le t$, so each has the
representation $G^{-1}(ZG(t))$ with $Z$ uniform on $[0,1]$.  Sorting these
values gives \eqref{eq: exact reduction coupling}.  This constructs the
regular conditional distribution of the ordered vector given its maximum,
with the auxiliary uniforms independent of $T$.
\end{proof}

\begin{lemma}[Pathwise Lipschitz property]
\label{lem: exact reduction Lipschitz}
Suppose that $G$ is CDF log-concave.  For every $z\in[0,1]$, the map
$\phi_z(t)\coloneqq G^{-1}(zG(t))$ satisfies
\begin{equation}
    0\le \phi_z(t)-\phi_z(s)\le t-s
    \qquad\text{for all }0\le s<t<\bar v.
    \label{eq: exact reduction Lipschitz}
\end{equation}
Consequently, for every $\omega$ and every $j$,
\begin{equation}
    0\le\eta_j(t,\omega)-\eta_j(s,\omega)\le t-s
    \qquad\text{for all }0\le s<t<\bar v.
    \label{eq: exact reduction coordinate Lipschitz}
\end{equation}
\end{lemma}

\begin{proof}
The claim is immediate for $z=0$.  Fix $z\in(0,1]$.  Since $G$ is increasing,
$\phi_z$ is increasing and $\phi_z(t)\le t$.  Let $L\coloneqq\log G$ on
$(0,\bar v)$.  CDF log-concavity means that $L$ is concave, and
\begin{equation}
    L(\phi_z(t))=L(t)+\log z.
    \label{eq: exact reduction log identity}
\end{equation}
Fix $0<s<t$, and put $d\coloneqq t-s$, $a\coloneqq\phi_z(t)$, and
$a'\coloneqq\phi_z(s)$.  Suppose, toward a contradiction, that
$a-a'>d$.  Then $a-d>a'>0$.  Since $a\le t$ and increments of a concave
function over a fixed distance are nonincreasing in their starting point,
\begin{equation}
    L(a)-L(a-d)
    \ge
    L(t)-L(t-d)
    =L(t)-L(s).
\end{equation}
On the other hand, strict monotonicity of $L$, $a-d>a'$, and
\eqref{eq: exact reduction log identity} give
\begin{equation}
    L(a)-L(a-d)
    <
    L(a)-L(a')
    =L(t)-L(s),
\end{equation}
which is a contradiction.  Thus $a-a'\le t-s$.  The case $s=0$ follows from
$\phi_z(0)=0$ and $0\le\phi_z(t)\le t$.  Applying this argument to every
coordinate in \eqref{eq: exact reduction coupling}, together with
$\eta_1(t,\omega)=t$, proves
\eqref{eq: exact reduction coordinate Lipschitz}.
\end{proof}

For a continuous distribution $H$ on $[0,\bar v)$ with finite mean and a
budget $b\in[0,1]$, call a pair of Borel functions
$(u,Y):[0,\bar v)\to\mathbb R^2$ feasible for the \emph{upward system} if
\begin{align}
    0\le Y(t)\le1,
    \qquad
    0\le u(t)\le tY(t)
    &\quad\text{for all }t,
    \label{eq: exact reduction upward 1}\\
    u(s)\ge u(t)-(t-s)Y(t)
    &\quad\text{for all }0\le s<t<\bar v,
    \label{eq: exact reduction upward 2}\\
    \int_0^{\bar v}Y(t)dH(t)&\le b.
    \label{eq: exact reduction upward 3}
\end{align}
Let
\begin{equation}
    \mathcal U(H,b)
    \coloneqq
    \sup\left\{
    \int_0^{\bar v}u(t)dH(t):(u,Y)\text{ satisfies }
    \eqref{eq: exact reduction upward 1}\text{--}\eqref{eq: exact reduction upward 3}
    \right\}.
    \label{eq: exact reduction upward value}
\end{equation}
Condition \eqref{eq: exact reduction upward 2} is precisely the family of
single-dimensional incentive constraints preventing a lower type from imitating
a higher type; the reverse constraints are not imposed.

\begin{lemma}[Pathwise reduction]
\label{lem: exact reduction pathwise}
Suppose that $G$ is CDF log-concave, and let $(y,p^r)$ be any feasible rank
mechanism.  For each $\omega$, define
\begin{align}
    u(t,\omega)
    &\coloneqq
    \sum_{j=1}^K y_j(\eta(t,\omega))\eta_j(t,\omega)
    -p^r(\eta(t,\omega)),\\
    Y(t,\omega)
    &\coloneqq
    \sum_{j=1}^K y_j(\eta(t,\omega)),\\
    b(\omega)
    &\coloneqq
    \int_0^{\bar v}Y(t,\omega)dH(t).
\end{align}
Then $(u(\cdot,\omega),Y(\cdot,\omega))$ is feasible for the upward system
with budget $b(\omega)$.  Moreover,
\begin{align}
    RS(y,p^r)
    &=
    \mathbb E_{\omega}
    \left[\int_0^{\bar v}u(t,\omega)dH(t)\right],
    \label{eq: exact reduction path RS}\\
    \mathbb E_{\omega}[b(\omega)]&\le\bar m.
    \label{eq: exact reduction path budget}
\end{align}
\end{lemma}

\begin{proof}
Individual rationality gives $u(t,\omega)\ge0$.  Since payments are
nonnegative, $\eta_j(t,\omega)\le\eta_1(t,\omega)=t$, and
$\sum_jy_j\le1$, we also have
\begin{equation}
    u(t,\omega)
    \le
    \sum_{j=1}^K y_j(\eta(t,\omega))\eta_j(t,\omega)
    \le tY(t,\omega).
\end{equation}
Thus \eqref{eq: exact reduction upward 1} holds.

Fix $0\le s<t<\bar v$.  Incentive compatibility of the rank mechanism,
applied to the true type $\eta(s,\omega)$ and the report $\eta(t,\omega)$,
gives
\begin{align}
    u(s,\omega)
    &\ge
    \sum_{j=1}^K y_j(\eta(t,\omega))\eta_j(s,\omega)
    -p^r(\eta(t,\omega))\\
    &=
    u(t,\omega)
    -
    \sum_{j=1}^K y_j(\eta(t,\omega))
    \bigl(\eta_j(t,\omega)-\eta_j(s,\omega)\bigr)\\
    &\ge
    u(t,\omega)-(t-s)Y(t,\omega),
\end{align}
where the last inequality follows from
Lemma~\ref{lem: exact reduction Lipschitz}.  This proves
\eqref{eq: exact reduction upward 2}; the budget condition holds by the
definition of $b(\omega)$.  Finally,
Lemma~\ref{lem: exact reduction coupling} and Tonelli's theorem give
\eqref{eq: exact reduction path RS} and
\begin{equation}
    \mathbb E_{\omega}[b(\omega)]
    =
    \int_{E_K}\sum_{j=1}^K y_j(\eta)d\mu_K(\eta)
    \le\bar m,
\end{equation}
which is \eqref{eq: exact reduction path budget}.  All integrands are jointly
measurable because the coupling map is measurable and $y,p^r$ are Borel.
\end{proof}

\subsubsection{Upward Reduction Is Without Loss}

We next prove the single-dimensional result used to aggregate the pathwise
problems.  For a feasible pair $(u,Y)$ and $t\in[0,\bar v)$, define
\begin{equation}
    m_t(s)\coloneqq\inf_{r\in[s,t]}Y(r),
    \qquad
    \psi(t)\coloneqq\int_0^t m_t(s)ds.
    \label{eq: exact reduction chain envelope}
\end{equation}

\begin{lemma}[Chain envelope]
\label{lem: exact reduction chain envelope}
Every pair $(u,Y)$ satisfying
\eqref{eq: exact reduction upward 1}--\eqref{eq: exact reduction upward 2}
satisfies $u(t)\le\psi(t)$ for every $t\in[0,\bar v)$.
\end{lemma}

\begin{proof}
Fix $\bar t>0$; the claim at $\bar t=0$ follows from
\eqref{eq: exact reduction upward 1}.  For $r\in[0,\bar t]$, define
\begin{equation}
    \psi_{\bar t}(r)
    \coloneqq
    \int_0^r m_{\bar t}(s)ds.
\end{equation}
The function $s\mapsto m_{\bar t}(s)$ is nondecreasing and takes values in
$[0,1]$, so $\psi_{\bar t}$ is continuous and
$\psi_{\bar t}(\bar t)=\psi(\bar t)$.

Fix $\epsilon>0$ and set
\begin{equation}
    A
    \coloneqq
    \{r\in[0,\bar t]:u(r)\le\psi_{\bar t}(r)+\epsilon r\}.
\end{equation}
By \eqref{eq: exact reduction upward 1}, $u(0)=0$, so $0\in A$.  Let
$\varrho\coloneqq\sup A$.  We first show that $\varrho\in A$.  Choose
$r_n\in A$ with $r_n\uparrow\varrho$.  If some $r_n=\varrho$, the claim is
immediate.  Otherwise, \eqref{eq: exact reduction upward 2} and $Y\le1$ give
\begin{equation}
    u(\varrho)
    \le
    u(r_n)+(\varrho-r_n)Y(\varrho)
    \le
    \psi_{\bar t}(r_n)+\epsilon r_n+(\varrho-r_n).
\end{equation}
Taking $n\to\infty$ gives
$u(\varrho)\le\psi_{\bar t}(\varrho)+\epsilon\varrho$.

Suppose, toward a contradiction, that $\varrho<\bar t$.  Let
$m\coloneqq\inf_{(\varrho,\bar t]}Y$, and choose
$r^*\in(\varrho,\bar t]$ such that $Y(r^*)\le m+\epsilon$.  For every
$s\in(\varrho,r^*]$, we have $[s,\bar t]\subseteq(\varrho,\bar t]$ and
therefore $m_{\bar t}(s)\ge m$.  Hence
\begin{equation}
    \psi_{\bar t}(r^*)
    \ge
    \psi_{\bar t}(\varrho)+(r^*-\varrho)m.
\end{equation}
Using \eqref{eq: exact reduction upward 2} once more,
\begin{align}
    u(r^*)
    &\le
    u(\varrho)+(r^*-\varrho)Y(r^*)\\
    &\le
    \psi_{\bar t}(\varrho)+\epsilon\varrho
    +(r^*-\varrho)(m+\epsilon)\\
    &\le
    \psi_{\bar t}(r^*)+\epsilon r^*.
\end{align}
Thus $r^*\in A$, contradicting $r^*>\varrho=\sup A$.  We conclude that
$\varrho=\bar t$, and hence
$u(\bar t)\le\psi(\bar t)+\epsilon\bar t$.  Letting
$\epsilon\downarrow0$ proves the result.
\end{proof}

For $c\in(0,1]$, let
\begin{equation}
    A_c\coloneqq\{t\in[0,\bar v):Y(t)\ge c\}.
\end{equation}
For a Borel set $A\subseteq[0,\bar v)$, define
\begin{align}
    \ell_A(t)
    &\coloneqq
    \operatorname{Leb}\{s\in[0,t]:[s,t]\subseteq A\},\\
    \mathcal V(A)&\coloneqq\int_0^{\bar v}\ell_A(t)dH(t),\\
    \mathcal Q(A)&\coloneqq H(A).
    \label{eq: exact reduction layer functionals}
\end{align}
The function $\ell_A$ is Borel measurable.  Indeed, if
\begin{equation}
    \sigma_A(t)
    \coloneqq
    \sup\bigl(\{0\}\cup([0,t]\setminus A)\bigr),
\end{equation}
then $\sigma_A$ is nondecreasing and
$\ell_A(t)=(t-\sigma_A(t))\mathbf 1_A(t)$.  Moreover,
$\ell_A(t)\le t$, so $\mathcal V(A)$ is finite whenever $H$ has a finite mean.

\begin{lemma}[Layer-cake representation]
\label{lem: exact reduction layer cake}
For $\psi$ defined by \eqref{eq: exact reduction chain envelope},
\begin{equation}
    \int_0^{\bar v}\psi(t)dH(t)
    =
    \int_0^1\mathcal V(A_c)dc,
    \qquad
    \int_0^{\bar v}Y(t)dH(t)
    =
    \int_0^1\mathcal Q(A_c)dc.
    \label{eq: exact reduction layer cake identities}
\end{equation}
\end{lemma}

\begin{proof}
Since $m_t(s)\in[0,1]$,
\begin{equation}
    m_t(s)=\int_0^1\mathbf 1\{m_t(s)\ge c\}dc.
\end{equation}
The condition $m_t(s)\ge c$ is equivalent to $[s,t]\subseteq A_c$.  Hence,
for every $t$,
\begin{equation}
    \psi(t)=\int_0^1\ell_{A_c}(t)dc.
    \label{eq: exact reduction pointwise layer cake}
\end{equation}
Similarly,
\begin{equation}
    Y(t)=\int_0^1\mathbf 1\{t\in A_c\}dc.
\end{equation}
The second identity in \eqref{eq: exact reduction layer cake identities}
therefore follows immediately from Tonelli's theorem.

For completeness, we verify the joint measurability needed to integrate
\eqref{eq: exact reduction pointwise layer cake}.  For fixed $c$,
$t\mapsto\ell_{A_c}(t)$ is Borel.  For fixed $t$,
$c\mapsto\ell_{A_c}(t)$ is nonincreasing and left-continuous on $(0,1]$:
if $c_n\uparrow c$, the sets
$\{s\in[0,t]:m_t(s)\ge c_n\}$ decrease to
$\{s\in[0,t]:m_t(s)\ge c\}$.  Define
$d_n(c)\coloneqq2^{-n}(\lceil2^nc\rceil-1)$ and set
$\ell_{A_0}(t)\coloneqq t$.  Then $d_n(c)<c$, $d_n(c)\uparrow c$, and
$(t,c)\mapsto\ell_{A_{d_n(c)}}(t)$ is jointly Borel because $d_n$ takes
only finitely many values.  Left-continuity implies pointwise convergence to
$(t,c)\mapsto\ell_{A_c}(t)$.  Tonelli's theorem now gives the first identity.
\end{proof}

For $a\in[0,\bar v)$, define
\begin{equation}
    \mathcal V_{\mathrm{th}}(a)
    \coloneqq
    \int_a^{\bar v}(1-H(t))dt
    =
    \int_{(a,\bar v)}(t-a)dH(t),
    \qquad
    \mathcal Q_{\mathrm{th}}(a)\coloneqq1-H(a).
    \label{eq: exact reduction threshold value budget}
\end{equation}
The upper-threshold allocation $x_a(t)=\mathbf 1\{t>a\}$ is feasible in the single-dimensional
problem with budget $\mathcal Q_{\mathrm{th}}(a)$ and value $\mathcal V_{\mathrm{th}}(a)$.  Therefore,
\begin{equation}
    \mathcal V_{\mathrm{th}}(a)\le\mathcal R(H,\mathcal Q_{\mathrm{th}}(a)).
    \label{eq: exact reduction threshold feasible}
\end{equation}

\begin{lemma}[Single-threshold exchange]
\label{lem: exact reduction single threshold}
For every Borel set $A\subseteq[0,\bar v)$ and every $\lambda\ge0$,
\begin{equation}
    \mathcal V(A)-\lambda \mathcal Q(A)
    \le
    \max\left\{0,
    \sup_{a\in[0,\bar v)}
    \bigl(\mathcal V_{\mathrm{th}}(a)-\lambda \mathcal Q_{\mathrm{th}}(a)\bigr)
    \right\}.
    \label{eq: exact reduction single threshold inequality}
\end{equation}
\end{lemma}

\begin{proof}
Let $C$ be the union of all nondegenerate closed intervals contained in $A$.
For every $t$, the sets
\begin{equation}
    \{s\in[0,t):[s,t]\subseteq A\}
    \quad\text{and}\quad
    \{s\in[0,t):[s,t]\subseteq C\}
\end{equation}
coincide.  The omitted point $s=t$ is Lebesgue-null, so
$\ell_A=\ell_C$ pointwise.  Since $C\subseteq A$, we have
$\mathcal Q(C)\le \mathcal Q(A)$.  It is therefore enough to bound
$\mathcal V(C)-\lambda \mathcal Q(C)$.

Every point of $C$ belongs to a nondegenerate interval contained in $C$.
Consequently, every connected component of $C$ is a nondegenerate interval
and contains a rational number.  There are therefore at most countably many
components; denote them by $(C_k)_k$ and let $a_k\coloneqq\inf C_k$.  In
particular, $C$ is Borel.  If $t\in C_k$, then any interval $[s,t]$ contained
in $C$ lies in the same connected component, while $(a_k,t]\subseteq C_k$.
Thus
\begin{equation}
    \ell_C(t)=t-a_k\quad\text{for }t\in C_k,
    \qquad
    \ell_C(t)=0\quad\text{for }t\notin C.
\end{equation}
With $\beta_a(t)\coloneqq t-a-\lambda$, we obtain
\begin{equation}
    \mathcal V(C)-\lambda \mathcal Q(C)
    =
    \sum_k \Delta_k,
    \qquad
    \Delta_k\coloneqq\int_{C_k}\beta_{a_k}(t)dH(t).
    \label{eq: exact reduction component decomposition}
\end{equation}
The series converges absolutely because the $C_k$ are disjoint,
$|\beta_{a_k}(t)|\le t+\lambda$ on $C_k$, and
$\int(t+\lambda)dH(t)<+\infty$.

If every $\Delta_k\le0$, then the left-hand side of
\eqref{eq: exact reduction single threshold inequality} is at most zero.
Otherwise, let $P\coloneqq\{k:\Delta_k>0\}$.  For every $k\in P$, the function
$\beta_{a_k}$ is increasing and has a positive integral over $C_k$.
Therefore,
\begin{equation}
    \sup C_k>a_k+\lambda.
    \label{eq: exact reduction profitable interval}
\end{equation}

Suppose first that $\lambda>0$.  If $C_j$ lies to the left of $C_k$, then
\eqref{eq: exact reduction profitable interval} gives
\begin{equation}
    a_k\ge\sup C_j>a_j+\lambda.
\end{equation}
Hence the set $\{a_k:k\in P\}$ is $\lambda$-separated, and its infimum is
attained.  Let $a\coloneqq\min_{k\in P}a_k$, and choose
$k_0\in P$ with $a_{k_0}=a$.  If $\lambda=0$, instead set
$a\coloneqq\inf_{k\in P}a_k$, whether or not the infimum is attained.

We claim that
\begin{equation}
    \beta_a(t)\ge0
    \qquad\text{for every }
    t\in(a,\bar v)\setminus\bigcup_{k\in P}C_k.
    \label{eq: exact reduction complement nonnegative}
\end{equation}
For $\lambda=0$, this is immediate.  Suppose that $\lambda>0$.  If
$\sup C_{k_0}=+\infty$, the set in
\eqref{eq: exact reduction complement nonnegative} is empty.  Otherwise, any
such $t$ must satisfy $t\ge\sup C_{k_0}$, because
$(a,\sup C_{k_0})$ is contained in the interval $C_{k_0}$.  Equation
\eqref{eq: exact reduction profitable interval} then gives
$t>a+\lambda$, proving the claim.

Since $a\le a_k$ for every $k\in P$, we have
$\beta_a\ge\beta_{a_k}$ pointwise.  Using
\eqref{eq: exact reduction complement nonnegative}, disjointness of the
$C_k$, and continuity of $H$, which gives $H(\{a\})=0$, we conclude that
\begin{align}
    \mathcal V(C)-\lambda \mathcal Q(C)
    &=\sum_k\Delta_k\\
    &\le\sum_{k\in P}\Delta_k\\
    &\le\sum_{k\in P}\int_{C_k}\beta_a(t)dH(t)\\
    &\le\int_{(a,\bar v)}\beta_a(t)dH(t)\\
    &=\mathcal V_{\mathrm{th}}(a)-\lambda \mathcal Q_{\mathrm{th}}(a).
\end{align}
This proves \eqref{eq: exact reduction single threshold inequality}.
\end{proof}

\begin{lemma}[The upward relaxation is free]
\label{lem: exact reduction upward relaxation}
For every continuous distribution $H$ on $[0,\bar v)$ with finite mean and
every $b\in[0,1]$,
\begin{equation}
    \mathcal U(H,b)=\mathcal R(H,b).
    \label{eq: exact reduction upward equals one dimensional}
\end{equation}
\end{lemma}

\begin{proof}
We first prove $\mathcal R(H,b)\le\mathcal U(H,b)$.  Let $x$ be feasible in
\eqref{eq: exact reduction one dimensional value}, and set
\begin{equation}
    Y(t)\coloneqq x(t),
    \qquad
    u(t)\coloneqq\int_0^t x(s)ds.
\end{equation}
Since $x$ is nondecreasing,
\begin{equation}
    u(t)\le tx(t),
    \qquad
    u(t)-u(s)=\int_s^t x(r)dr\le(t-s)x(t)
    \quad (s<t).
\end{equation}
Thus $(u,Y)$ is feasible for the upward system.  Tonelli's theorem gives
\begin{equation}
    \int_0^{\bar v}u(t)dH(t)
    =
    \int_0^{\bar v}(1-H(s))x(s)ds,
\end{equation}
so taking suprema proves this direction.

We next prove the reverse inequality.  If $b=0$, then
$\int YdH=0$ and $Y\ge0$, while $0\le u(t)\le tY(t)$; hence
$\int u dH=0\le\mathcal R(H,0)$.  Fix $b\in(0,1]$, a feasible pair
$(u,Y)$, and $\lambda\ge0$.  By
\eqref{eq: exact reduction threshold feasible},
Lemma~\ref{lem: exact reduction single threshold}, and
$\mathcal R(H,0)\ge0$, for every $c\in(0,1]$,
\begin{equation}
    \mathcal V(A_c)-\lambda \mathcal Q(A_c)
    \le
    \sup_{b'\in[0,1]}
    \bigl(\mathcal R(H,b')-\lambda b'\bigr).
    \label{eq: exact reduction layer Lagrangian bound}
\end{equation}
Indeed, every threshold pair $(\mathcal Q_{\mathrm{th}}(a),\mathcal V_{\mathrm{th}}(a))$ is dominated by the feasible
single-dimensional value at budget $\mathcal Q_{\mathrm{th}}(a)$.

Integrating \eqref{eq: exact reduction layer Lagrangian bound} over $c$ and
using Lemmas~\ref{lem: exact reduction chain envelope} and
\ref{lem: exact reduction layer cake}, we obtain
\begin{align}
    \int_0^{\bar v}u(t)dH(t)
    -\lambda\int_0^{\bar v}Y(t)dH(t)
    &\le
    \int_0^1\bigl(\mathcal V(A_c)-\lambda \mathcal Q(A_c)\bigr)dc\\
    &\le
    \sup_{b'\in[0,1]}
    \bigl(\mathcal R(H,b')-\lambda b'\bigr).
    \label{eq: exact reduction integrated Lagrangian bound}
\end{align}

By Lemma~\ref{lem: exact reduction one dimensional representation}, the map
$b'\mapsto\mathcal R(H,b')$ is finite, concave, and nondecreasing.  If
$b\in(0,1)$, choose a finite supergradient $\lambda^*\ge0$ at $b$.  If
$b=1$, take $\lambda^*\coloneqq0$, which is a supergradient by monotonicity.
Thus
\begin{equation}
    \mathcal R(H,b')
    \le
    \mathcal R(H,b)+\lambda^*(b'-b)
    \qquad\text{for every }b'\in[0,1].
\end{equation}
Substituting $\lambda=\lambda^*$ into
\eqref{eq: exact reduction integrated Lagrangian bound} gives
\begin{align}
    \int_0^{\bar v}u(t)dH(t)
    &\le
    \lambda^*\int_0^{\bar v}Y(t)dH(t)
    +\mathcal R(H,b)-\lambda^*b\\
    &\le\mathcal R(H,b),
\end{align}
where the last inequality uses $\int YdH\le b$ and $\lambda^*\ge0$.
Taking the supremum over feasible $(u,Y)$ proves
$\mathcal U(H,b)\le\mathcal R(H,b)$.
\end{proof}

\begin{proof}[Proof of Theorem~\ref{thm: exact reduction}]
If $G$ has an infinite mean, then $G_K$ also has an infinite mean because
$1-G_K(t)\ge1-G(t)$.  The reduced no-screening mechanism
$x(t)\equiv\bar m$, $p(t)\equiv0$ is feasible and has infinite residual
surplus.  Assigning every agent her favorite object with probability $\bar m$
and charging zero implements the same rule in the original market; by symmetry,
each object then uses capacity $\bar m/K$.  Thus both suprema are $+\infty$,
and the result follows.  We henceforth
suppose that $G$ has a finite mean.  Then $H=G_K$ also has a finite mean,
since the maximum of $K$ nonnegative draws is bounded above by their sum.

Take any feasible rank mechanism $(y,p^r)$.  By
Lemma~\ref{lem: exact reduction pathwise} and
Lemma~\ref{lem: exact reduction upward relaxation},
\begin{align}
    RS(y,p^r)
    &\le
    \mathbb E_{\omega}\bigl[\mathcal R(H,b(\omega))\bigr]\\
    &\le
    \mathcal R\bigl(H,\mathbb E_{\omega}[b(\omega)]\bigr)\\
    &\le
    \mathcal R(H,\bar m).
    \label{eq: exact reduction final upper bound}
\end{align}
The second inequality is Jensen's inequality for the concave function
$b\mapsto\mathcal R(H,b)$, and the third follows from
\eqref{eq: exact reduction path budget} and monotonicity.  By
Lemma~\ref{lem: exact reduction one dimensional representation},
$\mathcal R(H,\bar m)$ is exactly the value of the reduced problem.
Lemma~\ref{lem: exact reduction rank representation} therefore implies that
the unrestricted multidimensional value is no larger than the reduced value.

It remains to prove the reverse inequality.  Let $(x,p)$ be any feasible
reduced mechanism, and write
\begin{equation}
    R(\bv)\coloneqq\max_{k\in K}v_k,
    \qquad
    M(\bv)\coloneqq\argmax_{k\in K}v_k.
\end{equation}
Define a mechanism in the original market by
\begin{equation}
    x_k^o(\bv)
    \coloneqq
    \begin{cases}
        \dfrac{x(R(\bv))}{|M(\bv)|},&k\in M(\bv),\\[0.7em]
        0,&k\notin M(\bv),
    \end{cases}
    \qquad
    p^o(\bv)\coloneqq p(R(\bv)).
    \label{eq: exact reduction favorite lift}
\end{equation}
Unit demand and individual rationality are immediate.  Since ties occur with
probability zero and every object is the unique favorite with probability
$1/K$,
\begin{equation}
    \int_Vx_k^o(\bv)dF(\bv)
    =
    \frac{1}{K}\int_0^{\bar v}x(t)dG_K(t)
    \le\frac{\bar m}{K},
\end{equation}
so the resource constraints hold.

To verify strategy-proofness, fix a true type $\bv$ and a report $\bv'$, and
write $r\coloneqq R(\bv)$ and $s\coloneqq R(\bv')$.  Under report $\bv'$, the
total allocation probability is $x(s)$, and every allocated object has true
value at most $r$.  Therefore, reduced strategy-proofness gives
\begin{align}
    \sum_{k\in K}v_kx_k^o(\bv')-p^o(\bv')
    &\le rx(s)-p(s)\\
    &\le rx(r)-p(r)\\
    &=\sum_{k\in K}v_kx_k^o(\bv)-p^o(\bv).
\end{align}
Thus the lifted mechanism is feasible.  Its residual surplus is
\begin{equation}
    \int_V
    \bigl(R(\bv)x(R(\bv))-p(R(\bv))\bigr)dF(\bv)
    =
    \int_0^{\bar v}(tx(t)-p(t))dG_K(t),
\end{equation}
which is exactly the residual surplus of the reduced mechanism.  Taking
suprema proves that the unrestricted value is at least the reduced value.
Together with \eqref{eq: exact reduction final upper bound}, this establishes
the equality of the two values.
\end{proof}

\subsection{Non-Favorite Assignments Can Improve Residual Surplus}
\label{subsec:ex_post_efficiency_is_not_without_loss}

This subsection proves Theorem~\ref{thm:nonfavorite-screening}.
The proof first constructs a finite-support example and then perturbs the
marginal distribution smoothly while preserving the strict residual-surplus
gain. The construction also illustrates the screening role of the
non-favorite coordinates of the valuation vector.

\begin{proof}[Proof of Theorem~\ref{thm:nonfavorite-screening}]

Consider two object types with equal capacities, $m_1 = m_2 = 47/200$. Thus, the total capacity is $47/100$.
For each agent $i$ and object $k$, the value $v_k^i$ is drawn independently, and $v_k^i = 0$ with probability $2/5$, $v_k^i = 1$ with probability $1/5$, and $v_k^i = 4$ with probability $2/5$.
Write an ordered type as $(r,s)$, where $r=\max\{v_1,v_2\}$ and $s=\min\{v_1,v_2\}$.  By direct calculation, the ordered-type masses are computed as
\[
\begin{array}{c|cccccc}
 (r,s) & (4,4) & (4,1) & (4,0) & (1,1) & (1,0) & (0,0) \\
\hline
 \text{mass} & 4/25 & 4/25 & 8/25 & 1/25 & 4/25 & 4/25 .
\end{array}
\]
Thus, the highest value, $r$, is $4$ with probability $16/25$, $1$ with probability $1/5$, and $0$ with probability $4/25$.

We first compute the optimum of the reduced single-dimensional problem
with type space $\{0,1,4\}$. Let $x_r$, $p_r$, and $u_r=rx_r-p_r$
denote its allocation probability, payment, and utility at type $r$.
Setting the utility of the zero type to zero and choosing the lowest feasible payments, we first have $p_1 = 0$ and $u_1 = x_1$. Since agents with $r = 1$ must not have an incentive to report $r = 4$, we must have $u_1 = x_1 \ge x_4 - p_4$; thus, the lowest payment is $p_4 = x_4 - x_1$. Thus, $u_4 = 4 x_4 - p_4 = x_1 + 3 x_4$.
Hence, the reduced problem is
\[
\max_{0\le x_1\le x_4\le 1}
    \frac{21}{25}x_1+\frac{48}{25}x_4
    \quad\text{s.t.}\quad
    \frac{1}{5}x_1+\frac{16}{25}x_4\le \frac{47}{100}.
\]
The optimum of this reduced problem pools the positive types,
$r=1,4$, with $x_1=x_4=47/84$ and $p_1=p_4=0$.
Its residual surplus is
\[
    V^{\text{red}}
    =\frac{47}{84}\cdot\left(\frac{16}{25} \cdot 4 + \frac{1}{5} \cdot 1 \right)
    =\frac{1081}{700}.
\]

Now consider the following neutral menu with zero payments.  The null option is also available.
\begin{itemize}
    \item $C$ (concentrated): receive one's reported favorite object with probability $\alpha = 1/2$.
    \item $M$ (mixed): receive each object with probability $\beta/2 = 3/8$, so that the total allocation probability is $\beta=3/4$.
\end{itemize}
For a type $(r,s)$ with $r>0$, the utilities from $C$ and $M$ are
\[
    U_C(r,s)=\frac{r}{2},
    \qquad
    U_M(r,s)=\frac{3(r+s)}{8}.
\]
Thus $M$ is preferred to $C$ if and only if
\[
    \frac{3(r+s)}{8}\ge \frac{r}{2}
    \quad\Longleftrightarrow\quad
    \frac{s}{r}\ge \frac{1}{3}.
\]
Therefore the balanced types $(4,4)$ and $(1,1)$ choose $M$, while the lopsided positive types $(4,1)$, $(4,0)$, and $(1,0)$ choose $C$.  The zero type chooses the null option.  Since the mechanism assigns each type an option that maximizes its utility from the menu, the induced direct mechanism is strategy-proof.  Individual rationality is immediate, and neutrality follows from the symmetric definition of $M$ and from defining $C$ relative to the reported favorite object.

Feasibility is also immediate.  The mass of $C$-types is
\[
    Q_C=\frac{4}{25}+\frac{8}{25}+\frac{4}{25}=\frac{16}{25},
\]
and the mass of $M$-types is
\[
    Q_M=\frac{4}{25}+\frac{1}{25}=\frac{1}{5}.
\]
Total expected allocation is
\[
    Q_C\alpha+Q_M\beta
    =\frac{16}{25}\cdot\frac12+\frac15\cdot\frac34
    =\frac{47}{100}=\bar m.
\]
By symmetry, each object bears exactly half of this load, namely $47/200$, which equals its capacity.  Since $\alpha,\beta\le 1$, unit demand is satisfied.

The residual surplus of this mechanism is
\begin{align*}
    V^*
    & = \beta\left(4\cdot\frac{4}{25}+1\cdot\frac{1}{25}\right)
        +\alpha\left(4\left(\frac{4}{25}+\frac{8}{25}\right)+1\cdot\frac{4}{25}\right) \\
    & = \frac34\cdot\frac{17}{25}
        +\frac12\cdot\frac{52}{25}
      = \frac{31}{20}.
\end{align*}
Therefore
\[
    V^*-V^{\text{red}}
    =\frac{31}{20}-\frac{1081}{700}
    =\frac{1}{175}>0.
\]
The second coordinate is valuable because balance is informative about the level of the maximum.  Indeed,
\[
    \mathbb E[r\mid r = s > 0]
    =\frac{4\cdot(2/5)^2+1\cdot(1/5)^2}{(2/5)^2+(1/5)^2}
    =\frac{17}{5}
    >
    \frac{23}{7}
    =\mathbb E[r\mid r>0].
\]
The mixed option attracts this richer-than-average balanced subpopulation without requiring additional burned payments.


At this discrete stage, the menu assigns only favorite objects on the
support, since only tied types select $M$. The comparison above is
therefore with the reduced benchmark; smoothing below yields a strict
comparison with all favorite-only mechanisms.

Finally, smooth each marginal atom at $a\in\{0,1,4\}$ into a
$C^2$ density supported on $[a,a+\varepsilon)$, preserving its mass,
and denote the resulting CDF by $\widetilde G_\varepsilon$. Let
\[
    G_{\varepsilon,\eta}
    \coloneqq(1-\eta)\widetilde G_\varepsilon+\eta B_\varepsilon,
\]
where $B_\varepsilon$ is supported on $[0,4+\varepsilon)$ and has a
$C^2$ density strictly positive on its interior. 
In what follows, all limits are taken jointly as
$(\varepsilon,\eta)\to(0,0)$ through positive values.
This marginal satisfies
the maintained smoothness and full-support assumptions.

Use the same allocation options, now charging
$p_C=2\varepsilon\alpha$ and $p_M=2\varepsilon\beta$, with symmetric
tie-breaking. The bottom cloud chooses the null option, and for small
$\varepsilon$ all positive clouds retain the choices described above.
Let $Q_{\varepsilon,\eta}$ and $\widetilde V_{\varepsilon,\eta}$ denote
this menu's aggregate resource use and residual surplus. Because the
probability that at least one draw comes from $B_\varepsilon$ vanishes,
\[
    Q_{\varepsilon,\eta}\longrightarrow\bar m,
    \qquad
    \widetilde V_{\varepsilon,\eta}\longrightarrow\frac{31}{20}.
\]
Multiply every menu allocation and payment by
\[
    \lambda_{\varepsilon,\eta}
    \coloneqq\min\left\{1,\frac{\bar m}{Q_{\varepsilon,\eta}}\right\}.
\]
This preserves menu choices, strategy-proofness, individual rationality,
and unit demand. Symmetry makes each object's load at most $\bar m/2$,
so the scaled menu is feasible. Since $\lambda_{\varepsilon,\eta}\to1$,
its residual surplus $V^M_{\varepsilon,\eta}$ converges to $31/20$.

It remains to bound all favorite-only mechanisms in the smoothed
market. Symmetrization and conditional averaging over non-favorite
values, as in the proof of
Proposition~\ref{prop:neutrality-ex-post-correspondence}, 
map any such mechanism to a feasible reduced one with the same
residual surplus and aggregate resource use.
Favorite-only assignment is
imposed directly here, so ex post efficiency is not needed.
Let $T\sim H_{\varepsilon,\eta}\coloneqq G_{\varepsilon,\eta}^2$ and
$\mu_{\varepsilon,\eta}\coloneqq\mathbb E[T]$. Then
$\mu_{\varepsilon,\eta}\to69/25$. For $0\le t<1$,
$\Pr(T>t)\ge(1-\eta)^2(21/25)$, whereas for $1\le t<4+\varepsilon$
the remaining value is at most $3+\varepsilon$. Thus every threshold
satisfies
\[
    \mathbb E[T-t\mid T>t]
    \le c_{\varepsilon,\eta}
    \coloneqq\max\left\{
        \frac{\mu_{\varepsilon,\eta}}{(1-\eta)^2(21/25)},
        3+\varepsilon
    \right\}
    \longrightarrow\frac{23}{7}.
\]
The same bound applies to the constant allocation component. The
constant-and-threshold decomposition in the proof of
Theorem~\ref{thm: NBUE iff no screening} therefore implies
\[
    V^{\mathrm{fav}}_{\varepsilon,\eta}
    \le\mathcal R(H_{\varepsilon,\eta},\bar m)
    \le\bar m\,c_{\varepsilon,\eta}
    \longrightarrow\frac{1081}{700}<\frac{31}{20},
\]
where $V^{\mathrm{fav}}_{\varepsilon,\eta}$ is the supremum over
favorite-only mechanisms. Since $V^M_{\varepsilon,\eta}\to31/20$, the
scaled menu strictly outperforms every favorite-only mechanism for
all sufficiently small $\varepsilon$ and $\eta$.

\end{proof}

The construction also clarifies why ex post efficiency can be a substantive
restriction.
In the perturbed full-support environment, the mixed option assigns a non-favorite object with positive probability to a positive-measure set of near-diagonal types with $v_1\ne v_2$.  Reassigning the probability placed on the non-favorite object to the favorite object preserves feasibility by symmetry and strictly improves those types.  Hence, the mechanism that beats favorite-only mechanisms violates ex post efficiency.  Consequently, ex post efficiency is not without loss for residual-surplus maximization: it rules out a genuine screening channel based on the balance of the valuation vector.

\subsection{Neutrality, Ex Post Efficiency, and the Reduced Market}
\label{subsec:neutrality-ex-post-reduction}

This subsection proves
Theorem~\ref{thm:neutrality-ex-post-reduction}.
We first establish a stronger mechanism-level correspondence and then derive
the value equivalence stated in the main text. The surrounding discussion
also clarifies the distinct roles of neutrality and ex post efficiency.


For a permutation $\sigma$ of the object labels, define $(\sigma\bv)_k\coloneqq v_{\sigma^{-1}(k)}$, and use the same action on allocation vectors.
A mechanism $(\bx,p)$ is \emph{neutral} if 
\begin{equation} 
\bx(\sigma\bv)=\sigma\bx(\bv) 
\quad\text{and}\quad 
p(\sigma\bv)=p(\bv) 
\qquad\text{for every permutation }\sigma. 
\end{equation} 

An allocation rule $\bx$ is \emph{ex post efficient} if there is no Borel
measurable allocation rule $\by:V\to[0,1]^K$ such that
\begin{align}
    \int_V y_k(\bv)\,dF(\bv)&\le m_k
        &&\text{for every }k\in K,\label{eq:ee-alternative-resource}\\
    \sum_{k\in K}y_k(\bv)&\le1
        &&\text{for every }\bv\in V,\label{eq:ee-alternative-unit-demand}\\
    \sum_{k\in K}v_k y_k(\bv)&\ge
        \sum_{k\in K}v_k x_k(\bv)
        &&\text{for every }\bv\in V,\label{eq:ee-weak-improvement}
\end{align}
and there is a Borel set $A\subset V$ with $F(A)>0$ on which the inequality
in \eqref{eq:ee-weak-improvement} is strict.

The definition concerns only the allocation rule: the improving allocation need
not preserve the original mechanism's incentive properties or payment rule.
Thus, ex post efficiency rules out the use of a Pareto-dominated assignment as a
screening instrument.

\begin{proposition}[Mechanism-level correspondence]
\label{prop:neutrality-ex-post-correspondence}
Consider a continuous i.i.d.\ market. Let $\mech=(\bx,p)$ be a feasible mechanism and neutral mechanism whose allocation rule $\bx$ is ex post efficient. Then, the following
statements hold.
\begin{enumerate}[(i)]
    \item There is a Borel set $V^*\subset V$ with $F(V^*)=1$ such that, for
    every $\bv\in V^*$ and every $k\in K$,
    \begin{equation}
        x_k(\bv)>0
        \quad\Longrightarrow\quad
        v_k=\max_{\ell\in K}v_\ell.
        \label{eq:neutral-ee-favorite-only}
    \end{equation}

    \item There exists a feasible reduced mechanism $(x^r,p^r)$ satisfying
    \eqref{eq: reduced sp} and \eqref{eq: reduced ir}, with
    \begin{equation}
        \int_0^{\bar v}x^r(v)\,dG_K(v)=\bar m,
        \label{eq:neutral-ee-reduced-resource-equality}
    \end{equation}
    and
    \begin{equation}
        RS(\mech)
        =
        \int_0^{\bar v}
        \bigl(vx^r(v)-p^r(v)\bigr)\,dG_K(v).
        \label{eq:neutral-ee-rs-preservation}
    \end{equation}
\end{enumerate}

Conversely, every feasible reduced mechanism satisfying
\eqref{eq:neutral-ee-reduced-resource-equality} induces a feasible, neutral,
and ex post efficient mechanism in the original continuous i.i.d.\ market
that achieves the same residual surplus.
\end{proposition}

\begin{proof}
Write
\begin{equation}
    R(\bv)\coloneqq\max_{k\in K}v_k,
    \qquad
    X(\bv)\coloneqq\sum_{k\in K}x_k(\bv).
\end{equation}
Because $G$ is continuous, the favorite object is unique with probability one.
Let
\begin{equation}
    D\coloneqq
    \left\{\bv\in V:
    \left|\argmax_{k\in K}v_k\right|=1\right\},
\end{equation}
so that $F(D)=1$, and write $J(\bv)$ for the unique favorite label on $D$.

\subsubsection{Neutrality and Ex Post Efficiency Imply Favorite-Object Assignment}
Suppose, toward a contradiction, that
\begin{equation}
    F\bigl(\{\bv\in D:x_k(\bv)>0
    \text{ for some }k\ne J(\bv)\}\bigr)>0.
\end{equation}
For distinct labels $a,b\in K$, define
\begin{equation}
    A_{ab}\coloneqq
    \{\bv\in D:J(\bv)=b,\ x_a(\bv)>0\}.
\end{equation}
The preceding set is the finite union of the sets $A_{ab}$, so there exist
$a\ne b$ such that $F(A_{ab})>0$.  Fix such a pair, write $A\coloneqq A_{ab}$,
and let $\tau$ be the transposition of $a$ and $b$.  Since every type in $A$
has unique favorite $b$, whereas every type in $\tau A$ has unique favorite
$a$, the sets $A$ and $\tau A$ are disjoint.

Define an alternative allocation rule $\by$ as follows.  For $\bv\in A$, set
\begin{align}
    y_a(\bv)&\coloneqq0,
    &y_b(\bv)&\coloneqq x_a(\bv)+x_b(\bv),
    &y_k(\bv)&\coloneqq x_k(\bv)\quad(k\notin\{a,b\}).
    \label{eq:favorite-reallocation-A}
\end{align}
For $\bv\in\tau A$, set
\begin{align}
    y_a(\bv)&\coloneqq x_a(\bv)+x_b(\bv),
    &y_b(\bv)&\coloneqq0,
    &y_k(\bv)&\coloneqq x_k(\bv)\quad(k\notin\{a,b\}),
    \label{eq:favorite-reallocation-tau-A}
\end{align}
and set $\by(\bv)=\bx(\bv)$ elsewhere.  This reallocation preserves each
type's total allocation probability and therefore preserves unit demand.

It also preserves the aggregate use of every object.  For object $a$, using
exchangeability of $F$ and neutrality,
\begin{align}
    \int_V\bigl(y_a(\bv)-x_a(\bv)\bigr)\,dF(\bv)
    &=-\int_A x_a(\bv)\,dF(\bv)
      +\int_{\tau A}x_b(\bv)\,dF(\bv)\\
    &=-\int_A x_a(\bv)\,dF(\bv)
      +\int_A x_b(\tau\bv)\,dF(\bv)\\
    &=0.
    \label{eq:favorite-resource-a}
\end{align}
The last equality follows from $x_b(\tau\bv)=x_a(\bv)$.  The same calculation
shows that the aggregate use of object $b$ is unchanged, and all other objects
are unchanged pointwise.  Hence $\by$ satisfies the resource constraints.

For every $\bv\in A$,
\begin{equation}
    \sum_{k\in K}v_k y_k(\bv)
    -\sum_{k\in K}v_k x_k(\bv)
    =(v_b-v_a)x_a(\bv)>0.
    \label{eq:favorite-strict-improvement-A}
\end{equation}
The analogous inequality holds on $\tau A$.  All other types are unaffected.
Thus $\by$ is a feasible Pareto improvement that is strict on a set of positive
measure, contradicting ex post efficiency.  This proves
\eqref{eq:neutral-ee-favorite-only}.

\subsubsection{Ex Post Efficiency Exhausts Aggregate Capacity}
We claim that
\begin{equation}
    \int_V X(\bv)\,dF(\bv)=\bar m.
    \label{eq:neutral-ee-capacity-binds}
\end{equation}
Suppose instead that the left-hand side is strictly smaller than $\bar m$.
Then some object $h$ has strictly unused capacity.  Moreover,
$\int_V X\,dF<\bar m<1$, so the set $\{\bv:X(\bv)<1\}$ has positive measure.
Because $v_h>0$ almost surely, there exist $n\in\mathbb N$ and a Borel set
$C\subset V$ with $F(C)>0$ such that
\begin{equation}
    X(\bv)\le1-\frac1n
    \quad\text{and}\quad
    v_h\ge\frac1n
    \qquad\text{for every }\bv\in C.
\end{equation}
Choose $\varepsilon>0$ such that
\begin{equation}
    \varepsilon\le\frac1n
    \quad\text{and}\quad
    \varepsilon F(C)
    \le m_h-\int_Vx_h(\bv)\,dF(\bv).
\end{equation}
Increasing $x_h(\bv)$ by $\varepsilon$ on $C$ preserves unit demand and the
resource constraints, leaves every other type unchanged, and strictly raises
every type's value on $C$.  This again contradicts ex post efficiency and proves
\eqref{eq:neutral-ee-capacity-binds}.

\subsubsection{Conditional Averaging Produces a Reduced Mechanism}
Let $V^0\subset D$ be a full-measure Borel set on which
\eqref{eq:neutral-ee-favorite-only} holds.  For every $j\in K$, let
$\mu_{j,r}$ be a regular conditional distribution of $\bv$ given
$J(\bv)=j$ and $R(\bv)=r$.  Under the i.i.d.\ distribution, $J$ is uniform on
$K$, is independent of $R$, and $R$ has distribution $G_K$.  Since $K$ is
finite, there is a Borel set $S\subset[0,\bar v)$ with $G_K(S)=1$ such that
\begin{equation}
    \mu_{j,r}(V^0)=1
    \qquad\text{for every }j\in K\text{ and every }r\in S.
\end{equation}
Full support of $G_K$ implies that $S$ is dense in $[0,\bar v)$.

For $r\in S$, choose any $j\in K$ and define
\begin{equation}
    \widetilde x(r)
    \coloneqq\int_V X(\bv)\,d\mu_{j,r}(\bv),
    \qquad
    \widetilde p(r)
    \coloneqq\int_V p(\bv)\,d\mu_{j,r}(\bv).
    \label{eq:neutral-ee-conditional-averages}
\end{equation}
These definitions do not depend on $j$: a permutation carrying $j$ to another
label pushes the corresponding conditional distribution forward, while
neutrality leaves $X$ and $p$ unchanged.  We take Borel versions of these
conditional expectations.  Since $0\le X\le1$, we have
$\widetilde x(r)\in[0,1]$.  Equations
\eqref{eq:neutral-ee-capacity-binds} and
\eqref{eq:neutral-ee-favorite-only}, together with disintegration, yield
\begin{align}
    \int_0^{\bar v}\widetilde x(r)\,dG_K(r)
    &=\bar m,
    \label{eq:neutral-ee-averaged-resource}\\
    RS(\mech)
    &=\int_0^{\bar v}
      \bigl(r\widetilde x(r)-\widetilde p(r)\bigr)\,dG_K(r).
    \label{eq:neutral-ee-averaged-rs}
\end{align}
Individual rationality gives
\begin{equation}
    r\widetilde x(r)-\widetilde p(r)\ge0
    \qquad\text{for every }r\in S.
    \label{eq:neutral-ee-averaged-ir}
\end{equation}

We next verify the reduced incentive constraints on $S$.  Fix $r,s\in S$ and
$j\in K$.  For
$\mu_{j,r}\times\mu_{j,s}$-almost every pair $(\bv,\bw)$, both types lie in
$V^0$.  Strategy-proofness of the original mechanism, applied to true type
$\bv$ and report $\bw$, therefore gives
\begin{equation}
    r X(\bv)-p(\bv)
    \ge r X(\bw)-p(\bw),
\end{equation}
because both reports have favorite label $j$ and the allocation at either
report is concentrated on that favorite object.  Taking expectations gives
\begin{equation}
    r\widetilde x(r)-\widetilde p(r)
    \ge r\widetilde x(s)-\widetilde p(s)
    \qquad\text{for every }r,s\in S.
    \label{eq:neutral-ee-averaged-ic}
\end{equation}

It remains to define the reduced mechanism pointwise on the null set outside
$S$.  Put
\begin{equation}
    \widetilde u(r)
    \coloneqq r\widetilde x(r)-\widetilde p(r)
    \qquad(r\in S).
\end{equation}
Equation \eqref{eq:neutral-ee-averaged-ic} is equivalent to
\begin{equation}
    \widetilde u(r)
    \ge
    \widetilde u(s)+(r-s)\widetilde x(s)
    \qquad(r,s\in S).
    \label{eq:neutral-ee-subgradient-on-S}
\end{equation}
Define the convex extension
\begin{equation}
    u^r(v)
    \coloneqq
    \sup_{s\in S}
    \left\{\widetilde u(s)+(v-s)\widetilde x(s)\right\}
    \qquad(v\in[0,\bar v)).
    \label{eq:neutral-ee-convex-extension}
\end{equation}
This function is finite: nonnegative payments imply
$\widetilde u(s)-s\widetilde x(s)=-\widetilde p(s)\le0$, so every affine
function in the supremum is at most $v$.  It is also nondecreasing and
$1$-Lipschitz.  Equation \eqref{eq:neutral-ee-subgradient-on-S} implies
\begin{equation}
    u^r(r)=\widetilde u(r)
    \quad\text{and}\quad
    \widetilde x(r)\in\partial u^r(r)
    \qquad(r\in S),
\end{equation}
where $\partial u^r(v)$ denotes the subdifferential of $u^r$ at $v$.
Choose a Borel function $x^r:[0,\bar v)\to[0,1]$ such that
$x^r(v)\in\partial u^r(v)$ for every $v$ and
$x^r(r)=\widetilde x(r)$ for every $r\in S$; for example, use the prescribed
supporting slope on $S$ and a one-sided derivative of $u^r$ off $S$.  Define
\begin{equation}
    p^r(v)\coloneqq v x^r(v)-u^r(v).
    \label{eq:neutral-ee-extended-payment}
\end{equation}
The subgradient inequality gives, for all $v,w\in[0,\bar v)$,
\begin{equation}
    v x^r(v)-p^r(v)
    =u^r(v)
    \ge u^r(w)+(v-w)x^r(w)
    =v x^r(w)-p^r(w),
\end{equation}
which is the reduced strategy-proofness constraint.

We finally verify individual rationality and nonnegative payments.  From
\eqref{eq:neutral-ee-convex-extension},
\begin{equation}
    u^r(0)
    =\sup_{s\in S}
    \{\widetilde u(s)-s\widetilde x(s)\}
    =\sup_{s\in S}\{-\widetilde p(s)\}\le0.
\end{equation}
On the other hand, \eqref{eq:neutral-ee-averaged-ir} and
$\widetilde p\ge0$ imply
$0\le\widetilde p(s)\le s\widetilde x(s)\le s$.  Since $S$ is dense at zero,
the preceding supremum is at least zero.  Hence $u^r(0)=0$.  Because $u^r$ is
nondecreasing, $u^r(v)\ge0$ for every $v$, proving individual rationality.
The subgradient inequality at $v$, evaluated at zero, gives
\begin{equation}
    0=u^r(0)\ge u^r(v)-v x^r(v),
\end{equation}
so $p^r(v)\ge0$.  On $S$, $(x^r,p^r)$ coincides with
$(\widetilde x,\widetilde p)$.  Since $G_K(S)=1$,
\eqref{eq:neutral-ee-reduced-resource-equality} and
\eqref{eq:neutral-ee-rs-preservation} follow from
\eqref{eq:neutral-ee-averaged-resource} and
\eqref{eq:neutral-ee-averaged-rs}.

\subsubsection{Ex Post Efficiency of the Favorite-Object Lift}

Let $(x^r,p^r)$ be a feasible reduced mechanism satisfying
\eqref{eq:neutral-ee-reduced-resource-equality}.  Apply the favorite-object
lift in \eqref{eq: exact reduction favorite lift} to $(x^r,p^r)$, and denote
the resulting original-market mechanism by $(\bx^o,p^o)$.  As shown in the
final part of the proof of Theorem~\ref{thm: exact reduction}, this
construction is feasible and preserves residual surplus.
That verification does not use CDF log-concavity.  The equal treatment of
tied favorite objects also makes the lift neutral.  

It remains only to verify ex post efficiency. Suppose, toward a contradiction, that a feasible allocation rule $\by$
weakly improves every type and strictly improves types on a set of positive
measure.  Let
\begin{equation}
    a(\bv)\coloneqq\sum_{k\in K}y_k(\bv).
\end{equation}
Since no object has value greater than $R(\bv)$, weak improvement implies
\begin{equation}
    R(\bv)a(\bv)
    \ge
    \sum_{k\in K}v_k y_k(\bv)
    \ge
    \sum_{k\in K}v_kx^o_k(\bv)
    =
    R(\bv)x^r(R(\bv)).
\end{equation}
Because $R(\bv)>0$ almost surely, it follows that
$a(\bv)\ge x^r(R(\bv))$ almost surely.  Feasibility of $\by$ and
\eqref{eq:neutral-ee-reduced-resource-equality} then give
\begin{align}
    \bar m
    &=
    \int_V x^r(R(\bv))\,dF(\bv)\\
    &\le
    \int_V a(\bv)\,dF(\bv)
    =
    \sum_{k\in K}\int_V y_k(\bv)\,dF(\bv)\\
    &\le
    \sum_{k\in K}m_k
    =
    \bar m.
\end{align}
Hence $a(\bv)=x^r(R(\bv))$ almost surely.  Consequently,
\begin{equation}
    \sum_{k\in K}v_k y_k(\bv)
    \le
    R(\bv)a(\bv)
    =
    R(\bv)x^r(R(\bv))
    =
    \sum_{k\in K}v_kx^o_k(\bv)
\end{equation}
almost surely.  This rules out a strict improvement on a set of positive
measure.  Thus $\bx^o$ is ex post efficient, completing the proof of the
converse statement.
\end{proof}

\begin{proof}[Proof of Theorem~\ref{thm:neutrality-ex-post-reduction}]
The first direction of
Proposition~\ref{prop:neutrality-ex-post-correspondence} shows that the
restricted multidimensional value is no greater than the reduced value.
Conversely, as noted after \eqref{eq: reduced ir}, equality can be imposed in
\eqref{eq: reduced resource} without changing the reduced value. The
converse direction of the proposition then lifts every such reduced mechanism
to a neutral and ex post efficient multidimensional mechanism with the same
residual surplus. Hence the two values are equal.
\end{proof}

\section{Additional Analytical and Numerical Results}

\subsection{Property of the Weibull Distribution}
\label{sec: Weibull example}

This section details the example of the Weibull distribution, which is illustrated in Figure~\ref{fig: Weibull example}.

The Weibull distribution with shape parameter $\alpha > 0$ is given by
\begin{equation}
    G(v) = 1 - \exp(- v^{\alpha}),
\end{equation}
and its probability density function is
\begin{equation}
    g(v) = \alpha v^{\alpha - 1}\exp(- v^{\alpha}).
\end{equation}
Note that Weibull has an IHR if $\alpha \ge 1$ and a DHR if $\alpha \le 1$. When $\alpha = 1$, it is identical to a standard exponential distribution.

The inverse hazard rate is given by
\begin{equation}
    \vartheta(v; G) = \frac{1 - G(v)}{g(v)}= \frac{1}{\alpha}v^{1 - \alpha}
\end{equation}
and thus
\begin{equation}
    \vartheta'(v; G) = \frac{1 - \alpha}{\alpha}v^{- \alpha}\to 0 \text{ as }
    v \to \infty,
\end{equation}
implying that the Weibull distribution satisfies the von Mises condition for all $\alpha > 0$.
This then implies that the Weibull distribution belongs to the twice-differentiable domain of attraction of Gumbel.

When $G$ is the Weibull distribution with shape parameter $\alpha$, from
\eqref{eq: a_K b_K gumbel}, we have
\begin{equation}
    a_{K}= \frac{K}{\alpha}\Gamma\left(\frac{1}{\alpha}, \log K \right),\: b_{K}
    = \left( \log K \right)^{\frac{1}{\alpha}}.
\end{equation}

The probability density of the normalized largest-order statistic, which is illustrated in Figure~\ref{fig: weibull example probability density}, is given by $\hat{g}_{K}(w) = a_{K}g_{K}(a_{K}w + b_{K})$. The normalized hazard rate is given by
\begin{equation}
    r(w; \hat{G}_{K}) = \frac{a_{K}g_{K}(a_{K}w + b_{K})}{1 - G_{K}(a_{K}w +
    b_{K})},
\end{equation}
and its derivative, illustrated in Figure~\ref{fig: weibull example derivative of HR}, is given by
\begin{equation}
    r'(w; \hat{G}_{K}) = \frac{a_{K}^{2}g'_{K}(a_{K}w + b_{K}) (1 - G_{K}(a_{K}w + b_{K})) + a_{K}^{2}g_{K}^{2}(a_{K}w + b_{K})}{(1 - G_{K}(a_{K}w + b_{K}))^{2}}    .
\end{equation}

\subsection{Asymmetric Capacity}
\label{subsubsec: asymmetric capacity}

In Section~\ref{sec: continuous market}, for analytical tractability, we assumed that all object types have an equal capacity. This assumption is not always realistic; for example, there are only two weekend days compared to five weekdays. In such environments, an ex post efficient mechanism may assign non-favorite objects when an agent's favorite object is scarce, making it difficult to reduce the problem to a single-dimensional environment. This complexity makes it challenging to characterize an efficient mechanism.

In this section, we solve a linear program (LP) to numerically derive an efficient mechanism and compare its structure with that of the efficient mechanism in a continuous i.i.d.\ market. The residual surplus function $RS$ and constraints \eqref{eq: defn resource}, \eqref{eq: defn strategy proofness}, \eqref{eq: defn individual rationality}, and \eqref{eq: defn unit demand} are all linear in the allocation and payment rule, $(\bx,p)$, and thus this problem is an LP. To use a solver, we discretize the valuation space $V \subset \mathbb{R}^{K}_+$ by dividing each dimension into $n$ intervals with equal probability weights to produce $n^{K}$ grid points, each of which is specified as a $K$-tuple of lower endpoints. Throughout this appendix, we take $n=30$ intervals.
We consider the case of $K = 2$ and assume that the value for each object is drawn independently from a standard exponential distribution. If both objects had equal capacities, then SD would be efficient. Here, we instead assume that $m_{1}= 0.4$ and $m_{2}= 0.1$.

\begin{figure}[t]
    \begin{subfigure}{.32\linewidth}
        \centering
        \includegraphics[width=\linewidth]{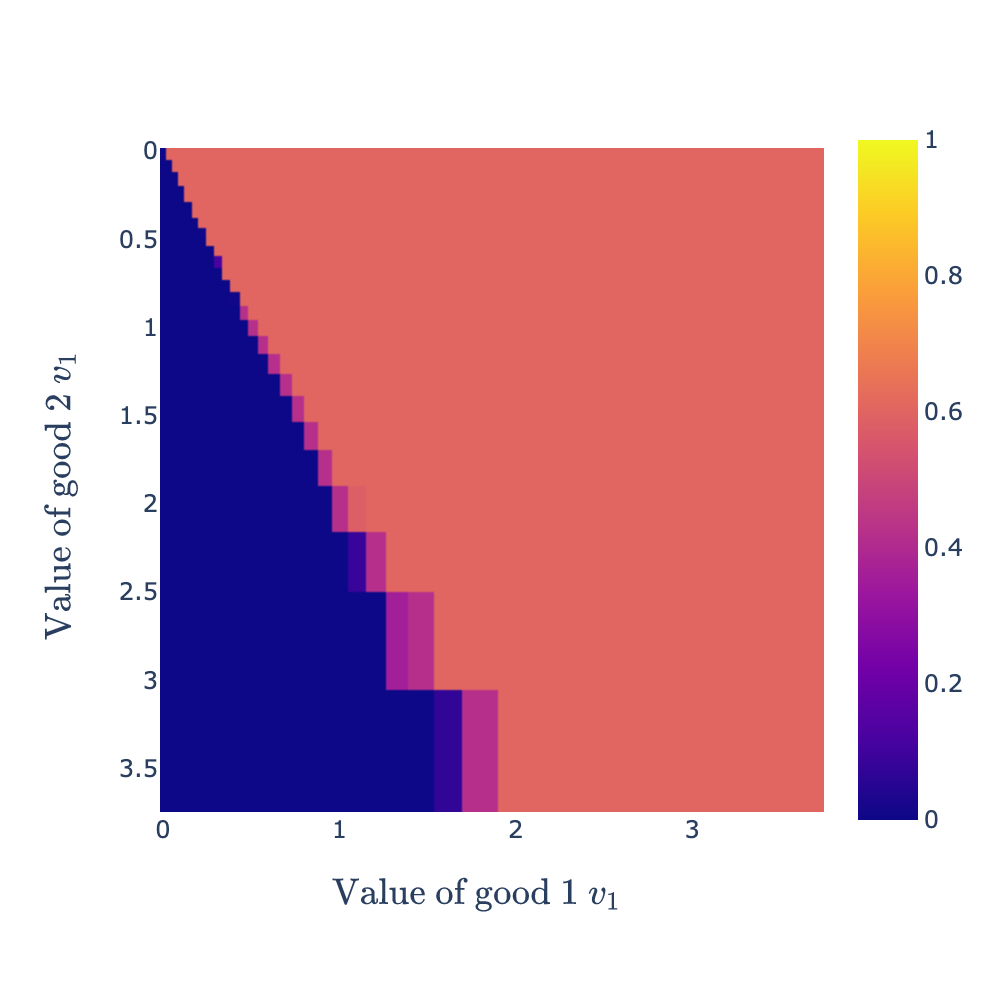}
        \subcaption{$x_{1}(v_{1}, v_{2})$}
    \end{subfigure}
    \begin{subfigure}{.32\linewidth}
        \centering
        \includegraphics[width=\linewidth]{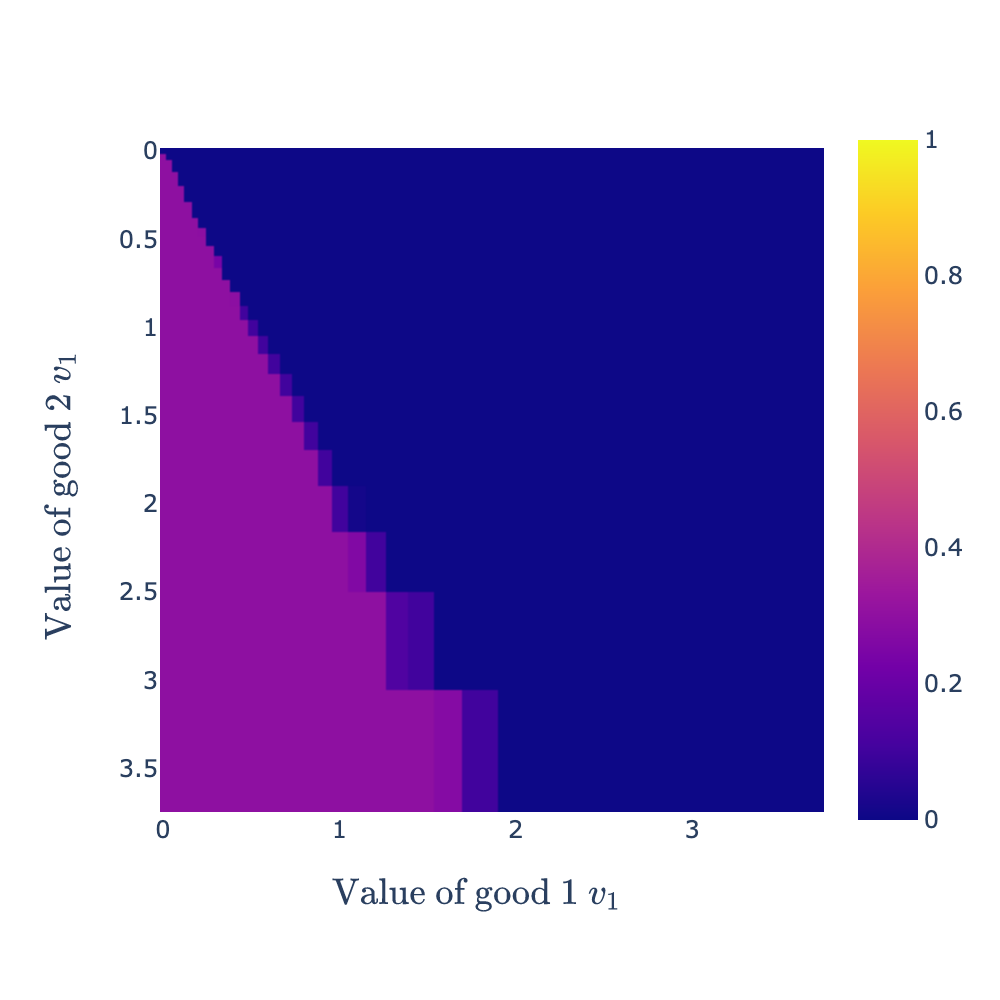}
        \subcaption{$x_{2}(v_{1}, v_{2})$}
    \end{subfigure}
    \begin{subfigure}{.32\linewidth}
        \centering
        \includegraphics[width=\linewidth]{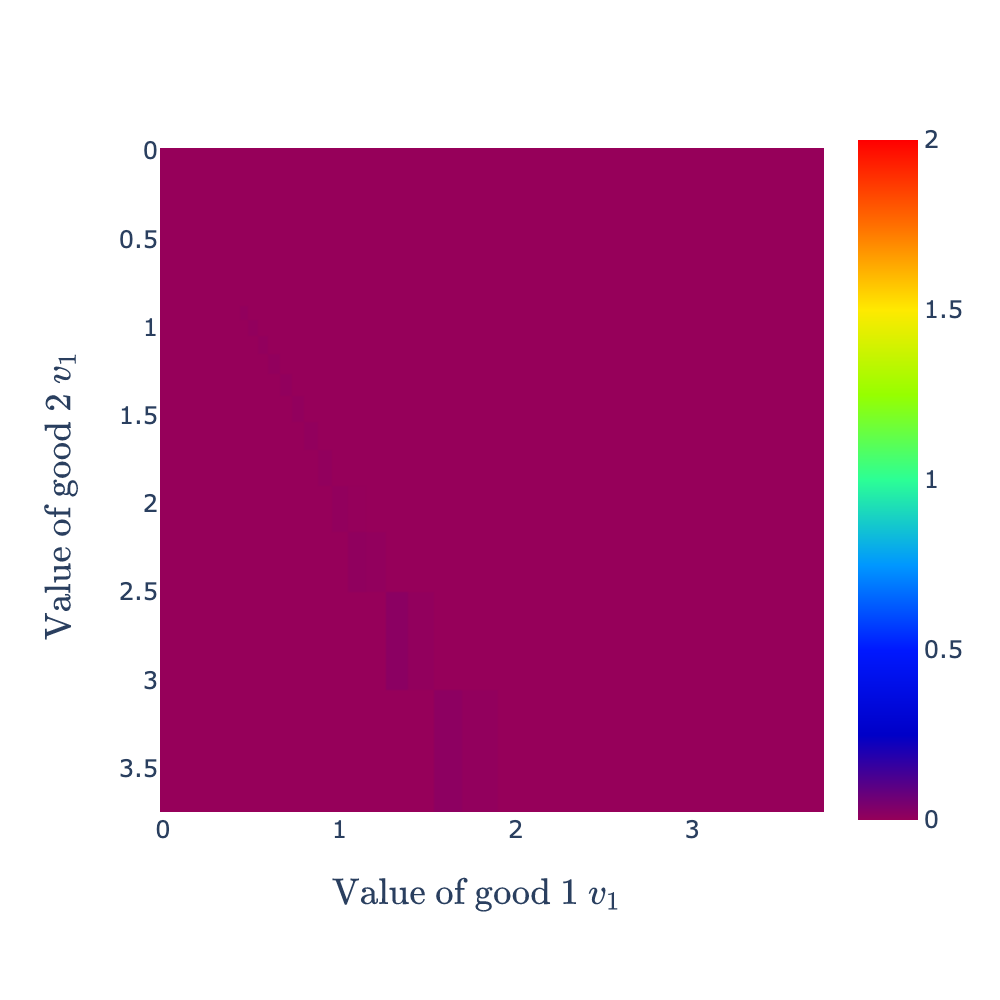}
        \subcaption{$p(v_{1}, v_{2})$}
    \end{subfigure}
    \caption{An efficient mechanism for the case of asymmetric capacity. The optimal solution can be achieved with RF.}
    \label{fig:exp_asym_cap_LP}
\end{figure}

Figure~\ref{fig:exp_asym_cap_LP} illustrates an optimal solution to the LP for the case of asymmetric capacity. In all three panels, the horizontal axis represents the value of object $1$ ($v_{1}$), and the vertical axis represents the value of object $2$ ($v_{2}$). The colors of panels (a), (b), and (c) indicate the probability of allocating objects $1$ and $2$ ($x_{1}(v_{1}, v_{2}), x_{2} (v_{1}, v_{2})$), and the payment ($p(v_{1}, v_{2})$), respectively.

The efficient mechanism is neither SD nor VCG. The mechanism offers two options: receiving object $1$ with high probability or receiving object $2$ with low probability, without requiring payments. Following the terminology of \citet{goldner2024simple}, we call this mechanism the \emph{random favorite mechanism} (RF).\footnote{\citet{goldner2024simple} study an (ex ante) symmetric environment, and therefore, each agent $i$ claims $k \in \argmax v_{k}^{i}$ as her favorite object in equilibrium. This property does not necessarily hold with asymmetric capacity because the probability of being assigned depends on which object to claim.} By requiring agents to determine which object to claim before drawing a lottery, this mechanism partially screens for agents' preference for their favorite object relative to the other one---only agents who strongly prefer object $2$ over object $1$ desire to claim object $2$, given that object $1$ is much easier to obtain. By contrast, under SD, claiming object $2$ when it is available is risk-free, and therefore, all agents who prefer object $2$ over object $1$ obtain object $2$ whenever possible. Consequently, RF outperforms SD with asymmetric capacities.

While RF outperforms SD in terms of residual surplus, it comes with several drawbacks. While RF can be implemented as a strategy-proof mechanism in a continuous market, to determine the allocation probability of each option, we need information about the value distribution, which is often considered challenging in the market design literature \citep{wilson1985game}. By contrast, SD is detail-free. Furthermore, RF cannot satisfy strategy-proofness in a finite market, even if the market is large.\footnote{Several studies on matching theory \citep[e.g.,][]{abdulkadirouglu2011resolving} have demonstrated that non-strategy-proof mechanisms may outperform strategy-proof mechanisms in equilibrium because their equilibrium strategies can reflect agents' preference intensity. Note also that RF may not satisfy even a weaker notion of truthfulness, such as Bayesian incentive compatibility (BIC). While \citet{goldner2024simple} prove that RF satisfies BIC when each object $k$ has unit capacity and each value $v_k^i$ is drawn i.i.d., these assumptions are rarely satisfied in practice.} This is because the object an agent should claim depends on which objects other agents are claiming. In addition, RF cannot satisfy the practically desirable properties that FCFS and the sequential-form SD meet (discussed in Section~\ref{sec: mass vaccination}).

\begin{figure}[t]
    \centering
    \includegraphics[width=0.5\linewidth]{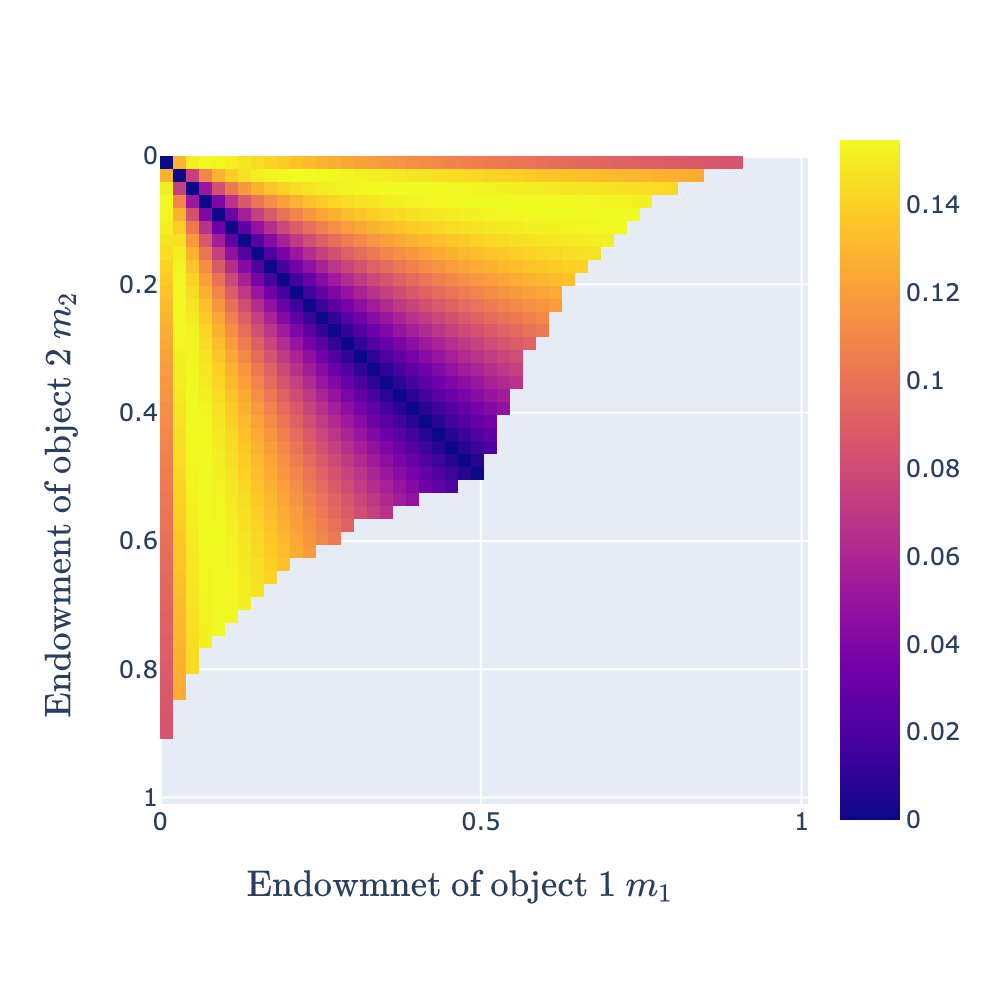}
    \caption{Percentage difference of the residual surplus achieved by RF and SD
    $(RS(\mech_{RF}) - RS(\mech_{SD}))/ RS(\mech_{RF})$ over parameter values for which the RF mechanism admits an interior solution.}
    \label{fig:exp_asym_endowment_utildiff}
\end{figure}

To evaluate the tradeoff between social welfare and the simplicity of the mechanism, we compare the performance of SD and RF. The allocation probability of RF should satisfy a market-clearing condition. That is, given the allocation probability, agents should optimally choose one option. This determines the demand for each object, and together with the resource constraint, the demand determines the allocation probability. The initially posited allocation probabilities must coincide with the probabilities implied by the resulting demands. For each pair of endowments $(m_{1}, m_{2})$, there exists a unique RF that satisfies the market-clearing condition, and its performance has a closed-form representation when values follow an exponential distribution. The derivation is presented in Appendix~\ref{subsec: RF Mechanism}.

Figure~\ref{fig:exp_asym_endowment_utildiff} displays the percentage difference of residual surplus between RF and SD (i.e., $(RS(\mech_{RF}) - RS (\mech_{SD}))/ RS(\mech_{RF})$) for various endowments, $(m_{1}, m_{2})$. For any endowments, RF performs as well as or better than SD. Indeed, when $m _{1}= m_{2}$, RF and SD return an identical allocation, and the performance difference is zero. The percentage difference is at most about 15\%, implying that the gain from using RF is not excessively large.

\subsection{Illustration of Mechanisms Returned by RegretNet}\label{subsec: illustration of mechanisms returned by regretnet}

\begin{figure}[t]
    \begin{subfigure}{.32\linewidth}
        \centering
        \includegraphics[width=\linewidth]{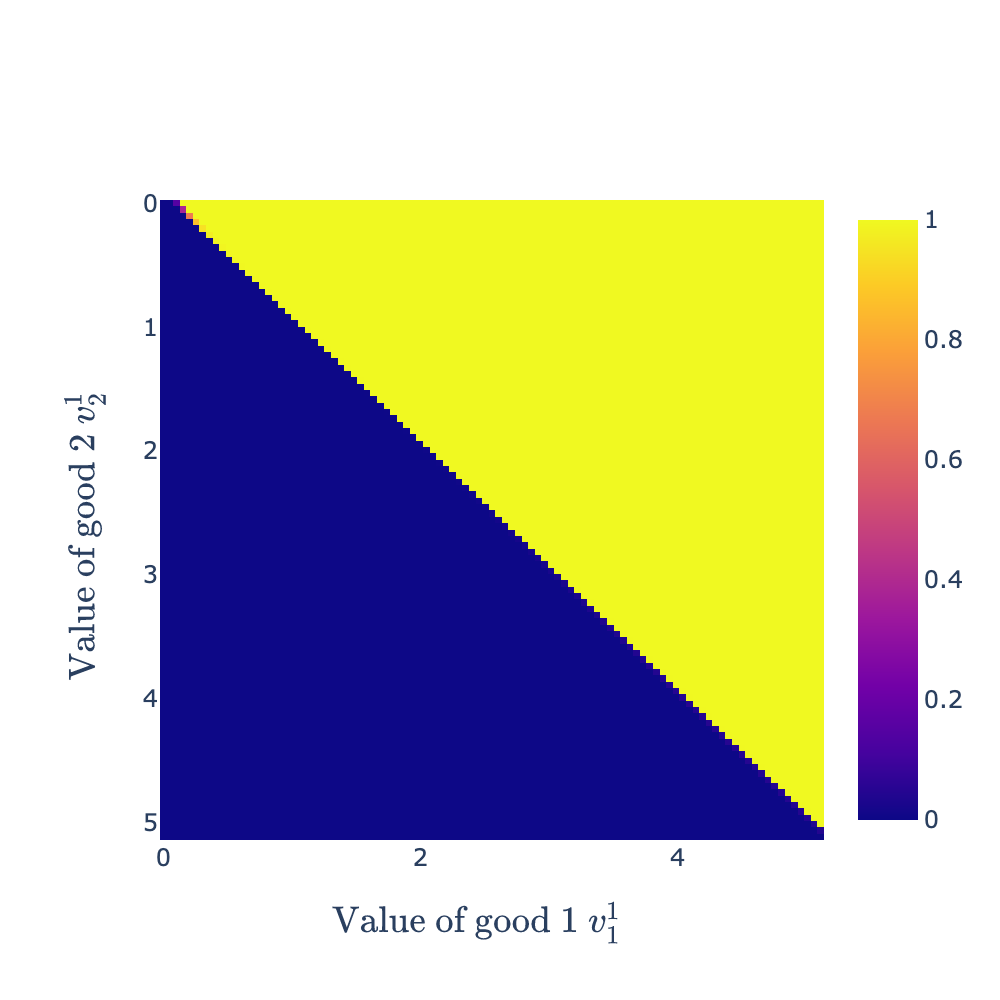}
        \subcaption{$x_{1}^{1}(v_{1}^{1}, v_{2}^{1}, 2, 1)$}
    \end{subfigure}
    \begin{subfigure}{.32\linewidth}
        \centering
        \includegraphics[width=\linewidth]{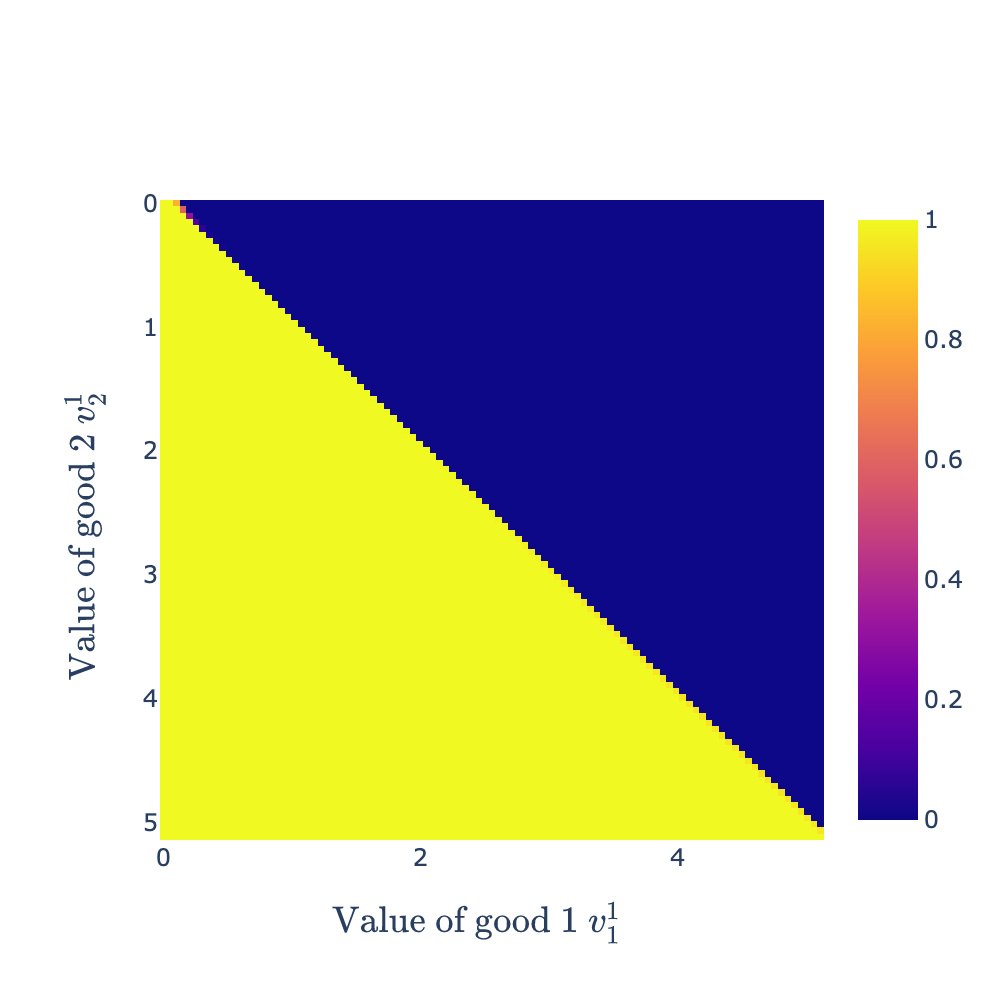}
        \subcaption{$x_{2}^{1}(v_{1}^{1}, v_{2}^{1}, 2, 1)$}
    \end{subfigure}
    \begin{subfigure}{.32\linewidth}
        \centering
        \includegraphics[width=\linewidth]{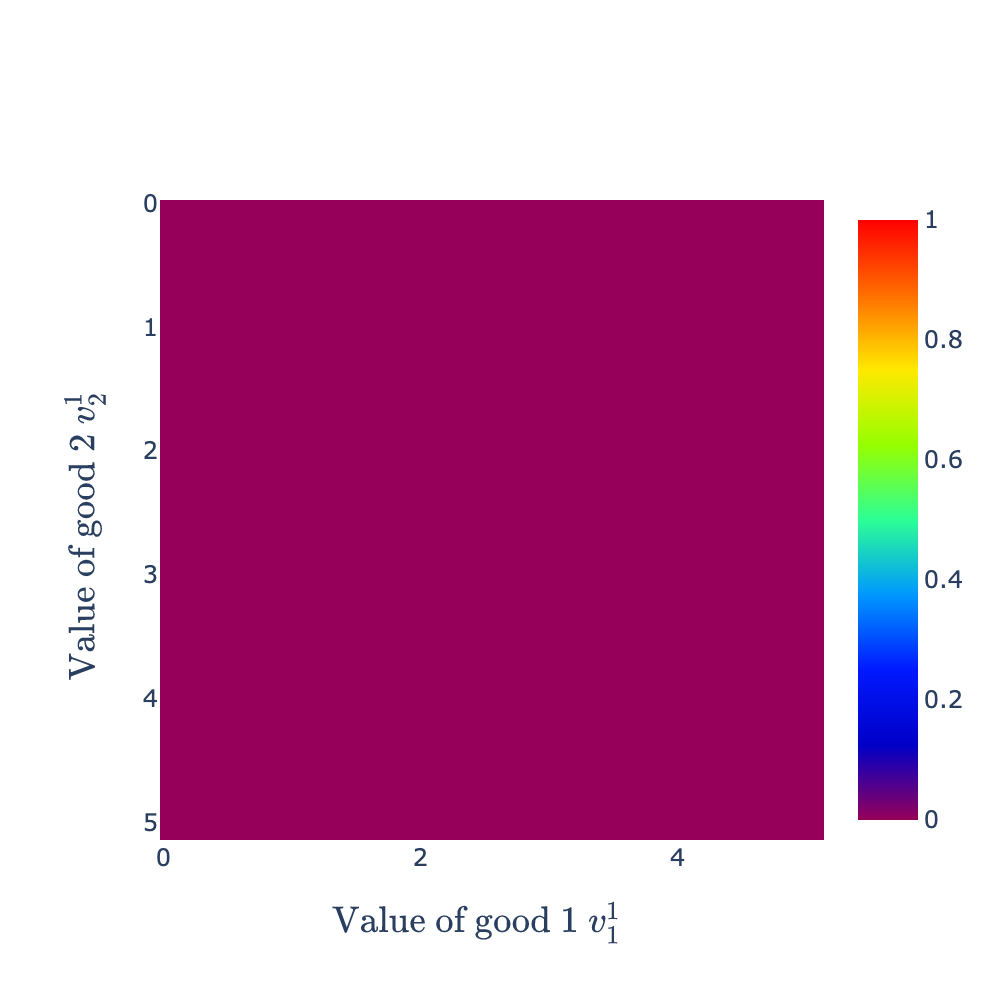}
        \subcaption{$p^{1}(v_{1}^{1}, v_{2}^{1}, 2, 1)$}
    \end{subfigure}
    \caption{Allocation and payment for agent 1 from an efficient mechanism
    learned by RegretNet. The marginal value distribution is the Weibull distribution with parameter $0.8$. Agent 2's value is fixed to $(v^{2}_{1},v^{2}_{2})=
    (2,1)$.}
    \label{fig:regretnet_iid_weibull08}
\end{figure}

\begin{figure}[t]
    \begin{subfigure}{.32\linewidth}
        \centering
        \includegraphics[width=\linewidth]{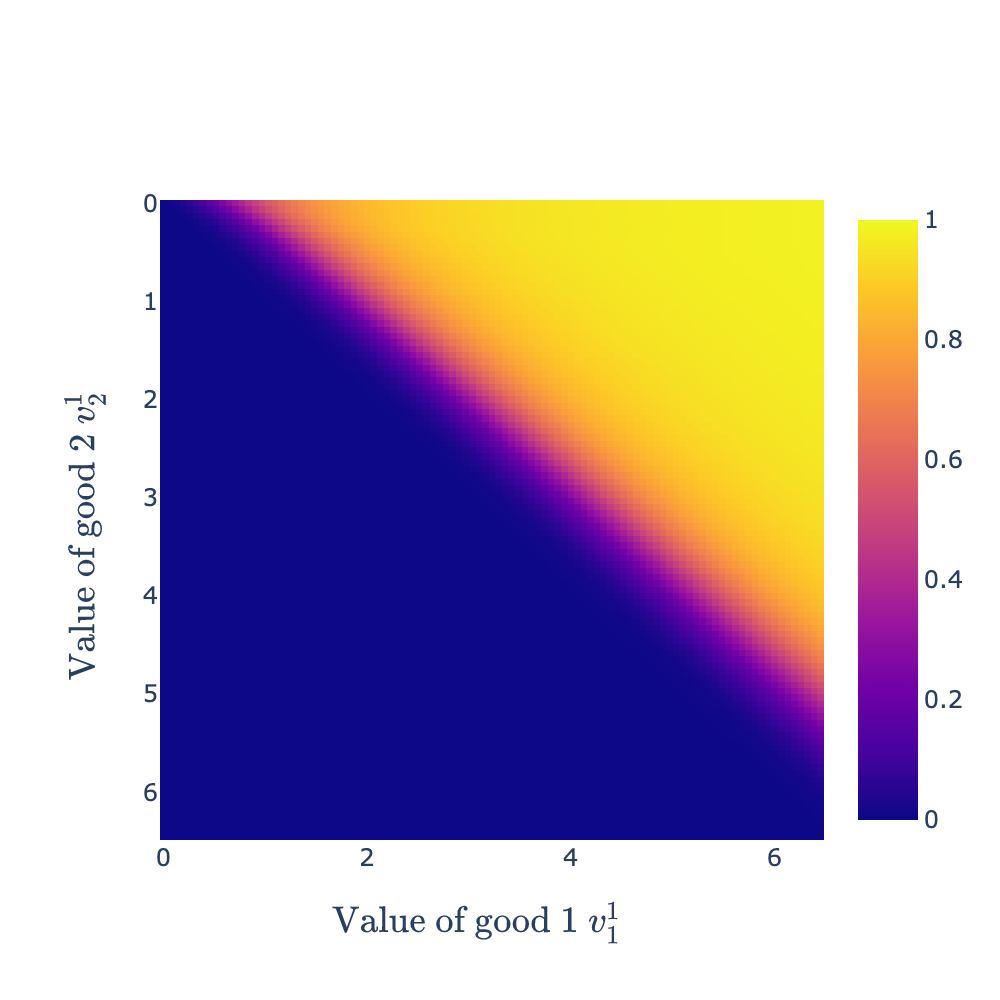}
        \subcaption{$x_{1}^{1}(v_{1}^{1}, v_{2}^{1}, 2, 1)$}
    \end{subfigure}
    \begin{subfigure}{.32\linewidth}
        \centering
        \includegraphics[width=\linewidth]{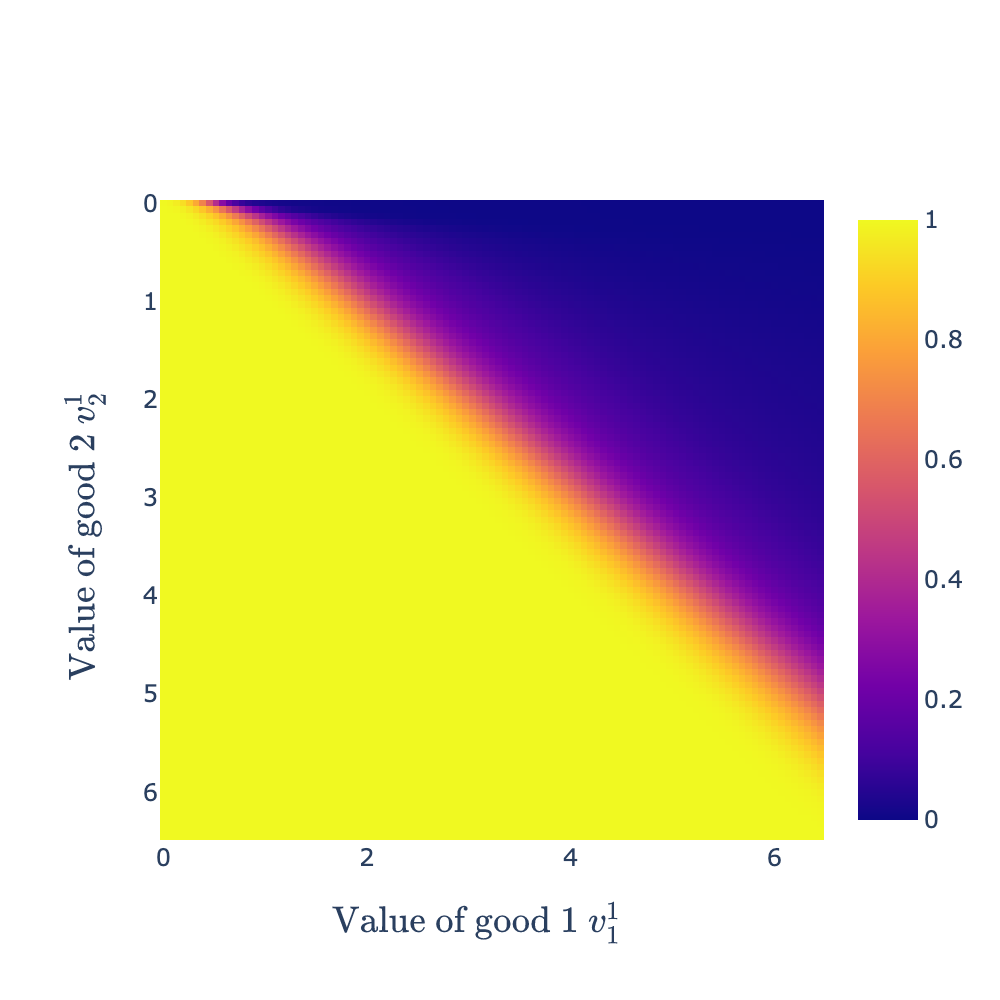}
        \subcaption{$x_{2}^{1}(v_{1}^{1}, v_{2}^{1}, 2, 1)$}
    \end{subfigure}
    \begin{subfigure}{.32\linewidth}
        \centering
        \includegraphics[width=\linewidth]{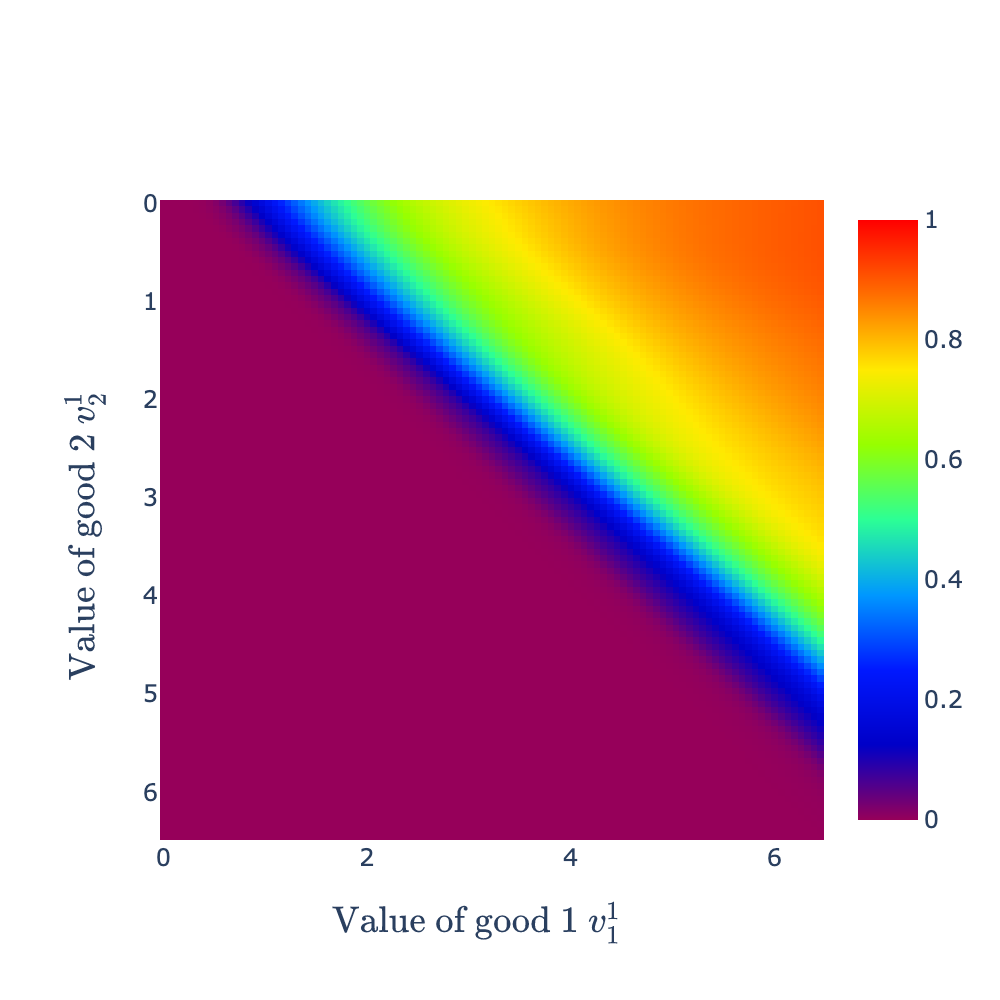}
        \subcaption{$p^{1}(v_{1}^{1}, v_{2}^{1}, 2, 1)$}
    \end{subfigure}
    \caption{Allocation and payment for agent 1 from an efficient mechanism
    learned by RegretNet. The marginal value distribution is the Weibull distribution with parameter $0.7$. Agent 2's value is fixed to $(v^{2}_{1},v^{2}_{2})=
    (2,1)$.}
    \label{fig:regretnet_iid_weibull07}
\end{figure}

Figures~\ref{fig:regretnet_iid_weibull08} and \ref{fig:regretnet_iid_weibull07} show the approximately efficient mechanisms learned by RegretNet for a two-agent, two-object setting where the value distribution is i.i.d.\ Weibull. The horizontal and vertical axes represent $v_{1}^{1}$ and $v_{2}^{1}$, agent $1$'s value for objects $1$ and $2$. Agent $1$'s allocation probability and the amount of agent $1$'s payment are represented by color. While RegretNet returns the full shape of the mechanism, for the illustration's sake, we fix agent $2$'s valuation to $(v_{1}^{2}, v_{2} ^{2}) = (2, 1)$ to draw these figures.

Figure~\ref{fig:regretnet_iid_weibull08} shows the case for a Weibull distribution with parameter 0.8. The mechanism allocates agent $1$ to her favorite object, without payment and regardless of agent $2$'s value, implying that the mechanism is SD, which prioritizes agent $1$ over agent $2$.

Figure~\ref{fig:regretnet_iid_weibull07} shows the case for a Weibull distribution with parameter 0.7. The value distribution has a thicker tail, and the learned mechanism roughly allocates the contested object with a price equal to the other agent's value. Since agent $2$ prefers object $1$ over object $2$ by $v_{1}^{2} - v_{2}^{2}= 2 - 1 = 1$, agent $1$ obtains object $1$ only if her excess value $v_{1}^{1}- v_{2}^{1}$ is larger than $1$, and agent $1$ has to pay the price of $1$ in such cases. This is how VCG allocates objects.

\subsection{Performance of SD, VCG, and the (Numerically) Efficient Mechanisms}\label{subsec: performance of SD VCG and Efficient appendix}

\begin{figure}[t!]
    \centering
    \begin{subfigure}[t]{0.48\textwidth}
        \includegraphics[width=\textwidth]{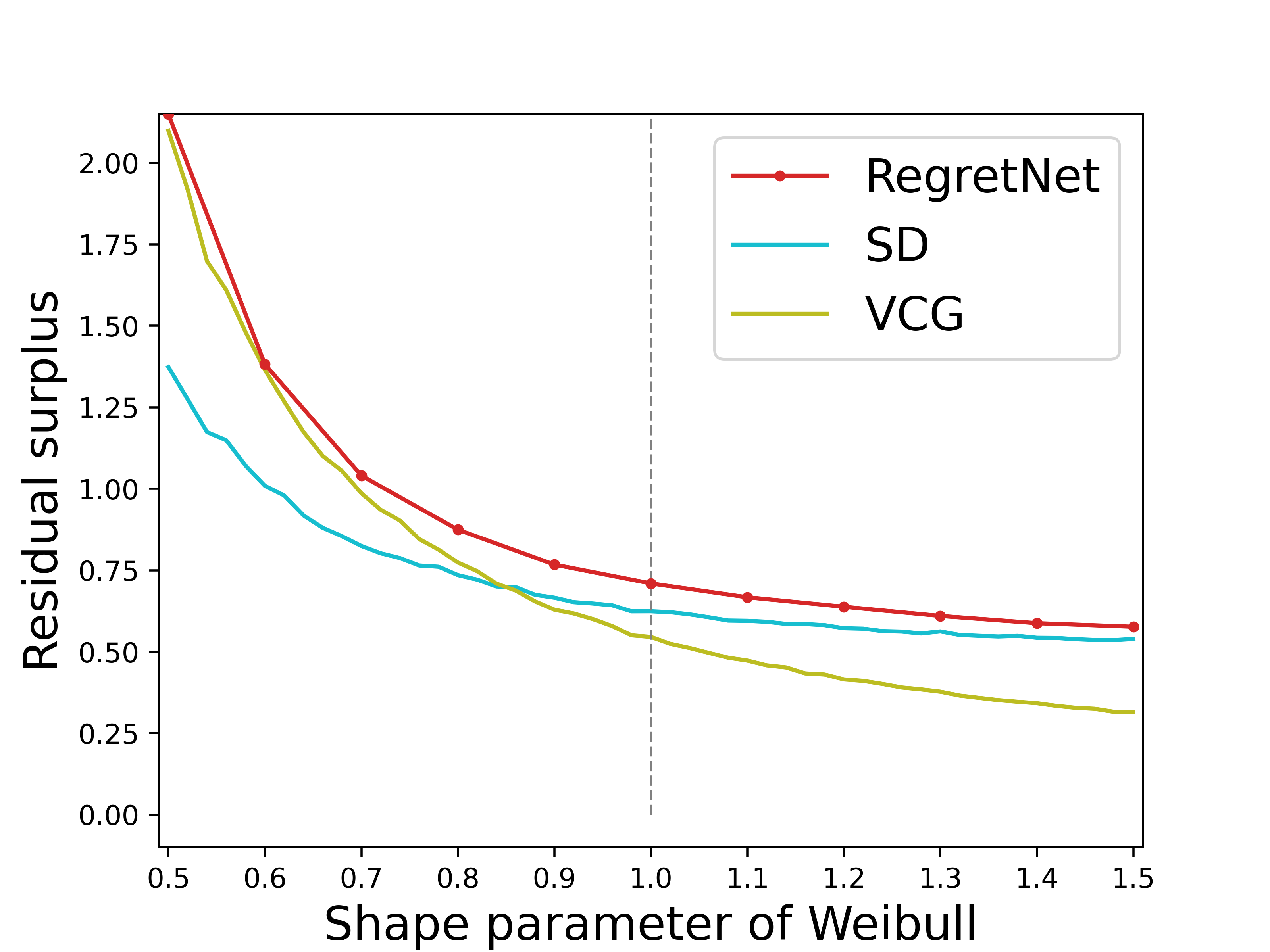}
        \subcaption{$c = 1$}
        \label{fig: finite weibull c1 appendix}
    \end{subfigure}
    \hspace{0.02\textwidth}
    \begin{subfigure}[t]{0.48\textwidth}
        \includegraphics[width=\textwidth]{EC/figures_ec/finite_weibull_c=5.png}
        \subcaption{$c = 5$}
        \label{fig: finite weibull c5 appendix}
    \end{subfigure}\\
    \begin{subfigure}[t]{0.48\textwidth}
        \includegraphics[width=\textwidth]{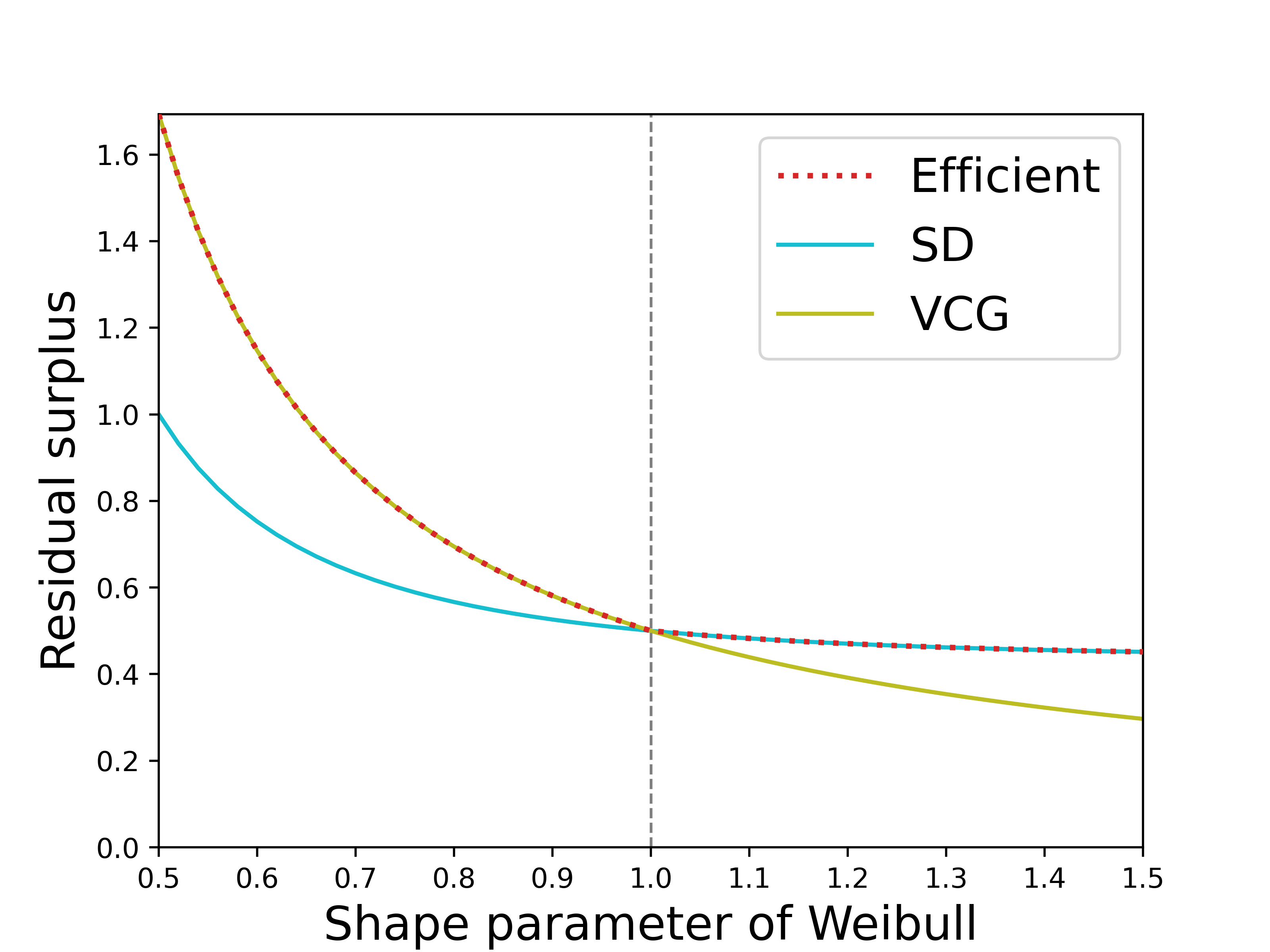}
        \subcaption{Single-good case}
        \label{fig: finite weibull single good appendix}
    \end{subfigure}
    \hspace{0.02\textwidth}
    \begin{subfigure}[t]{0.48\textwidth}
        \includegraphics[width=\textwidth]{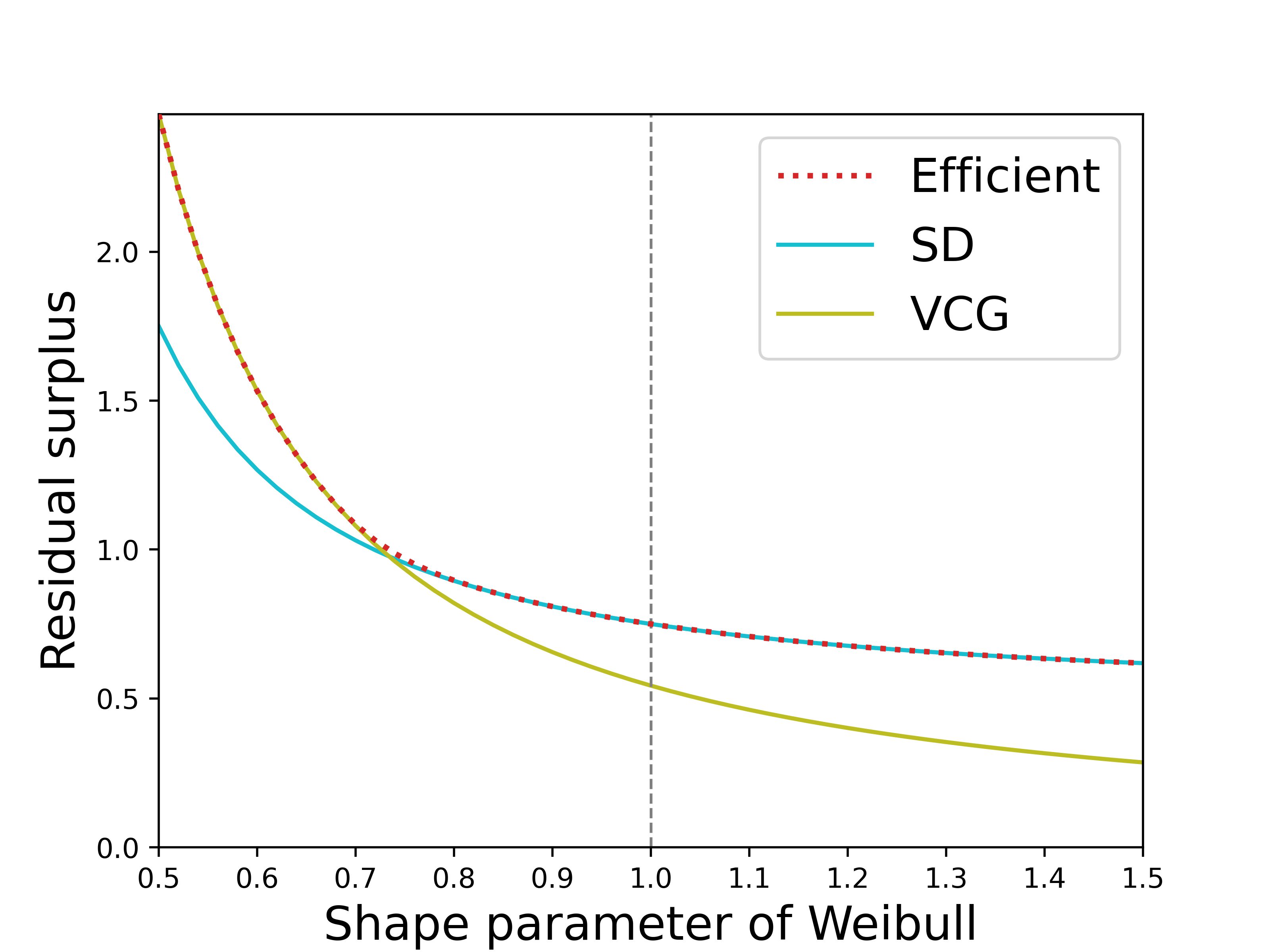}
        \subcaption{Continuous case}
        \label{fig: finite weibull continuous appendix}
    \end{subfigure}
    \caption{The performance of SD, VCG, and the (numerically) efficient mechanism}
    \label{fig: finite weibull all appendix}
\end{figure}

Figure~\ref{fig: finite weibull all appendix} illustrates the residual surplus achieved by SD, VCG, and the (numerically) efficient mechanism across four cases: $c = 1$, $c = 5$, the single-good case, and the continuous case. (Note that the $c = 5$ case is identical to Figure~\ref{fig: finite weibull c5} in the main text but is included here for comparison.)

For the single-good and continuous market cases, the efficient mechanism is derived analytically using the results from \citet{Hartline2008} and Section~\ref{sec: continuous market}. In contrast, for the $c = 1$ and $c = 5$ cases, the mechanism is obtained numerically using RegretNet.

In the single-good case, the efficient mechanism precisely aligns with whichever of SD and VCG performs better across all values of $\alpha$. In the continuous case, the same pattern holds visually, although a closer numerical inspection reveals that the efficient mechanism can outperform both SD and VCG by a small but strictly positive margin. We also note that, in the continuous case, where there is minimal but non-zero heterogeneity among goods (i.e., $K = 2$), the threshold value of the Weibull parameter shifts from 1.0 to below 0.8. As $K$ increases, this threshold approaches $\alpha = 0$.

The case with $c = 1$ exhibits characteristics that lie between those of the single-good case and the $c = 5$ case. Even for $c = 1$, the effect of having multiple goods is clearly evident, as reflected in the observed performance difference.

\subsection{Simulation Settings for the Case of Correlated Values}
\label{subsec: supp finite market}

This section illustrates how we construct the joint distribution $F$ for the case of correlated values. Let $\Sigma \in [-1, 1]^{(I \times K) \times (I \times K)}$ be an $(I \times K)$-dimensional correlation matrix. The \emph{Gaussian copula} $C_{\Sigma}: [0, 1]^{I \times K}\to [0, 1]$ is defined as 
\begin{equation}\label{eq: Gaussian Copula}
    C_{\Sigma}(\bw) = N_{\Sigma}(N^{-1}(w_{1}^{1}),\dots,N^{-1}(w_{K}^{I})),
\end{equation}
where $N_{\Sigma}$ is the joint cumulative distribution function of an $(I \times K)$-dimensional multivariate normal distribution with mean vector zero and covariance matrix $\Sigma$, and $N^{-1}$ is the inverse cumulative distribution
function of a (univariate) standard normal.

By construction of $C_{\Sigma}$, for all $(i,k)$, the marginal distribution of $w_{k}^{i}$ is a uniform distribution over $[0, 1]$. Accordingly, by defining 
\begin{equation}\label{eq: joint distribution from copula}
    F(\bv) = C_{\Sigma}(G(v_{1}^{1}), \dots, G(v_{K}^{I})),
\end{equation}
we can construct a joint cumulative distribution $F$ of $\bv$ such that for all $(i, k)$, the marginal value distribution of $v_{k}^{i}$ is $G$. Furthermore, by imposing symmetry on the covariance matrix $\Sigma$ used in the construction of the copula, such as keeping within-agent correlation constant or between-agent correlation constant, the symmetry will be preserved in the covariance matrix of the output joint distribution $F$. Although covariance and correlation coefficients are not necessarily preserved during the transformation process, the relative strength of the correlations is maintained. Therefore, if stronger correlations are set in the covariance matrix $\Sigma$, the resulting joint distribution $F$ will also exhibit stronger correlations. As a result, by varying the covariance matrix $\Sigma$, we can generate a joint distribution $F$ with the desired strengths of within-agent correlation and between-agent correlation.

In the numerical analysis, instead of directly computing the joint distribution $F$ using \eqref{eq: Gaussian Copula} and \eqref{eq: joint distribution from copula}, we generate random variables $\bv$ following the intended joint distribution
$F$ by generating $\bu = (u_{1}^{1},\dots,u_{K}^{I})$ following a multivariate normal distribution with mean $0$ and covariance matrix $\Sigma$, and then transforming it as $\bv = (G^{-1}(N(u_{1}^{1})), \dots, G^{-1} (N(u_{K}^{I})))$.

We use the following covariance matrices $\Sigma$ for our simulations. To evaluate the effect of the within-agent correlation, we employ (i) $\cov(u_{k}^{i}, u_{k}^{i}) = 1$ for all $(i, k) \in I \times K$, (ii) $\cov(u_{k}^{i}, u_{l}^{i}) = c$ for all $i \in I$, for all distinct $k, l \in K$, for some $c \in [0, 1]$, and (iii) $\cov(u_{k}^{i}, u_{l}^{j}) = 0$ otherwise. To evaluate the effect of the between-agent correlation, we employ (i) $\cov(u_{k}^{i}, u_{k}^{i}) = 1$ for all $(i, k) \in I \times K$, (ii) $\cov(u_{k}^{i}, u_{k}^{j}) = c$ for all $k \in K$, for all distinct $i, j \in I$, for some $c \in [0, 1]$, and (iii) $\cov(u_{k}^{i}, u_{l}^{j}) = 0$ otherwise.

Finally, we describe how we calculate the correlation coefficient of the generated distribution $F$, which is represented on the horizontal axis of Figures~\ref{fig: within-agent correlation} and \ref{fig: between-agent correlation}. The method used is similar for both cases; thus, we will only describe the within-agent correlation case. Noting that all $v_{k}^{i}$ share the
same marginal distribution, the correlation coefficient between $v_{k}^{i}$ and $v_{l}^{i}$ for $k \neq l$ is defined as follows:
\begin{equation}
    \label{eq: defn correlation coefficient}\corr(v_{k}^{i}, v_{l}^{i}) = \frac{\cov(v_{k}^{i},
    v_{l}^{i})}{\var(v_{k}^{i})}.
\end{equation}
Here, $\var(v_{k}^{i})$ is known since we specify the marginal distribution as Weibull. However, the joint distribution $F$ derived using the Gaussian copula does not have a tractable closed-form representation; thus $\cov(v_{k}^{i},v_{l}^{i})$ cannot be derived in closed form. Therefore, we estimate it by sampling. Specifically, let the sample size be $T$, and denote the value of $v_{k}^{i}$ in the $t$-th sample as $v_{k}^{i}(t)$. The sample covariance $S(i, k, l)$ between $v_{k}^{i}$ and $v_{l}^{i}$ is expressed as follows:
\begin{equation}
    S(i, k, l) = \frac{1}{T}\sum_{t = 1}^{T}\left(v_{k}^{i}(t) - \mu \right)\left(v_{l}^{i}(t) - \mu \right),
\end{equation}
where $\mu \coloneqq \mathbb{E}[v_k^i(t)]$. Note that we can calculate $\mu$ from the marginal value distribution $G$ that each $v_{k}^{i}(t)$ follows, and thus there is no need to compute its sample mean. Due to the symmetry of the covariance matrix $\Sigma$ used to generate $\bu$, $\cov(v_{k}^{i},v_{l}^{i}) = \cov(v^{i}_{\hat{k}},v^{i}_{\hat{l}})$ holds as long as $k \neq l$ and $\hat{k}\neq \hat{l}$. Using this property, we aggregate the samples of within-agent correlation across different $(i, k, l)$ to obtain the following sample covariance:
\begin{equation}
    \label{eq: sample covariance}\widehat{\cov}(v_{k}^{i}, v_{l}^{i}) = \frac{1}{I}
    \frac{2}{K(K-1)}\sum_{i = 1}^{I}\sum_{k = 1}^{K}\sum_{l = k + 1}^{K}S(i ,
    k, l).
\end{equation}
In the simulations, we used a sample size of $T = 100{,}000$ for each setting. By substituting the sample covariance obtained from \eqref{eq: sample covariance} into \eqref{eq: defn correlation coefficient}, we derive the correlation coefficients plotted on the horizontal axis of Figure~\ref{fig: within-agent correlation}.

\section{Proofs of Remaining Results}
\label{sec: proofs}

\subsection{Proof of Theorem~\ref{thm: NBUE iff no screening}}

\begin{proof}
Let $\bar G_K(t)=1-G_K(t)$. Choose a right-continuous version of a nondecreasing allocation $x$ and let $dx$ denote its Stieltjes measure. At every continuity point,
\begin{equation}
    x(v)=x(0)+\int_{(0,\bar v)}\mathbf1\{t\le v\}\,dx(t).
\end{equation}
Because $G_K$ is continuous and a monotone function has only countably many discontinuities, Tonelli's theorem gives
\begin{align}
    \int_0^{\bar v}x(v)dG_K(v)
    &=x(0)+\int_{(0,\bar v)}\bar G_K(t)\,dx(t),
    \label{eq: NBUE proof resource decomposition}\\
    \int_0^{\bar v}\bar G_K(v)x(v)dv
    &=x(0)\mathbb{E}_{G_K}[v]
      +\int_{(0,\bar v)}\left(\int_t^{\bar v}\bar G_K(v)dv\right)dx(t).
    \label{eq: NBUE proof value decomposition}
\end{align}

Suppose first that $G_K$ is NBUE. Then, for every $t$,
\begin{equation}
    \int_t^{\bar v}\bar G_K(v)dv
    \le\mathbb{E}_{G_K}[v]\bar G_K(t).
\end{equation}
Substituting this inequality into \eqref{eq: NBUE proof value decomposition} and using \eqref{eq: NBUE proof resource decomposition}, every feasible allocation with resource use at most $\bar{m}$ satisfies
\begin{equation}
    \int_0^{\bar{v}}(1- G_K(v))x(v)dv
    \le
    \mathbb{E}_{G_K}[v]\int x\,dG_K
    \le
    \bar{m}\mathbb{E}_{G_K}[v].
\end{equation}
The constant allocation $x\equiv \bar{m}$ attains $\bar{m}\mathbb{E}_{G_K}[v]$, so no screening is efficient for every $\bar{m}\in(0,1)$.

Conversely, suppose NBUE fails at some $t$. Let
\begin{equation}
    \bar q=\bar G_K(t)\in(0,1),
    \qquad
    b(t)=\frac{1}{\bar q}\int_t^{\bar v}\bar G_K(v)dv>\mathbb{E}_{G_K}[v].
\end{equation}
If $\bar{m}\le \bar q$, the allocation
\begin{equation}
    x_{\bar{m}}(v)=\frac{\bar{m}}{\bar q}\mathbf1\{v>t\}
\end{equation}
uses resource $\bar{m}$ and yields residual surplus $\bar{m} b(t)>\bar{m}\mathbb{E}_{G_K}[v]$. If $\bar{m}>\bar q$, set
\begin{equation}
    c=\frac{\bar{m}-\bar q}{1-\bar q}\in(0,1)
\end{equation}
so that
\begin{equation}
    x_{\bar{m}}(v)=c+(1-c)\mathbf1\{v>t\}
\end{equation}
uses resource $c+(1-c)\bar q=\bar{m}$. Its residual surplus exceeds that of no screening by
\begin{equation}
    (1-c)\bar q\bigl(b(t)-\mathbb{E}_{G_K}[v]\bigr)>0.
\end{equation}
Thus no screening is inefficient for every $\bar{m}\in(0,1)$ whenever NBUE fails.
\end{proof}

\subsection{Proof of Theorem~\ref{thm: DHR iff full screening}}

\begin{proof}
Define $h(q)\coloneqq \vartheta(G_K^{-1}(q);G_K)$ for $q \in (0, 1)$, and, for an allocation rule $x$, write $z(q)\coloneqq x(G_K^{-1}(q))$.
As in the standard ironing argument, the reduced problem can be written as
maximizing $\int_0^1 h(q)z(q)\,dq$
over nondecreasing $z:[0,1]\to[0,1]$ satisfying
$\int_0^1z(q)\,dq\le\bar m$. The full-screening allocation is
$z^{FS}(q)=\mathbf 1\{q>1-\bar m\}$.

Suppose first that $G_K$ has a DHR. Then
$\vartheta=1/r$ is nondecreasing, and hence so is $h$. Therefore, the
displayed objective is maximized by assigning the resource to the highest
quantiles. Thus, $z^{FS}$ is efficient for every
$\bar m\in(0,1)$.

Conversely, suppose that $G_K$ does not have a DHR. Then $h$ is not
nondecreasing. Under the maintained smoothness assumptions, there exists
$q_0\in(0,1)$ such that $h'(q_0)<0$. Choose
$\epsilon\in(0,\min\{q_0,1-q_0\})$ sufficiently small that $h$ is strictly
decreasing on $[q_0-\epsilon,q_0+\epsilon]$, and set
$\bar m=1-q_0$. Consider
\[
    \tilde z(q)=
    \begin{cases}
        0, & q<q_0-\epsilon,\\
        1/2, & q_0-\epsilon\le q\le q_0+\epsilon,\\
        1, & q>q_0+\epsilon.
    \end{cases}
\]
This allocation is nondecreasing and uses the same amount of resource as
$z^{FS}$.
Moreover,
\begin{align*}
    \int_0^1h(q)\bigl(\tilde z(q)-z^{FS}(q)\bigr)dq
    &=
    \frac12
    \left(
        \int_{q_0-\epsilon}^{q_0}h(q)dq
        -
        \int_{q_0}^{q_0+\epsilon}h(q)dq
    \right)\\
    &=
    \frac12\int_0^\epsilon
        \bigl(h(q_0-u)-h(q_0+u)\bigr)du
    >0.
\end{align*}
Hence, full screening is not efficient for $\bar m=1-q_0$.
\end{proof}

\subsection{Proof of Theorem~\ref{thm: NBUE expands}}

\begin{proof}
By direct calculation, we can verify that a cumulative distribution function $F$ is NBUE if and only if
\begin{equation}\label{eq: NBUE target equation}
    \int_t^{\bar{v}} \frac{F(t)(1 - F(v))}{1 - F(t)} dv \le \int_0^t (1 - F(v))dv
\end{equation}
for all $t \in (0, \bar{v})$. We will show that if \eqref{eq: NBUE target equation} is satisfied with $F = G^K$, then \eqref{eq: NBUE target equation} is also satisfied with $F = G^{K + 1}$.

Since the right-hand side of \eqref{eq: NBUE target equation} with $F = G^{K + 1}$ is larger than that with $F = G^K$, it suffices to show that the left-hand side of \eqref{eq: NBUE target equation} with $F = G^{K + 1}$ is smaller than that with $F = G^{K}$. Thus, the desired equation is
\begin{equation}
    \int_t^{\bar{v}} \left[\frac{G^K(t)(1 - G^K(v))}{1 - G^K(t)} - \frac{G^{K +1 }(t)(1 - G^{K + 1}(v))}{1 - G^{K + 1}(t)}\right] dv \ge 0.
\end{equation}
To this end, we will prove 
\begin{equation}\label{eq: NBUE target equation 2}
    \frac{G^K(t)(1 - G^K(v))}{1 - G^K(t)} - \frac{G^{K +1 }(t)(1 - G^{K + 1}(v))}{1 - G^{K + 1}(t)} \ge 0
\end{equation}
for all $v \in [t, \bar{v})$.

For notational simplicity, let $a = G(t)$ and $b = G(v)$. Then, \eqref{eq: NBUE target equation 2} can be expressed as
\begin{equation}
    \frac{a^K(1 - b^K)}{1 - a^K} - \frac{a^{K + 1}(1 - b^{K + 1})}{1 - a^{K + 1}} \ge 0.
\end{equation}
This inequality is equivalent to
\begin{equation}
    (1 - b^K)(1 - a^{K + 1}) - a(1 - b^{K + 1})(1 - a^K) \ge 0,
\end{equation}
or
\begin{equation}\label{eq: NBUE target equation 3}
    (1 + b + \dots + b^{K - 1})(1 + a + \dots + a^K) - a(1 + b + \dots + b^K)(1 + a + \dots + a^{K - 1}) \ge 0.
\end{equation}
Let $A = 1 + a + \dots + a^{K - 1}$, and $B = 1 + b + \dots + b^{K - 1}$. Then, \eqref{eq: NBUE target equation 3} becomes
\begin{equation}
    B(1 + aA) - a(1 + bB)A \ge 0,
\end{equation}
or equivalently,
\begin{equation}\label{eq: NBUE target equation 4}
    (B - aA) + aAB (1 - b) \ge 0.
\end{equation}
It follows from $a = G(t)$, $b = G(v)$, and $v \ge t$ that $a, b \in [0, 1]$ and $B \ge A \ge aA \ge 0$. Accordingly, \eqref{eq: NBUE target equation 4} is indeed satisfied, as desired.
\end{proof} 

\subsection{Proof of Theorem~\ref{thm: hazard rate increase}}
\begin{proof}
    Recall that
    \begin{equation}
        r(v; G_{K}) = \frac{g_{K}(v)}{1 - G_{K}(v)}= \frac{K g(v) (G(v))^{K
        - 1}}{1 - (G(v))^{K}}.
    \end{equation}
    By direct calculation, we can show that
    \begin{equation}
        r'(v; G_{K}) = \frac{K (G(v))^{K - 2}}{(1 - (G(v))^{K})^{2}}\left(K (
        g(v))^{2}- ((g(v))^{2}- g'(v)G(v))(1 - (G(v))^{K}) \right).
    \end{equation}
    Thus, $r'(v ; G_{K}) \ge 0$ if and only if
    \begin{equation}
        \label{eq: IHR proof}K (g(v))^{2}\ge ((g(v))^{2}- g'(v)G(v))(1 - (G (
        v))^{K}).
    \end{equation}

    The inequality \eqref{eq: IHR proof} trivially holds for all $K\ge 1$ if $g'(v) \ge 0$. In the following, we consider the case of $g'(v) < 0$.
    We show that, for any fixed $v$, if \eqref{eq: IHR proof} holds with $K$,
    then \eqref{eq: IHR proof} also holds with $K + 1$.

    Suppose that \eqref{eq: IHR proof} holds with $K$. Then, we have
    \begin{equation}
        K (g(v))^{2}\ge ((g(v))^{2}- g'(v)G(v))(1 - (G(v))^{K}).
    \end{equation}
    Thus,
    \begin{equation}
        (K + 1) (g(v))^{2}\ge ((g(v))^{2}- g'(v)G(v)) \frac{K + 1}{K}(1 - ( G
        (v))^{K}).
    \end{equation}
    Therefore, it suffices to show that
    \begin{equation}
        \frac{K + 1}{K}(1 - x^{K}) \ge 1 - x^{K + 1}
    \end{equation}
    for all $x \in [0, 1)$, or equivalently,
    \begin{equation}
        \label{eq: IHR proof 2}(K + 1) (1 - x^{K}) - K (1 - x^{K+1}) \ge 0.
    \end{equation}
    It follows from $1 - x^{K}= (1 - x)(x^{K-1}+ x^{K-2}+ \dots + x + 1)$ that
    \eqref{eq: IHR proof 2} holds if
    \begin{equation}
        (K + 1) (x^{K-1}+ x^{K-2}+ \dots + x + 1) - K (x^{K}+ x^{K-1}+ \dots
        + x + 1) \ge 0,
    \end{equation}
    or equivalently,
    \begin{equation}
        \label{eq: IHR proof 3}x^{K - 1}+ \dots + x + 1 \ge K x^{K}.
    \end{equation}
    Indeed, \eqref{eq: IHR proof 3} holds for all $x \in [0, 1]$ because $x^{k}
    \ge x^{K}$ holds for all $k = 0, 1, \dots, K - 1$, as desired.

    We next show that for all $v \in (0, \bar{v})$, there exists $K_{0}$ such
    that for all $K > K_{0}$, $r'(v; G_{K}) > 0$. The right-hand side of
    \eqref{eq: IHR proof} is bounded above by $((g(v))^{2}- g'(v)G(v))$.
    Since we assume that $G$ has full support, i.e., $g(v) > 0$ for all $v \in
    (0, \bar{v})$, there exists $K_{0}$ such that
    \begin{equation}
        K_{0}(g(v))^{2}> (g(v))^{2}- g'(v)G(v).
    \end{equation}
    Clearly, for all $K > K_{0}$, we have \eqref{eq: IHR proof}, implying
    that $r'(v; G_K) > 0$.
\end{proof}

\subsection{Proof of Theorem~\ref{thm: bounded support}}

\begin{proof}
    Take $\epsilon > 0$ and $\delta > 0$ arbitrarily. Let
    $\bar{\delta}= \min\{\delta, (1 - \bar{m})/2 \}$. Since $\bar{v}= \sup \operatorname{support}
    (G) < + \infty$, we have $G(\bar{v}- \epsilon) < 1$. Let $K_{0}$ be the smallest
    integer $K$ satisfying $G_{K}(\bar{v}- \epsilon) \coloneqq (G(\bar{v}- \epsilon
    ))^{K}< \bar{\delta}$, i.e.,
    $K_{0}= \lfloor \log\bar{\delta}/ \log G(\bar{v}- \epsilon) \rfloor + 1$. By
    construction of $K_{0}$, for any $K > K_{0}$, we have $G_{K}(\bar{v}- \epsilon
    ) < \bar{\delta}$, or equivalently, $\Pr(v \in (\bar{v}- \epsilon, \bar{v}
    )) > 1 - \bar{\delta}$.

    Given such $K_{0}$, for any $K > K_{0}$,
    \begin{align}
        RS(\mech_{SD}) & = \bar{m}\int_{0}^{\bar{v}}v dG_{K}(v)                                    \\
                       & > \bar{m}\int_{\bar{v} - \epsilon}^{\bar{v}}(\bar{v}- \epsilon) dG_{K}(v) \\
                       & = \bar{m}\Pr(v \in (\bar{v}- \epsilon, \bar{v})) (\bar{v}- \epsilon)      \\
                       & > \bar{m}(1 - \bar{\delta}) (\bar{v}- \epsilon)                           \\
                       & \ge \bar{m}(1 - \delta) (\bar{v}- \epsilon).
    \end{align}

    Next, we evaluate $RS(\mech_{VCG})$ with the same $K_{0}$. For any
    $K > K_{0}$, we have $G_{K}(\bar{v}- \epsilon) < \bar{\delta}< 1 - \bar{m}$.
    Accordingly, we have $q \coloneqq G_{K}^{-1}(1 - \bar{m}) > \bar{v}- \epsilon$.
    Accordingly,
    \begin{align}
        RS(\mech_{VCG}) & = \int_{q}^{\bar{v}}\left(v - q\right)dG_{K}(v)                                          \\
                        & < \int_{q}^{\bar{v}}\left(\bar{v}- (\bar{v}- \epsilon) \right)dG_{K}(v) \\
                        & = \epsilon (1 - G_{K}(q))                                                \\
                        & = \bar{m}\epsilon.    
    \end{align}
\end{proof}

\subsection{Proof of Theorem~\ref{cor: Gumbel conclusion 2}}

We first introduce two lemmas.

\begin{lemma}
    \label{lem: gumbel case hazard rate convergence} 
    If $G \in \vmc{0}$, then for any finite interval $[\underline{c}, \bar{c}]$, there exists $K_{0}$ such that for all $K > K_{0}$, we have $\vartheta'(w; \hat{G}_{K}) < 0$ for $w \in [\underline{c}, \bar{c}]$.
\end{lemma}

\begin{proof}
    Recall that $\vartheta(w; \hat{G}_{K}) \coloneqq (1 - \hat{G}_{K}(w))/\hat
    {g}_{K}(w)$.
    \begin{equation}
        \vartheta'(w; \hat{G}_{K}) = \left(\frac{1 - \hat{G}_{K}(w)}{\hat{g}_{K}(w)}
        \right)' = \frac{-(\hat{g}_{K}(w))^{2}- \hat{g}'_{K}(w)(1 - \hat{G}_{K}(w))}{(\hat{g}_{K}(w))^{2}}
        .
    \end{equation}
    $G$ has full support and is twice differentiable. Also, $\hat{G}_{K}, \hat{g}_{K}, \hat{g}'_{K}$ converge
    uniformly to $\Lambda, \lambda, \lambda'$ in $[\underline{c}, \bar{c}]$.
    Accordingly, for large enough $K$, there exists $\underline
    {g}$ such that $\hat{g}_{K}(w) > \underline{g}$ for all $w \in [\underline
    {c}, \bar{c}]$. Furthermore, for all $w \in [\underline{c}, \bar{c}]$,
    \begin{equation}
        \vartheta'(w; \hat{G}_K) \to \frac{-(\lambda(w))^{2}- \lambda'(w)(1 -
        \Lambda(w))}{(\lambda(w))^{2}}= \vartheta'(w; \Lambda) < 0,
    \end{equation}
    where the convergence is uniform in $[\underline{c}, \bar{c}]$. Hence,
    there exists $K_{0}$ such that for all $K > K_{0}$, we have $\vartheta'(w
    ; \hat{G}_{K}) < 0$ for $w \in [\underline{c}, \bar{c}]$.
\end{proof} 

\begin{lemma}\label{lem: flat allocation} 
    Let $[a,b]$ be an interval on which the virtual
    valuation $\vartheta(\cdot; \hat{G}_{K})$ is nonincreasing, and $x$ be
    the allocation rule of any strategy-proof mechanism. Then, the
    allocation rule
    \begin{equation}
        \tilde{x}(w) \coloneqq
        \begin{cases}
            \dfrac{\displaystyle \int_a^b x(w')\hat{g}_K(w')dw'}{\hat{G}_K(b)-\hat{G}_K(a)} & \text{ if } w \in [a,b], \\
            x(w)                                                                            & \text{otherwise},
        \end{cases}
    \end{equation}
    is a monotonic allocation rule that satisfies the resource constraint. Furthermore, its residual surplus is no less than that of $x$.
\end{lemma}

This lemma is a basic step in deriving revenue-maximizing mechanisms and is well known. We include it here for completeness. The proof follows the approach
outlined in \citet{Fu2017myerson}.

\begin{proof}
    Since $x$ is an allocation rule of a strategy-proof mechanism, $x$ is
    monotonic; thus, $\tilde{x}$ is also monotonic.

    Next, we show that $\tilde{x}$ satisfies the resource constraint.
    \begin{align}
         & \int_{-\infty}^{\infty}\tilde{x}(w)\hat{g}_{K}(w) dw                                                                  \\
         & = \int_{a}^{b}\tilde{x}(w)\hat{g}_{K}(w) dw + \int_{w \notin [a, b]}x(w)\hat{g}_{K}(w) dw                              \\
         & = (\hat{G}_{K}(b) - \hat{G}_{K}(a))\dfrac{\int_a^b x(w)\hat{g}_K(w)dw}{\hat{G}_{K}(b)-\hat{G}_{K}(a)}+ \int_{w \notin [a, b]}x(w)\hat{g}_{K}(w) dw \\
         & = \int_{-\infty}^{\infty}x(w)\hat{g}_{K}(w) dw.
    \end{align}
    Since $x$ satisfies the resource constraint, $\tilde{x}$ also satisfies it.

    For $w \notin [a, b]$, $x$ and $\tilde{x}$ are identical and hence
    generate the same residual surplus. On $[a, b]$, let
    $\hat{G}_{K}(\cdot | w \in [a, b])$ denote the conditional distribution of $w$
    given $w \in [a, b]$, which has density $\hat{g}_{K}/(\hat{G}_{K}(b) - \hat{G}_{K}(a))$.
    Then,
    \begin{align}
         & \int_{a}^{b}\tilde{x}(w)\vartheta(w; \hat{G}_{K})\hat{g}_{K}(w)dw \\
         & = \dfrac{\int_a^b x(w)\hat{g}_K(w)dw}{\hat{G}_K(b)-\hat{G}_K(a)}\int_{a}^{b}\vartheta(w; \hat{G}_{K}) \hat{g}_{K}(w)dw                                                 \\
         & = \mathbb{E}_{w \sim \hat{G}_K(\cdot|w \in [a, b])}[x(w)]\mathbb{E}_{w \sim \hat{G}_K(\cdot|w \in [a, b])}[\vartheta(w; \hat{G}_{K})](\hat{G}_{K}(b) - \hat{G}_{K}(a)) \\
         & \ge \mathbb{E}_{w \sim \hat{G}_K(\cdot|w \in [a, b])}[x(w)\vartheta(w; \hat{G}_{K})](\hat{G}_{K}(b) - \hat{G}_{K}(a))                                            \\
         & = \int_{a}^{b}x(w)\vartheta(w; \hat{G}_{K})\hat{g}_{K}(w)dw.
    \end{align}
    The inequality is an application of the Harris inequality, which states that
    for any nondecreasing function $\zeta$ and nonincreasing function $\xi$ on
    $\mathbb{R}$ and any probability measure on $\mathbb{R}$, we have $\mathbb{E}
    [\zeta\xi] \le \mathbb{E}[\zeta]\mathbb{E}[\xi]$.
\end{proof} 

\begin{proof}[Proof of Theorem~\ref{cor: Gumbel conclusion 2}]
    Given any $\epsilon > 0$, take $\bar{c}$ and $\underline{c}$ to satisfy $1 - \epsilon/2 < (\Lambda(\bar{c}) - \Lambda(\underline{c}))$. Then, since $\hat{G}_{K}(w) \to \Lambda (w)$ as $K \to \infty$, there exists $K_{1}$ such that for all $K > K_{1}$, we have $1 - \epsilon < \hat{G}_{K}(\bar{c}) - \hat{G}_{K}(\underline{c})$. Lemmas~\ref{lem: gumbel case hazard rate convergence} and \ref{lem: flat allocation} imply that there also exists $K_{2}$ such that for all $K > K_{2}$, for all $w \in [\underline{c}, \bar{c}]$, $\hat{x}(w) = \bar{x}_K$ for some $\bar{x}_K \in [0, 1]$. Let $K_{0} = \max\{K_1, K_2\}$ and fix $K>K_0$. Then, clearly we have $\Pr(\hat{x}(w) = \bar{x}_K) > 1 - \epsilon$. Furthermore, the resource constraint implies that
    \begin{equation}
        \bar{m}= \int_{-\infty}^{\infty}\hat{x}(w)\hat{g}_{K}(w)dw \ge \int_{\underline{c}}
        ^{\bar{c}}\hat{x}(w)\hat{g}_{K}(w)dw > \bar{x}_K(1 - \epsilon),
    \end{equation}
    or equivalently, $\bar{x}_K< \bar{m}/(1 - \epsilon)$. The resource constraint also implies
    \begin{equation}
        \bar{m}= \int_{-\infty}^{\infty}\hat{x}(w)\hat{g}_{K}(w)dw \le \int_{\underline{c}}
        ^{\bar{c}}\hat{x}(w)\hat{g}_{K}(w)dw + \int_{w \notin [\underline{c},\bar{c}]}
        \hat{g}_{K}(w)dw = (1 - p)\bar{x}_K+ p,
    \end{equation}
    where $p = \Pr(w \notin [\underline{c}, \bar{c}]) < \epsilon$. The above inequality can be rewritten as $\bar{x}_K\ge (\bar{m}- p)/(1 - p)$. The right-hand side is larger than $(\bar{m}- \epsilon)/(1 - \epsilon)$ because it is decreasing in $p$ and $p < \epsilon$. Accordingly,
    $\bar{x}_K> (\bar{m}- \epsilon)/(1 - \epsilon)$.
\end{proof}

\subsection{Proof of Proposition~\ref{prop: IDHR efficient mechanism}}\label{subsec: IDHR efficient mechanism}

\begin{proof}
We apply the construction of the ironed virtual value established in Section~6 of \citet{myerson1981optimal} to a distribution function $G_{K}$.

\begin{definition}[Ironed Virtual Value]\label{defn: ironed virtual value}
    The ironed virtual value $\bar{\vartheta}$ is constructed as follows:
    \begin{enumerate}
        \item For $q \in [0, 1]$, define
            $h(q) = \vartheta(G_{K}^{-1}(q); G_{K})$.
        \item Define $\mathcal H(q) = \int_{0}^{q}h(r) dr$.
        \item Define $\check{\mathcal H}$ as the convex hull of $\mathcal H$, the largest convex
            function bounded above by $\mathcal H$ for all $q \in [0, 1]$.
        \item Define $\check{h}(q)$ as the derivative of $\check{\mathcal H}(q)$, where defined, and extend
            to all of $[0, 1]$ by right continuity.
        \item $\bar{\vartheta}(v; G_{K}) = \check{h}(G_{K}(v))$.
    \end{enumerate}
\end{definition}

    If $\hat{G}_K$ is NBUE, then by Theorem~\ref{thm: NBUE iff no screening}, a no-screening mechanism is efficient. In this case, we may set $w^{**}=+\infty$ and take $\hat{x}(w)\equiv \bar{m}$. 
    
    In the following, we consider the case in which $\hat{G}_K$ is not NBUE. We first prove the following lemma.
    \begin{lemma}
    \label{lem: idhr virtual value} 
    Suppose that the reduced value distribution $\hat{G}_{K}$ has an IDHR and is not NBUE. Then, the ironed virtual value $\bar{\vartheta}$ is given by
    \begin{equation}
        \label{eq: ironed virtual value}
        \bar{\vartheta}(w; \hat{G}_{K}) =
        \begin{cases}
            \vartheta(w^{**}; \hat{G}_{K}) & \text{ for }w \in [0, w^{**}]       \\
            \vartheta(w; \hat{G}_{K})      & \text{ for }w \in (w^{**}, +\infty)
        \end{cases}
    \end{equation}
    where $w^{**}$ is the solution of the following equation:
    \begin{equation}
        \label{eq: vdstar}
        \frac{\displaystyle \int_{0}^{w}\vartheta(w'; \hat{G}_{K})\hat{g}_{K}(w')dw'}{\hat{G}_{K}(w)}= \vartheta(w; \hat{G}_{K}).
    \end{equation}
    \end{lemma}

    \begin{proof}
    Since $\hat{G}_{K}$ has an IDHR, $\mathcal H$ is concave on $(0,\hat{G}_{K}(w^*))$ and convex on $(\hat{G}_{K}(w^{*}), 1)$. Since $\hat{G}_K$ is not NBUE, the convex hull $\check{\mathcal H}$ cannot coincide with the line segment connecting $(0, 0)$ and $(1, \mathcal H(1))$, and therefore the supporting line from the origin touches $\mathcal H$ at an interior point $q^{**} \in (0, 1)$; thus, we have the following:
    \begin{equation}
        \check{\mathcal H}(q) =
        \begin{cases}
            \dfrac{\mathcal H(q^{**})}{q^{**}}\cdot q & \text{ for }q < q^{**},   \\
            \mathcal H(q)                             & \text{ for }q\ge q^{**}.
        \end{cases}
    \end{equation}
    This implies that the ironed virtual value is given by \eqref{eq: ironed virtual value}, where $w^{**}= \hat{G}_{K}^{-1}(q^{**})$.

    Furthermore, since $q^{**}$ is defined as the point of tangency, it
    solves $\mathcal H(q)/q = h(q)$, or equivalently,
    \begin{equation}
        \frac{\displaystyle \int_{0}^{q}\vartheta(\hat{G}_{K}^{-1}(r); \hat{G}_K)dr}{q}= \vartheta(\hat{G}_{K}^{-1}(q); \hat{G}_K).
    \end{equation}
    Finally, we obtain \eqref{eq: vdstar} by replacing $q$ with $\hat{G}_{K}(w)$
    and applying the substitution $r = \hat{G}_{K}(w')$ to the integral in the left-hand
    side.
    \end{proof} 

    By Theorem 2.8 of \citet{Hartline2008}, an efficient mechanism maximizes the value of $\hat{x}(w)$ in regions where $\bar{\vartheta}(w; \hat{G}_K)$ is larger while fulfilling the resource constraint and the unit demand condition. The conclusion is immediate from this.
\end{proof}

\subsection{Proof of Theorem~\ref{thm: frechet conclusion}}

\begin{proof}
    By Proposition~\ref{prop: Pickands twice diff convergence}, for any $0< \underline{c} < \bar{c}<\infty$, $\hat{G}_K \to \Phi_\alpha$, $\hat{g}_K \to \phi_\alpha$, $\hat{g}'_K \to \phi'_\alpha$ uniformly on $[\underline c,\bar c]$.
    Since the hazard-rate derivative $r'(w;F)$ is a continuous functional of $(F,f,f')$ as long as $1 - F$ is bounded away from $0$, we obtain $r'(w;\hat{G}_K)\to r'(w;\Phi_\alpha)$ uniformly on $[\underline{c}, \bar{c}]$.

    Since $\Phi_\alpha$ has an IDHR, there exists $w^* > 0$ such that $r'(w; \Phi_\alpha) > 0$ for $w < w^*$ and $r'(w; \Phi_\alpha) < 0$ for $w > w^*$.
    We choose $\underline{c}$ and $\bar{c}$ such that $0 < \underline{c} < \min\{\epsilon, w^*\}$ and $\bar{c} > w^{**} > w^*$, where $w^{**}$ is the unique solution to \eqref{eq: frechet vdstar equation}.
    Then, for any small $\delta>0$, for all sufficiently large $K$, $r'(w; \hat{G}_K) > 0$ on $[\underline{c}, w^*-\delta]$ and $r'(w; \hat{G}_K)<0$ on $[w^*+\delta, \bar{c}]$.
    
    Since $\vartheta(\cdot; \hat{G}_{K})$ is decreasing on $[\underline{c}, w^{*}- \delta)$, the efficient allocation rule $\hat{x}$ must be constant on an interval starting from (or before) $\underline{c}$. 
    Specifically, there exist $w_K^{\dagger} > w^* - \delta$ and a constant $\bar{x}_K \in [0, 1]$ such that $\hat{x}(w) = \bar{x}_K$ for all $w \in [\underline{c}, w_K^{\dagger})$, and $\hat{x}(w) > \bar{x}_K$ for $w > w_K^{\dagger}$.

    To evaluate the location of $w_K^{\dagger}$, we define the auxiliary function $\Theta$. For a distribution $\hat{G}$, let
    \begin{equation}
        \Theta(w; \hat{G}) \coloneqq \dfrac{\displaystyle \int_{0}^{w}\vartheta(t; \hat{G})\hat{g}(t) dt}{\hat{G}(w)} - \vartheta(w; \hat{G}).
    \end{equation}
    Note that $w^{**}$ is the unique solution to $\Theta(w; \Phi_\alpha) = 0$.
    Furthermore, for a lower bound $\underline{c} > 0$, define the truncated version $\Theta_{\underline{c}}$ by
    \begin{equation}
        \Theta_{\underline{c}}(w; \hat{G}) \coloneqq \dfrac{\displaystyle \int_{\underline{c}}^{w}\vartheta(t; \hat{G})\hat{g}(t) dt}{\hat{G}(w)-\hat{G}(\underline{c})} - \vartheta(w; \hat{G}) = \dfrac{\displaystyle \int_{\underline{c}}^w (1 - \hat{G}(t))dt}{\hat{G}(w)-\hat{G}(\underline{c})} - \vartheta(w; \hat{G}).
    \end{equation}
    
    Due to the uniform convergence of $\Theta_{\underline{c}}(\cdot; \hat{G}_K)$ and its derivative to those of $\Theta_{\underline{c}}(\cdot; \Phi_\alpha)$ on $(\underline{c}, \bar{c}]$, combined with the IDHR property of $\Phi_\alpha$, $\Theta_{\underline{c}}(w; \hat{G}_K) = 0$ has a unique solution $w_{K, \underline{c}}^{\dagger\dagger}$ for sufficiently large $K$.
    Analogous to the proof of Lemma~\ref{lem: idhr virtual value}, the definition of $w_{K, \underline{c}}^{\dagger\dagger}$ ensures that the line segment connecting $(\hat{G}_{K}(\underline{c}), \mathcal H(\hat{G}_{K}(\underline{c})))$ and $(\hat{G}_{K}(w_{K, \underline{c}}^{\dagger\dagger}), \mathcal H(\hat{G}_{K}(w_{K, \underline{c}}^{\dagger\dagger})))$ is tangent to $\mathcal H$ at the latter point. Due to the IDHR structure, this line segment lies below $\mathcal H$ on the interval $[\underline{c}, w_{K, \underline{c}}^{\dagger\dagger}]$. Since this line segment connects two points on the graph of $\mathcal H$, the ironed cumulative virtual value $\check{\mathcal H}$, being the global convex hull of $\mathcal H$, must lie below it. Consequently, $\check{\mathcal H}$ is strictly less than $\mathcal H$ on the interior of the interval. By the property of the convex hull, $\check{\mathcal H}$ must be a line segment on any interval where it strictly detaches from $\mathcal H$. Thus, the ironed virtual value is constant on $[\underline{c}, w_{K, \underline{c}}^{\dagger\dagger}]$.
    Since $w_K^{\dagger}$ is the right endpoint of the constant allocation interval starting from (or before) $\underline{c}$ for the original distribution, it must satisfy $w_K^{\dagger} \ge w_{K, \underline{c}}^{\dagger\dagger}$.

    We now show that $w_{K, \underline{c}}^{\dagger\dagger}$ approaches $w^{**}$ for sufficiently small $\underline{c}$ and large $K$.
    Since $\Phi_\alpha$ has an IDHR, $\Theta(\cdot; \Phi_\alpha)$ is strictly decreasing in a neighborhood of $w^{**}$. 
    Thus, for the fixed $\epsilon$, there exists $\eta > 0$ such that
    \begin{equation}
        \Theta(w^{**} - \epsilon/2; \Phi_\alpha) > \eta \quad \text{and} \quad \Theta(w^{**} + \epsilon/2; \Phi_\alpha) < - \eta.
    \end{equation}

    First, consider the approximation by $\Theta_{\underline{c}}$. 
    As $\underline{c} \to 0$, $\int_{\underline{c}}^w (1 - \Phi_\alpha(t))dt \to \int_0^w (1 - \Phi_\alpha(t))dt$ and $\Phi_\alpha(\underline{c}) \to 0$. 
    For $w$ in the neighborhood of $w^{**}$, $\Phi_\alpha(w) - \Phi_\alpha(\underline{c})$ is strictly positive and bounded away from zero for sufficiently small $\underline{c}$.
    Thus, $\Theta_{\underline{c}}(w; \Phi_\alpha) \to \Theta(w; \Phi_\alpha)$ for any $w$ in the neighborhood of $w^{**}$.
    We can choose $\underline{c} < \min\{\epsilon, w^*\}$ small enough that for $w \in \{w^{**} - \epsilon/2, w^{**} + \epsilon/2\}$,
    \begin{equation}
        |\Theta_{\underline{c}}(w; \Phi_\alpha) - \Theta(w; \Phi_\alpha)| < \eta/3.
    \end{equation}
    Fix such $\underline{c}$.

    Next, consider the convergence as $K\to\infty$. 
    Since $\hat{G}_K \to \Phi_\alpha$ uniformly on $[\underline{c}, \bar{c}]$ and $\Phi_\alpha(w) - \Phi_\alpha(\underline{c})$ is bounded away from zero for $w$ in the neighborhood of $w^{**}$, $\Theta_{\underline{c}}(\cdot; \hat{G}_K)$ converges uniformly to $\Theta_{\underline{c}}(\cdot; \Phi_\alpha)$ on the neighborhood of $w^{**}$.
    Consequently, there exists $K_0$ such that for all $K > K_0$,
    \begin{equation}
        |\Theta_{\underline{c}}(w; \hat{G}_K) - \Theta_{\underline{c}}(w; \Phi_\alpha)| < \eta/3 \quad \text{for } w \in \{w^{**} - \epsilon/2, w^{**} + \epsilon/2\}.
    \end{equation}

    Combining these evaluations, for $K > K_0$:
    \begin{align}
        \Theta_{\underline{c}}(w^{**} - \epsilon/2; \hat{G}_K) & > \Theta(w^{**} - \epsilon/2; \Phi_\alpha) - 2\eta/3 > \eta/3 > 0, \\
        \Theta_{\underline{c}}(w^{**} + \epsilon/2; \hat{G}_K) & < \Theta(w^{**} + \epsilon/2; \Phi_\alpha) + 2\eta/3 < -\eta/3 < 0.
    \end{align}
    By the intermediate value theorem, the root $w_{K, \underline{c}}^{\dagger\dagger}$ lies in $(w^{**} - \epsilon/2, w^{**} + \epsilon/2)$.

    Finally, we have
    \begin{equation}
        w_K^{\dagger} \ge w_{K, \underline{c}}^{\dagger\dagger} > w^{**} - \epsilon/2 > w^{**} - \epsilon.
    \end{equation}
    Since $\hat{x}(w) = \bar{x}_K$ for $w \in [\underline{c}, w_K^{\dagger})$ and $\underline{c} < \epsilon$, it follows that $\hat{x}(w) = \bar{x}_K$ for all $w \in (\epsilon, w^{**} - \epsilon)$.
\end{proof}

\subsection{Proof of Proposition~\ref{thm: between-agent correlation first best}}

\begin{proof}
First, it is straightforward to see that, for all $\rho$,
\begin{align}
    RS(\mech_{SD}; F^\rho) &\ge \frac{1}{I}\mathbb{E}\left[\sum_{k=1}^K m_k \min_{i\in I} v^i_k\right]
      = \frac{1}{I}\sum_{k=1}^K m_k \mathbb{E}\left[\min_{i \in I} v^i_k\right], \\
    RS(\mech_{FB}^\rho; F^\rho) &\le \frac{1}{I}\mathbb{E}\left[\sum_{k=1}^K m_k \max_{i\in I} v^i_k\right] = \frac{1}{I}\sum_{k=1}^K m_k \mathbb{E}\left[\max_{i \in I} v^i_k\right].
\end{align}
Furthermore, since we assume that the goods are scarce (i.e., $I > \sum_{k \in K}m_k$), for each $\bv$, there exists an agent $i^*(\bv) \in I$ who is not allocated. Under VCG, each agent who obtains object $k$ must pay at least $v_k^{i^*(\bv)}$. Accordingly,
\begin{equation}
    \mathbb{E}[\text{(VCG payment)}] \ge \frac{1}{I}\sum_{k=1}^K m_k \mathbb{E}\left[v^{i^*(\bv)}_k\right] \ge \frac{1}{I}\sum_{k=1}^K m_k \mathbb{E}\left[\min_{i\in I} v^i_k\right]
\end{equation}

Therefore,
\begin{gather}
    \frac{\sum_{k \in K} m_k \mathbb{E}[\min_i v^i_k]}{\sum_{k \in K} m_k \mathbb{E}[\max_i v^i_k]} \le \frac{RS(\mech_{SD}; F^\rho)}{RS(\mech_{FB}^\rho; F^\rho)} \le 1,\\
    0 \le RS(\mech_{VCG}; F^\rho) \le \frac{1}{I}\sum_{k=1}^K m_k \left(\mathbb{E}\left[\max_{i\in I} v^i_k\right] - \mathbb{E}\left[\min_{i\in I} v^i_k\right] \right).
\end{gather}
Accordingly, it suffices to show that for each object $k$, $\mathbb{E}[\min_{i \in I} v_k^i] \to \mu_k$ and $\mathbb{E}[\max_{i \in I} v_k^i] \to \mu_k$ as $\rho \to 1$.

For each $k \in K$, let $F_k^{\max,\rho}$ denote the distribution of $\max_{i \in I} v^i_k$ and $F_k^{\min,\rho}$ that of $\min_{i \in I} v^i_k$.  
Using standard integral identities,
\begin{equation}
    \mathbb{E}\left[\max_{i \in I} v^i_k\right] = \int_{0}^{\infty} (1-F_k^{\max,\rho}(w))dw, \qquad \mathbb{E}\left[\min_{i \in I} v^i_k\right] = \int_{0}^{\infty} (1-F_k^{\min,\rho}(w))dw.
\end{equation}

Let $S_k(v) \coloneqq I(1 - F_k(v))$. Since $\int_0^\infty|S_k(w)|dw = I \mu_k < \infty$, $S_k$ is integrable. Furthermore, we have (i) $1 - F_k^{\min,\rho}(w) \le 1 - F_k^{\max,\rho}(w)$ by definition, and (ii) $1 - F_k^{\max,\rho}(w) \le S_k(w)$ by the union bound. Thus, once we show $F_k^{\max,\rho}(w)\to F_k(w)$ and $F_k^{\min,\rho}(w)\to F_k(w)$ at all continuity points of $F_k$, then the dominated convergence theorem implies $\mathbb{E}[\max_i v^i_k]\to \mu_k$ and $\mathbb{E}[\min_i v^i_k]\to \mu_k$.

Finally, we prove pointwise convergence. Let $X^i_k \coloneqq (v^i_k-\mu_k)/\sigma_k$ so that for all $i \in I$, $k \in K$, and $j \neq i$, we have $\mathbb{E}[X^i_k]=0$, $\mathrm{Var}(X^i_k)=1$, and $\mathrm{Cov}(X^i_k,X^j_k)\ge \rho$. Define the differences $D^i_k \coloneqq X^1_k-X^i_k$. Then, $\mathbb{E}[D^i_k]=0$ and $ \mathrm{Var}(D^i_k)=2-2\mathrm{Cov}(X^1_k,X^i_k)\le 2(1-\rho)$.

By Chebyshev's inequality, for all $k$ and $\varepsilon'>0$,
\begin{align}
    \Pr\left(|D^i_k|\ge \varepsilon'\right)
    &\le\frac{\mathrm{Var}(D^i_k)}{\varepsilon'^2} \\
    &\le \frac{2(1-\rho)}{\varepsilon'^2}
      \xrightarrow[\rho\to 1]{} 0 .
\end{align}

Let $M_k \coloneqq \max_{i\ne 1}|v^1_k-v^i_k|$. Then, by the union bound, 
\begin{align}
    \Pr\left(M_k\ge \varepsilon\right)
    &\le \sum_{i\ne 1}
        \Pr\left(|v^1_k-v^i_k|\ge\varepsilon\right) \\
    &= \sum_{i\ne 1}
        \Pr\left(|D^i_k|\ge \frac{\varepsilon}{\sigma_k}\right)
      \xrightarrow[\rho\to 1]{} 0 .
\end{align}

Fix a continuity point $v$ of $F_k$. Then
\begin{align}
    |F_k^{\max,\rho}(v)-F_k(v)|
    &= \Pr\!\left(\max_i v^i_k>v \text{ and } v^1_k\le v\right) \\
    &\le \Pr(M_k\ge \varepsilon)
       + \Pr\!\left(M_k<\varepsilon \text{ and } v^1_k\in(v-\varepsilon,v+\varepsilon)\right)\\
    &\le \Pr(M_k\ge \varepsilon)
       + \big(F_k(v+\varepsilon)-F_k(v-\varepsilon)\big).
\end{align}
The first term vanishes as $\rho\to 1$, and the second term can be made arbitrarily small by choosing $\varepsilon$. Thus $F_k^{\max,\rho}(v)\to F_k(v)$. We can prove that $F_k^{\min,\rho}(v)\to F_k(v)$ in a similar manner.
\end{proof}

\subsection{Derivation of the Efficient Threshold for a Fr\'echet Case}
\label{subsec: derivation of frechet vdstar equation}

We transform \eqref{eq: vdstar} to derive \eqref{eq: frechet vdstar equation}
as follows.
\begin{align}
    0 & = \dfrac{\int_0^w \vartheta(w'; \Phi_\alpha) \phi_\alpha(w')dw'}{\Phi_\alpha(w)}- \vartheta(w; \Phi_{\alpha}),                                                       \\
    0 & = \dfrac{\int_0^w (1-\Phi_\alpha(w'))dw'}{\Phi_\alpha(w)}- \frac{1-\Phi_{\alpha}(w)}{\phi_{\alpha}(w)},                                                             \\
    0 & = \int_{0}^{w}(1-e^{-w'^{-\alpha}})dw' - \frac{1-e^{-w^{-\alpha}}}{\alpha w^{-\alpha-1}},                                                                           \\
    0 & = \left[w'-\frac{1}{\alpha}\Gamma\left(-\frac{1}{\alpha},w'^{-\alpha}\right)\right]^{w}_{0}-\frac{1-e^{-w^{-\alpha}}}{\alpha w^{-\alpha-1}},                        \\
    0 & = w-\frac{1}{\alpha}\Gamma\left(-\frac{1}{\alpha},w^{-\alpha}\right)-\frac{1-e^{-w^{-\alpha}}}{\alpha w^{-\alpha-1}},                                             \\
    0 & = w - \frac{1}{\alpha}(-\alpha)\left[\Gamma\left(1-\frac{1}{\alpha}, w^{-\alpha}\right)-we^{-w^{-\alpha}}\right]-\frac{1-e^{-w^{-\alpha}}}{\alpha w^{-\alpha-1}}, \\
    0 & = (1-e^{-w^{-\alpha}})w + \Gamma\left(\frac{\alpha-1}{\alpha}, w^{-\alpha}\right)-\frac{1}{\alpha}w^{\alpha+1}(1-e^{-w^{-\alpha}}).
\end{align}

\subsection{Random Favorite (RF) Mechanism}\label{subsec: RF Mechanism}
We characterize the RF mechanism for the setting with a continuous i.i.d.\ market and two object types: $K = 2$.
The marginal value distribution is standard exponential: $G(v) = 1 - e^{- v}$, and the capacities of objects $1$ and $2$ are given by $(m_{1}, m_{2})$, where $m_{1}\ge m_{2}> 0$ and $m_{1}+ m_{2}\le 1$.

RF offers two options: without making any payments, (i) obtain object $1$ with
probability $a \in (0, 1]$ and obtain object $2$ with zero probability, or
(ii) obtain object $2$ with probability $b \in (0, 1]$ and obtain object $1$
with zero probability. An agent claims object $1$ if $a v_{1}\ge b v_{2}$,
or equivalently, $a v_{1}/b \ge v_{2}$, and claims object $2$ otherwise.

Conditional on $v_{1}$, the probability that an agent claims object $2$ is
$1 - G(a v_{1}/ b) = e^{- a v_1/b}$. Thus, the total mass of agents who
claim object $2$ is
\begin{equation}
    \int_{0}^{\infty}e^{- \frac{a}{b} v_1}e^{- v_1}dv_{1}= \frac{b}{a + b}.
\end{equation}
Accordingly, given $(a, b)$, the demand for object $1$ is $a^{2}/(a + b)$
and the demand for object $2$ is $b^{2}/(a + b)$. The market-clearing condition
requires that $a^{2}/(a + b) = m_{1}$ and $b^{2}/(a + b) = m_{2}$. Solving these equations, we obtain
\begin{equation}
a = m_{1}+ \sqrt{m_{1}m_{2}}, \quad b = m_{2}+ \sqrt{m_{1}m_{2}},
\end{equation}
which constitute the unique solution in $(0,1]^2$ satisfying the agent's
optimality condition and the market-clearing condition whenever
$m_1 + \sqrt{m_1 m_2} \le 1$. We focus on the case in which such an interior solution exists.

In this interior region, the residual surplus achieved by RF, denoted by $\mech_{RF}$, is given by
\begin{align}
     & RS(\mech_{RF})                                                                                                                             \\
     & = a \int_{0}^{\infty}v_{1}(1 - e^{- \frac{a}{b} v_1}) e^{- v_1}dv_{1}+ b \int_{0}^{\infty}v_{2}(1 - e^{- \frac{b}{a} v_2}) e^{- v_2}dv_{2} \\
     & = a \left(1 - \frac{b^{2}}{(a + b)^{2}}\right) + b \left( 1 - \frac{a^{2}}{(a + b)^{2}}\right)                                             \\
     & = (a + b) - \frac{ab(a + b)}{(a + b)^{2}}                                                                                                  \\
     & = m_{1}+ m_{2}+ 2 \sqrt{m_{1}m_{2}}- \sqrt{m_{1}m_{2}}                                                                                     \\
     & = m_{1}+ m_{2}+ \sqrt{m_{1}m_{2}}.
\end{align}

Next, we compute the residual surplus achieved by SD. Since $m_{1}\ge m_{2}$
and objects $1$ and $2$ are equally popular, in the beginning, all agents
obtain their favorite objects, and object $2$ is exhausted after a mass $2 m_2$ of agents has made a choice. Afterward, all agents
obtain object $1$, and after an additional mass $m_1 - m_2$ agents has made a choice, object $1$ is also exhausted. Accordingly, the residual surplus
achieved by SD is
\begin{align}
     & RS(\mech_{SD})                                                                                \\
     & = 2 m_{2}\int_{0}^{\infty}v g_{2}(v) dv + (m_{1}- m_{2}) \int_{0}^{\infty}v g(v) dv           \\
     & = 2 m_{2}\int_{0}^{\infty}v (2e^{-v}- 2 e^{-2v})dv + (m_{1}- m_{2}) \int_{0}^{\infty}v e^{-v}dv \\
     & = 3 m_{2}+ (m_{1}- m_{2})                                                                     \\
     & = m_{1}+ 2 m_{2}.
\end{align}

Since we assume $m_{1}\ge m_{2}$, within the interior region, $RS(\mech_{RF}) \ge RS(\mech_{SD})$ always
holds, and the equality holds if and only if $m_{1}= m_{2}$. In Figure~\ref{fig:exp_asym_endowment_utildiff},
we plot
\begin{equation}
    \frac{RS(\mech_{RF}) - RS(\mech_{SD})}{RS(\mech_{RF})}= \frac{\sqrt{m_1
    m_2}- \min\{m_{1}, m_{2}\}}{m_{1}+ m_{2}+ \sqrt{m_1 m_2}}
\end{equation}
to illustrate the performance difference between RF and SD.

\end{document}